\documentclass[a4paper,12pt]{article}
\usepackage{jheppub} 
\usepackage{latexsym,graphicx,amssymb,color,amsmath}
\usepackage{relsize,ccaption,url,latexsym,epsfig,a4wide}
\usepackage{pdfsync}
\usepackage[numbers,sort&compress]{natbib}
\usepackage{hyperref}
\usepackage{relsize}
\usepackage{epstopdf}
\usepackage{color}
\definecolor{cherry}{rgb}{0.9,.1,.2}
\definecolor{orange}{rgb}{1,0.5,0}

\usepackage{tikz}
\usetikzlibrary{matrix,arrows,decorations.pathmorphing}

\usepackage[all]{xy}
\numberwithin{equation}{section}
\newcommand{\be}{\begin{equation}}
\newcommand{\ee}{\end{equation}}
\newcommand{\ba}{\begin{eqnarray}}
\newcommand{\ea}{\end{eqnarray}}
\newcommand{\hf}{\frac{1}{2}}
 
\newcommand\req[1]{(\ref{#1})}

\newcommand\nop[1]{\mathop{:\!#1\!\!:}}

\newcommand\F{\mathbb{F}}
\newcommand\GGG{\mathcal G}
\newcommand\HHH{\mathcal H}
\newcommand\C{\mathbb{C}}

\newcommand\R{\mathbb{R}}
\newcommand\Z{\mathbb{Z}}
\newcommand{\NNN}{{\cal N}}

\def\Tr{\,{\rm Tr}\, }
\DeclareMathOperator{\ch}{ch}

\newcommand\eps{\varepsilon}

\newcommand\fa{\forall\,}
\newcommand\m{\mathcal}
\newcommand\qu{\overline}

\newcommand\wt{\widetilde}
\newcommand\wh{\widehat}

\def\sqr#1#2{{\vcenter{\vbox{\hrule height.#2pt
           \hbox{\vrule width.#2pt height#1pt \kern#1pt
                 \vrule width.#2pt}\hrule height.#2pt}}}}
\renewcommand\square{
\mathop{\mathchoice{\sqr{20}{25}}{\sqr{12}{15}}{\sqr{8}{10}}{\sqr{4}{5}}}}
\def\orbox#1#2{{\scriptstyle{#1}}\square_{#2}\limits}

\captionnamefont{\bfseries}
\title{%\mbox{}\hfill {\small DCPT-13/**}\\[40pt]
\boldmath A K3 sigma model with $\Z_2^8:\mathbb{M}_{20}$ symmetry}

\author[a]{Matthias R.\ Gaberdiel,}
\author[b]{Anne Taormina,}
\author[c]{Roberto Volpato,}
\author[d]{and Katrin Wendland}

\affiliation[a]{Institut f\"ur Theoretische Physik, ETH Zurich, 8093 Zurich, Switzerland}
\affiliation[b]{Centre for Particle Theory, Durham University, Durham, DH1 3LE, U.K. }
\affiliation[c]{Max-Planck-Institut f\"ur Gravitationsphysik, Am M\"uhlenberg\,1, D-14476 Golm, Germany}
\affiliation[d]{Mathematics Institute, University of Freiburg, D-79104 Freiburg, Germany}

\emailAdd{gaberdiel@itp.phys.ethz.ch}
\emailAdd{anne.taormina@durham.ac.uk}
\emailAdd{roberto.volpato@aei.mpg.de}
\emailAdd{katrin.wendland@math.uni-freiburg.de}

\abstract{
The K3 sigma model based on the $\mathbb{Z}_2$-orbifold of the $D_4$-torus theory is studied. It is shown
that it has an equivalent description in terms of twelve free Majorana fermions, 
or as a rational conformal field theory based on the affine algebra $\widehat{\mathfrak s\mathfrak u}(2)^6$. By combining these
different viewpoints we show that the ${\cal N}=(4,4)$ preserving symmetries of this theory
are described by the discrete symmetry group $\Z_2^8:\mathbb{M}_{20}$. This model therefore accounts for
one of the largest maximal symmetry groups of K3 sigma models. The symmetry group involves also
generators that, from the orbifold point of view, map untwisted and twisted sector states into one another.
}

\begin{document}
\begin{flushright}DCPT-13/33\\ AEI-2013-244\end{flushright}

\maketitle
\flushbottom
%%%%%%%%%%%%%%%%%%%%%%
\section{Introduction}

Recently, Eguchi, Ooguri and Tachikawa made the intriguing observation \cite{eot10} that the 
expansion coefficients of the elliptic genus of K3 can be naturally interpreted in terms of 
representations of the finite sporadic Mathieu group $\mathbb{M}_{24}$. This conjecture has now been
established. In particular, the \emph{twining genera}, i.e.\ the elliptic genera 
with the insertion of a group element $g\in \mathbb{M}_{24}$, have been determined combining different 
viewpoints \cite{ch10, ghv10a,ghv10b,eghi11}. The knowledge of all  twining genera fixes the decomposition
of every expansion coefficient in terms of $\mathbb{M}_{24}$ representations, and it was shown in 
\cite{ga12} (see also \cite{chm12}) that the resulting multiplicities are indeed (non-negative) integers. Thus
a consistent decomposition of all expansion coefficients in terms of $\mathbb{M}_{24}$ representations 
is possible. More recently, evidence was also obtained that the same is true for the \emph{twisted 
twining genera} \cite{Gaberdiel:2012gf,Gaberdiel:2013nya}, i.e.\ the analogues of Norton's generalised moonshine functions \cite{N}. 
The ideas underlying Mathieu moonshine have also now been extended in other directions, see in particular
\cite{Eguchi:2011aj,Cheng:2012tq,Eguchi:2012ye,Eguchi:2013es,Cheng:2013kpa,Cheng:2013wca,Harvey:2013mda,%
Harrison:2013bya}.

Since the elliptic genus counts the net contribution of BPS states of string theory on K3, these results suggest
that $\mathbb{M}_{24}$ has a natural action on the BPS spectrum of these sigma models. 
Obviously, the simplest way to realise such an action would be if it came from a genuine 
symmetry of the full sigma model. The `geometrical' symmetries of K3 surfaces were classified
some time ago by Mukai \cite{mu88}  (with additional insights by Kondo \cite{ko98}), 
who established that 
the holomorphic symplectic automorphism groups of K3 surfaces are all proper subgroups of the sporadic group 
$\mathbb{M}_{23}$ (which stabilises one element in the standard representation
of $\mathbb{M}_{24}$ as permutation group on $24$ symbols). Since the elliptic genus
is constant along {each connected component of}
the moduli space of ${\cal N}=(4,4)$ superconformal field theories at central charge 
$(c,\qu c)=(6,6)$, one may then hope that the symmetries at different points in moduli space
may be put together. This led two of us to suggest that an `overarching symmetry group' based on the classical 
geometric symmetries of K3 non-linear sigma models could be defined in this manner \cite{tawe11};
indeed already under restriction to symmetries of Kummer K3\!\! s one obtains the group $\Z_2^4:A_8$, which is a 
maximal subgroup of $\mathbb M_{24}$ not contained in $\mathbb M_{23}$ \cite{tawe12,tawe12b}.

In order to enhance the symmetry group to $\mathbb{M}_{24}$, 
stringy symmetries of K3 sigma models may need to be included. Mukai's Theorem was 
generalised to the case of supersymme\-try-preserving automorphism groups of non-linear sigma 
models on K3 \cite{Gaberdiel:2011fg}. Somewhat surprisingly it was found that the possible
supersymmetry-preserving automorphism groups of K3 sigma models are not all subgroups of $\mathbb{M}_{24}$,
but instead form groups that fit inside the Conway group $Co_1$, which contains $\mathbb{M}_{24}$. 
In fact, the result of \cite{Gaberdiel:2011fg} gave
a rather concrete description of the possible symmetry groups of all K3 sigma models. For example,
it predicted the existence of a K3 sigma model with symmetry group $G=5^{1+2} : \mathbb{Z}_4$ that 
was subsequently identified with a certain asymmetric $\mathbb{Z}_5$-orbifold of a torus theory that realises
a K3 theory \cite{Gaberdiel:2012um}. It also predicted the existence of a K3 sigma model with 
symmetry group $\Z_2^8 : \mathbb{M}_{20}$, one of the largest maximal symmetry groups of 
K3 sigma models.

It is the purpose of the present article to identify the microscopic realisation of this
sigma model. As it turns out the relevant K3 sigma model can be described as the usual
$\mathbb{Z}_2$-orbifold of a torus theory at the special 
$D_4$-point, such that the bosonic theory before orbifolding has an $\widehat{\mathfrak{so}}(8)_{1}$ current
symmetry, both for left- and right-movers. From this geometric viewpoint, the model
is a nonlinear sigma model on the so-called tetrahedral Kummer K3 studied in detail in \cite{tawe11}.

We should stress that the presence of the $\Z_2^8 : \mathbb{M}_{20}$ symmetry group is not entirely obvious
in this description. In fact, $\Z_2^8 : \mathbb{M}_{20}$ contains $A_5$, the alternating group of five symbols,
and it is a priori not clear how this 5-fold permutation symmetry should arise from the 
$\mathbb{Z}_2$-orbifold of the $D_4$-torus theory. The key idea behind our
paper is that the bosonic theory has, after orbifolding, the chiral symmetry
\be
\widehat{\mathfrak{so}}(4)_{1} \oplus \widehat{\mathfrak{so}}(4)_{1} \cong 
\widehat{\mathfrak{su}}(2)_{1}^{\oplus\, 4} \ . 
\ee
Furthermore, the  four free fermions of the supersymmetric torus theory give rise to the chiral symmetry
\be
\widehat{\mathfrak{so}}(4)_{1} \cong 
\widehat{\mathfrak{su}}(2)_{1}^{\oplus\, 2} 
\ee
that survives the orbifold projection. Taken together,
the $\mathbb{Z}_2$-orbifold of the $D_4$-torus theory therefore has the chiral symmetry \cite{nawe01}
\be
\widehat{\mathfrak{so}}(4)_{1} \oplus \widehat{\mathfrak{so}}(4)_{1} \oplus \widehat{\mathfrak{so}}(4)_{1} 
\cong 
\widehat{\mathfrak{su}}(2)_{1}^{\oplus\, 6} \ . 
\ee
One of these $\widehat{\mathfrak{su}}(2)_1$ algebras can be identified with the R-symmetry of the
${\cal N}=4$ superconformal algebra that must remain invariant under the supersymmetry-preserving
automorphisms. However, the other five factors may be permuted, and this is the origin of the $A_5$ symmetry
of our torus orbifold.\footnote{Only $A_5$ rather than $S_5$ emerges since
the states in the  $(\frac{1}{4},\frac{1}{2};\frac{1}{4},\frac{1}{2})$ multiplet of the left- and 
right-moving ${\cal N}=4$ algebras also have to be preserved, which requires the permutations to be even.} 
Remarkably, in the description of the model as a torus orbifold, some generators of this  $A_5$ symmetry mix 
states in the twisted and untwisted sectors. Our analysis also shows that the $\mathbb{Z}_2$-orbifold of the 
$D_4$-torus theory can be alternatively
realised as an asymmetric $\mathbb{Z}_4$-orbifold of the same $D_4$-torus. 
\smallskip

The paper is organised as follows. In Section~\ref{sec2} we discuss the $\mathbb{Z}_2$-orbifold of the 
$D_4$-torus theory
in detail, and explain how it may also be described in terms of twelve free Majorana fermions, both
for the left- and the right-movers. In Section~\ref{sec:su2} we provide yet another description of the same theory,
now as a rational conformal field theory (RCFT) based on the current algebra 
$\widehat{\mathfrak{su}}(2)_{1}^{\oplus\, 6}$. This approach exhibits the aforementioned permutation
symmetry most clearly; however, the structure of the ${\cal N}=(4,4)$ supercharges is rather involved
from that viewpoint. We  therefore construct the supersymmetry-preserving symmetries of our
orbifold model using the orbifold description, and then translate them into the 
$\widehat{\mathfrak{su}}(2)_{1}^{\oplus\, 6}$ language, using the free fermion description as an intermediate
step; this is done in Section~\ref{sec4}. 
Actually, as is explained in Section~\ref{s:newtorus}, there are at least fifteen different ways in which
one may write our K3 sigma model as a $\mathbb{Z}_2$-orbifold of a toroidal model --- obviously,
all these descriptions are equivalent, but differ in their distribution of
states into the twisted and untwisted
sector, respectively. Of these fifteen descriptions, five have the same expression for the four supercharges, and hence
their supersymmetry-preserving symmetries can be directly combined. In Section~\ref{thegroup}  we 
analyse the structure of the resulting group, and demonstrate that it contains at least
$\Z_2^8 : \mathbb{M}_{20}$; together with the results of \cite{Gaberdiel:2011fg} this then shows that the 
supersymmetry-preserving symmetry group of this model must be precisely equal to $\Z_2^8 : \mathbb{M}_{20}$.
Section~\ref{sec6} explains why our $\Z_2$-orbifold model has another realisation
as an asymmetric $\mathbb{Z}_4$-orbifold of the same $D_4$-torus theory. In fact, the $\mathbb{Z}_4$
action can be described in terms of two consecutive $\mathbb{Z}_2$ actions, namely 
the $\mathbb{Z}_2$-orbifold by T-duality (which is a consistent symmetry and leads again
to the $D_4$-torus theory), followed by the usual $\mathbb{Z}_2$ inversion action. 
Finally, we close with some conclusions in Section~\ref{sec:concl}. There are six appendices,
where some of the more technical material has been collected.

%%%%%%%%%%%%%%%%%%%%%%

%%%%%%%%%%%%%%%%%%%%%%
\section{K3 sigma model as $\mathbb{Z}_2$-orbifold of the $D_4$-torus model}\label{sec2}
%%%%%%%%%%%%%%%%%%%%%%
In this section we describe the K3 sigma model based on the $\mathbb{Z}_2$-orbifold of the $D_4$-torus model.
In particular, we recall that it possesses an
\begin{equation}
\widehat{\mathfrak{su}}(2)_{L,1}^ 6 \oplus \widehat{\mathfrak{su}}(2)_{R,1}^6
\end{equation}
affine symmetry algebra\footnote{We use 
the notation $\widehat{\mathfrak{su}}(2)_1^n:=\widehat{\mathfrak{su}}(2)_1^{\oplus n}$ 
throughout. The indices $L/R$ stand for left- and right-moving, respectively.} \cite[proof of Thm. 3.7]{nawe01}. 
We also explain how it may be described in terms of twelve left- and twelve right-moving  
Majorana fermions.

%%%%%%%%%%%%%%%%%%%%%%
\subsection{Geometric description of the  $D_4$-torus model and its $\Z_2$-orbifold}\label{geomD4}

%%%%%%%%%%%%%%%%%%%%%%
A generic $d=4$ \emph{bosonic} torus model contains an affine $\widehat{\mathfrak u}(1)^4$ 
algebra of left-moving currents $j_k(z)=i\partial \phi_k(z),\,k=1,\ldots,d=4$ and its right-moving counterpart 
(see Appendix~\ref{app:conv} for our conventions and notations). At particular points in the moduli space (i.e.\ 
at special values for the metric and the $B$-field), the 
$\widehat{\mathfrak u}(1)^4$ current algebra is enhanced to a non-abelian affine algebra of rank $4$. In the 
case  where the lattice $L$ underlying the torus ${\mathbb T}=\R^4/ L$ is the 
$D_4$-lattice $L_{D_4}\subset \mathbb{R}^4$, the B-field can be chosen such that the 
extended bosonic current algebra is $\widehat{\mathfrak s\mathfrak o}(8)_1$. To see this,
consider the following basis of $L_{D_4}$,
\be
\boldsymbol{l}_{1,2} := {1\over\sqrt2} \left(\begin{array}{c} 1\\1\\0\\0 \end{array} \right)\ , \quad
\boldsymbol{l}_{1,-2} := {1\over\sqrt2} \left(\begin{array}{c} 1\\-1\\0\\0 \end{array} \right)\ ,\quad
\boldsymbol{l}_{3,4} := {1\over\sqrt2} \left(\begin{array}{c} 0\\0\\1\\1 \end{array} \right)\ ,\quad
\boldsymbol{l}_{1,3} := {1\over\sqrt2} \left(\begin{array}{c} 1\\0\\1\\0 \end{array} \right)\ ,
\ee
as well as the B-field 
\begin{equation}\label{Bfield}
B:=\left( \begin{array}{cccc} 0&-1&0&0\\1&0&0&0\\0&0&0&-1\\0&0&1&0 \end{array} \right)\ ,
\end{equation}
and  set, for every pair of indices $j,\, k\in\{1,\ldots, 4\}$ with 
$j\neq k$,
\begin{equation}\label{lambdaij}
\boldsymbol{l}_{\pm j, \pm k}:={1\over\sqrt2} (\pm {\bf e}_j\pm {\bf e}_k)\in L_{D_4}\ , \quad
\boldsymbol{m}_{\pm j,\pm k}:=B\boldsymbol{l}_{\pm j, \pm k}+\boldsymbol{l}_{\pm j, \pm k} \in L^\ast_{D_4}\ ,
\end{equation} 
where ${\bf e}_1,\ldots,{\bf e}_4$ is the canonical orthonormal basis of $\R^4$.
Then, the model contains states with momentum-winding 
$(\boldsymbol{m}_{\pm j,\pm k},\boldsymbol{l}_{\pm j,\pm k})\in L_{D_4}^*\oplus L_{D_4}$, which have
left and right $\widehat{\mathfrak u}(1)^4_L$ charges (see Appendix~\ref{app:conv})
\be\label{windingcurrents}
{\bf Q}_{\pm j, \pm k}:=
{\bf Q}(\boldsymbol{m}_{\pm j,\pm k},\boldsymbol{l}_{\pm j, \pm k}) 
= \sqrt 2 \boldsymbol{l}_{\pm j, \pm k} = \pm {\bf e}_j\pm {\bf e}_k\ ,
\quad
{\bf\qu Q}(\boldsymbol{m}_{\pm j,\pm k}, \boldsymbol{l}_{\pm j, \pm k}) =0 \ .
\ee
Thus the momentum-winding fields 
$V_{({\bf Q}_{\pm j,\pm k};0)}(z)$ define twenty-four
$(1,0)$-fields which, together with $j_1(z),\ldots, j_4(z)$, form the standard 
$\widehat{\mathfrak s\mathfrak o}(8)_1$-current
algebra. 
The right-moving $\widehat{\mathfrak s\mathfrak o}(8)_1$-current algebra is analogously obtained 
using the twenty-four
$(0,1)$-fields 
\be
V_{(0;{\bf\qu Q}_{\pm j,\pm k})}(\qu z)\ ,\quad \text{ where }\quad
{\bf\qu Q}_{\pm j,\pm k}:=\mp {\bf e}_j\mp {\bf e}_k\ ,\quad
j,\, k\in\{1,\ldots, 4\} \text{ with }
j\neq k.
\ee
The full spectrum of the bosonic $D_4$-torus model
can be decomposed into representations of the left- and right-moving $\widehat{\mathfrak s \mathfrak o}(8)_1$ algebras as
\be\label{D4torus}
{\cal H}_{D_4 {\rm -torus}} = ({\cal H}_{L,0} \otimes {\cal H}_{R,0}) \oplus 
({\cal H}_{L,v} \otimes {\cal H}_{R,v}) \oplus ({\cal H}_{L,s} \otimes {\cal H}_{R,s}) \oplus ({\cal H}_{L,c} \otimes {\cal H}_{R,c})\ ,
\ee
where ${\cal H}_{L,0}$ is the left-moving $\widehat{\mathfrak s \mathfrak o}(8)_1$ vacuum representation, 
while ${\cal H}_{L,v}, {\cal H}_{L,s}$ and ${\cal H}_{L,c}$  are the vector and the two spinor representations, 
respectively. The   ${\cal H}_{R,*}$ denote the corresponding right-moving representations. The vector representation ${\cal H}_{L,v} \otimes {\cal H}_{R,v}$ is generated by OPEs of the holomorphic and antiholomorphic currents with the winding-momentum fields 
\be
{\bf Q}(\boldsymbol{m},\boldsymbol{l})
=\pm {\bf e}_i\ ,\quad {\bf\qu Q}(\boldsymbol{m},\boldsymbol{l})=\pm {\bf e}_j
\quad\quad \mbox{(64 fields)}\ , 
\ee
while 
the spinor representations  are generated by
\begin{align}
{\bf Q}(\boldsymbol{m},\boldsymbol{l})= {1\over2} \sum_{j=1}^4 \eps_j {\bf e}_j\ , \quad 
{\bf\qu  Q}(\boldsymbol{m},\boldsymbol{l})= {1\over2} \sum_{k=1}^4 \delta_k {\bf e}_k\ , 
\quad\quad \mbox{(128 fields)}  \label{bosgen} \\ \mbox{where } \eps_j,\, \delta_k\in\{\pm1\} \mbox{ and } 
\sum_{k=1}^4 (Q_k+\qu Q_k)\equiv 0\mod 2\ .
\nonumber \end{align}
In particular,  $\sum_{k=1}^4 Q_k$ and $\sum_{k=1}^4 \qu Q_k$ are even for ${\cal H}_{L,s} \otimes {\cal H}_{R,s}$ 
and odd for ${\cal H}_{L,c} \otimes {\cal H}_{R,c}$.
Thus, the lattice of $\widehat{\mathfrak u}(1)^4_L\oplus \widehat{\mathfrak u}(1)^4_R$ charges 
(see \eqref{chargelattice}) of the $D_4$-torus model equals
\be\label{Ddcharge}
\Gamma_{d,d}:= \left\{ ({\bf Q};{\bf\qu Q}) \in \Z^d\oplus\Z^d
\cup \left({\textstyle{1\over2}}+\Z\right)^d\oplus \left({\textstyle{1\over2}}+\Z\right)^d
\left| \sum_{k=1}^d \left(Q_k+\qu Q_k\right)\equiv 0 \mod 2\right.\right\}
\ee
with $d=4$.
\medskip

The supersymmetric torus model is obtained by adjoining
$d=4$ free Majorana fermions $\psi_k(z), \,k=1,\ldots,\,4$, related to  the ${\rm U}(1)$-currents $j_k(z)$ by 
world-sheet supersymmetry,  and their right-moving counterparts $\qu\psi_k(\bar z), \,k=1,\ldots,\,4$. These 
holomorphic fermions give rise to the affine symmetry
\begin{equation}\label{affer}
\widehat{\mathfrak{so}}(4)_1 \cong \widehat{\mathfrak s\mathfrak u}(2)_1^2 \ .
\end{equation}
More specifically, we can construct affine generators of $\wh{\mathfrak{su}}(2)_1^2$ by
\begin{align} J^{3,1}(z)  
& := \frac{1}{2} \Bigl(\nop{\chi^*_{1}(z) \chi_{1} (z)} + \nop{ \chi^*_{2} (z) \chi_{2}(z)} \Bigr)\label{Rsymm}\\[2pt]
J^{+,1}(z) &:= i\nop{\chi^*_{1}(z) \, \chi^*_{2}(z) } \ , \qquad
J^{-,1}(z) :=  i  \nop{\chi_{1}(z) \, \chi_{2}(z)} \ ,\label{Rsymm1}\\[4pt]
J^{3,2}(z)  & := \frac{1}{2} \Bigl(\nop{ \chi^*_{1}(z) \chi_{1} (z)}-\nop{ \chi^*_{2} (z) \chi_{2}(z)}\Bigr)\\[2pt]
J^{+,2}(z) &:= i\nop{\chi^*_{1}(z) \, \chi_{2}(z) } \ , \qquad
J^{-,2}(z) := i  \nop{\chi_{1}(z) \, \chi^*_{2}(z) } \ , \label{Rsymm2}
\end{align}
where $\chi_{j}$ and $\chi^*_{j}$, $j=1, 2$, are Dirac fermions  defined as
\begin{align} \label{Diracblock1}
\chi_{j}(z)&:=\frac{1}{\sqrt 2} (\psi_{2j-1}(z)+i\psi_{2j}(z))\ ,&  
\chi^*_{j}(z)&:=\frac{1}{\sqrt 2} (\psi_{2j-1}(z)-i\psi_{2j}(z))\ , \qquad j=1,\,2\ .
\end{align}
These currents (for $k=1,2$) satisfy the OPEs
\be\label{su2generators}
\begin{array}{rclrcl}
J^{3,k}(z)J^{3,k}(w)&\sim&\displaystyle\frac{1}{2(z-w)^2}\ , \qquad
&J^{3,k}(z)J^{\pm,k}(w)&\sim&\displaystyle \pm \frac{J^{\pm,k}(w)}{(z-w)}\ ,\\[0.75em]
J^{\pm,k}(z)J^{\pm,k}(w)&\sim&0\ , 
&J^{\pm,k}(z)J^{\mp,k}(w)&\sim&\displaystyle\frac{1}{(z-w)^2}\pm \frac{2 J^{3,k}(w)}{(z-w)}\  . 
\end{array}
\ee
Altogether, the affine symmetry (both for left- and right-movers) of the supersymmetric $D_4$-torus model  is 
\be
\widehat{\mathfrak s\mathfrak u}(2)_1^2  \oplus \widehat{\mathfrak{so}}(8)_1 \ . 
\ee
\medskip

In order to obtain a K3 sigma model we now consider a $\mathbb{Z}_2$-orbifold of this torus model. 
The group $\Z_2$ acts in the 
usual manner on the bosonic degrees of freedom, i.e.\ it maps 
$j_k(z)\mapsto -j_k(z)$, $\qu\jmath_k(\qu z)\mapsto -\qu\jmath_k(\qu z)$, $k=1,\ldots,4$,
and $V_{\lambda}\mapsto V_{-\lambda}$ for all $\lambda\in\Gamma_{4,4}$. 
In order for this to be a well-defined
symmetry we need  to choose our operators $c_\lambda$ of equation (\ref{vdef}) such that 
$c_{\lambda} = c_{-\lambda}$ for all $\lambda\in\Gamma_{4,4}$.  

On the fermions, the group $\Z_2$ acts as $\psi_k \mapsto - \psi_k$ and
likewise for the right-movers, $\overline\psi_k \mapsto -\overline\psi_k$, $k=1,\ldots,4$. 
In particular, the orbifold  leaves the 
$\widehat{\mathfrak{su}}(2)_{L,1}\oplus \widehat{\mathfrak{su}}(2)_{R,1}$
algebra in (\ref{affer}) invariant, since it is generated by the bilinear fermion combinations
\eqref{Rsymm} -- \eqref{Rsymm2}. 

In the left-moving sector\footnote{The treatment of the right-moving sector is analogous.}, 
the orbifold procedure projects out the four Cartan generators 
$j_1(z),\ldots,j_4(z)$
of the bosonic $\widehat{\mathfrak{so}}(8)_{L,1}$ algebra, and
maps positive and negative roots into one another; the invariant subalgebra is therefore that of
the Cartan involution of $\mathfrak{so}(8)_{L,1}$, i.e.  
$\mathfrak{so}(4)_{L,1}\oplus \mathfrak{so}(4)_{L,1}$. 
Altogether, the orbifold thus has an affine current algebra of type
\begin{equation}\label{so4cubecurrent}
\widehat{\mathfrak{so}}(4)_{L,1} \oplus \widehat{\mathfrak{so}}(4)_{L,1} \oplus  \widehat{\mathfrak{so}}(4)_{L,1} \cong
\widehat{\mathfrak{su}}(2)_{L,1}^{6} \ .
\end{equation}
Since the central charge of the (left) current algebra \eqref{so4cubecurrent}
equals $c=6$, the full K3 sigma model must be 
a rational theory with respect to this symmetry algebra. Hence we will be able to describe the whole theory
quite succinctly in terms of an 
$\widehat{\mathfrak{su}}(2)_{L,1}^{6} \oplus \widehat{\mathfrak{su}}(2)_{R,1}^{6}$ RCFT; 
this will be further developed in Section~\ref{sec:su2}. In order
to understand the equivalence between that description and the $D_4$-torus orbifold better, 
it is convenient to reformulate the torus orbifold in terms of free fermions.

%%%%%%%%%%%%%%%%%%%%%%%%%%%%%%%%%
\subsection{Free fermion description of the $D_4$-torus and its orbifold} \label{torferm}
%%%%%%%%%%%%%%%%%%%%%%%%%%%%%%%%%
The \emph{bosonic} $D_4$-torus model may be described in terms of eight free left- and right-moving 
Majorana fermions $\psi_j(z)$ and $\qu \psi_j(\qu z)$,  $j=5,\ldots 12$, all of whose boundary conditions are 
coupled. In Appendix \ref{appB}, we state the correspondence in detail in terms of the four 
left- and four right-moving Dirac fermions
\be \label{Diracblock23}
\begin{array}{rclrcll}
x_k &:=& {1\over\sqrt2} ( \psi_{k+4} + i \psi_{k+8} )\ , \qquad 
&x_k^\ast &:=& {1\over\sqrt2} ( \psi_{{k+4}} - i \psi_{{k+8}} )\ , \\[4pt]
\qu x_k &:=& {1\over\sqrt2} ( \qu\psi_{{k+4}} + i \qu\psi_{{k+8}} )\ , \qquad 
&\qu x_k^\ast &:=& {1\over\sqrt2} ( \qu\psi_{{k+4}} - i \qu\psi_{{k+8}} )\ ,
& k\in\{1,\ldots,4\}\ ,
\end{array}
\ee
which satisfy the OPEs%
\be\label{Diracfermionope}
x_k(z) \, x_k^\ast(w)\sim {1\over (z-w)}\sim x_k^\ast(z) \, x_k(w) \ .
\ee
By our choice of fermionisation  \eqref{bosonization}, we identify the holomorphic
U(1)-currents of the bosonic $D_4$-torus model as 
\be\label{boscurrents}
j_k(z) = i\partial\phi_k(z) = \nop{x_k(z) x_k^\ast(z)}= - i\nop{\psi_{{k+4}}(z)\psi_{{k+8}}(z)} \ . 
\ee
All other generating fields of the theory as determined in Subsection \ref{geomD4} are expressed
in terms of the Dirac fermions
$x_k(z),\, x_k^\ast(z),\, \qu x_k(\qu z),\, \qu x_k^\ast(\qu z),\, k=1,\ldots,4$,
and the `meromorphic building blocks' $\xi_k^\pm(z) = \nop{\exp\left( \pm{i\over2}\phi_k(z)\right)}$
that are defined in Appendix~\ref{appB}.
\medskip

Next we observe that the
$\mathbb{Z}_2$-action on the  $D_4$-torus model is induced by  the transformation that 
leaves $\psi_{{5}}(z),\ldots, \psi_{{8}}(z)$ invariant, 
while mapping $\psi_k(z)\mapsto -\psi_k(z)$ where $k\in\{{9,\ldots,12}\}$.
In other words, we have $x_k(z)\leftrightarrow x_k^\ast(z)$, and analogously for the right-moving 
fermions. 
Using the notations \eqref{windingcurrents},
the untwisted sector of the $\Z_2$-orbifold
is hence generated by the $\Z_2$-invariant $(1,0)$-fields with $\C$-basis 
\be
\begin{array}{rcl}
\mbox{for } j<k,\quad
\nop{x_j(z) x_k(z)} +  \nop{x_j^\ast(z) x_k^\ast(z)} 
&=& V_{({\bf Q}_{j,k};0)}(z) + V_{({\bf Q}_{-j,-k};0)}(z)\ ,\\[0.5em]
\nop{x_j(z) x_k^\ast(z)} +  \nop{x_j^\ast(z)x_k(z) } 
&=& V_{({\bf Q}_{j,-k};0)}(z)+V_{({\bf Q}_{-j,k};0)}(z)\ ,
\end{array}
\ee
along with the $\Z_2$-invariant $({1\over2},{1\over2})$-fields
of the form $V_{({\bf Q};{\bf \qu Q})}(z,\qu z)+ V_{(-{\bf Q};-{\bf \qu Q})}(z,\qu z)$
with $\C$-basis 
\be\label{invarianthalfhalf}
\begin{array}{rclrcll}
\mbox{if } {\bf Q} &=& {\bf e}_j, & {\bf\qu Q} &=& {\bf e}_k\colon
&  i\nop{ x_j(z)\qu x_k^\ast(\qu z)}+ i\nop{ x_j^\ast(z)\qu x_k(\qu z)}\ ,\\[5pt]
&&& {\bf\qu Q} &=& -{\bf e}_k\colon
& i\nop{ x_j(z)\qu x_k(\qu z)}+i\nop{ x_j^\ast(z)\qu x_k^\ast(\qu z)}\ ,\\[12pt]
\mbox{if } {\bf Q} &=& {1\over2}\sum\limits_{j=1}^4\eps_j{\bf e}_j, & 
{\bf\qu Q} &=& {1\over2}\sum\limits_{k=1}^4\delta_k{\bf e}_k\colon\\
&&&\multicolumn{4}{l}{\nop{ \prod\limits_{j=1}^4\xi_j^{\eps_j}(z)\prod\limits_{k=1}^4\qu\xi_k^{\delta_k}(\qu z)}
+\nop{ \prod\limits_{j=1}^4\xi_j^{-\eps_j}(z)\prod\limits_{k=1}^4\qu\xi_k^{-\delta_k}(\qu z)}\ ,}
\end{array}
\ee
where $\eps_j,\,\delta_k\in\{\pm\}$.
To describe the twisted sector, recall that the $\Z_2$-orbifolding of our eight 
Majorana fermions with coupled boundary conditions decouples effectively
the boundary conditions of  the fermions $\psi_{{5}}(z),\ldots,\psi_{{8}}(z)$ from those of 
$\psi_{{9}}(z),\ldots,\psi_{{12}}(z)$. 
For $\eps_k\in\{\pm\},\, k\in\{1,\ldots,4\}$, we define
\be
\Xi^+_{\eps_1,\ldots,\eps_4}(z) 
:= \nop{\prod_{k=1}^4 \eta_k^{\eps_k}(z)} \quad\quad\mbox{and}\quad\quad
\Xi^-_{\eps_1,\ldots,\eps_4}(z) 
:=  \nop{\prod_{k=1}^4 \eta_{k+4}^{\eps_k}(z)}\ ,
\ee
where the $\eta_k^\pm$ fields are introduced in Appendix \ref{appB}.
Then the twisted ground states of our $\Z_2$-orbifold are described by 
those $\Xi^\pm_{\eps_1,\ldots,\eps_4}(z)$ for which an even number of the
$\eps_k$ are equal to $+1$. 

Recall that the decoupling of the eight Majorana fermions into two sets of four implies that
our $\Z_2$-action breaks the $\widehat{\mathfrak{so}}(8)_1$-symmetry of the underlying toroidal
theory to $\widehat{\mathfrak{su}}(2)_1^4$. Indeed, 
by \eqref{invarianthalfhalf} a basis for the  $\Z_2$-invariant 
$(1,0)$-fields is given by
\be
\nop{\psi_{{j+4}}(z)\psi_{{k+4}}(z)}\ ,\qquad \nop{\psi_{{j+8}}(z)\psi_{{k+8}}(z)}\ ,\qquad 1\leq j<k\leq 4 \ ,
\ee
where the $\nop{\psi_{{j+4}}(z)\psi_{{k+4}}(z)}$ with $j,\, k\in\{1,\ldots,4\}$ generate an 
$\widehat{\mathfrak s\mathfrak o}(4)_1=\widehat{\mathfrak s\mathfrak u}(2)_1\oplus \widehat{\mathfrak s\mathfrak u}(2)_1 $
current algebra, and so do the $\nop{\psi_{{j+8}}(z)\psi_{{k+8}}(z)}$ with $j,\, k\in\{1,\ldots,4\}$.

\smallskip

Note  that our expressions and normalisations
of the U(1)-currents $J^{3,k}(z)$ and $j_k(z)$, respectively, are different (compare 
\eqref{Rsymm} and \eqref{su2generators} to \eqref{boscurrents} and \eqref{u1ope}).
In what follows, we shall use both choices of fermionisation conventions, since
both of them are sometimes convenient. %, depending on the application  we have in mind. 
We will carefully distinguish the two choices
in terms of our notations, not just for the U(1)-currents $J^{3,k}(z)$ and $j_k(z)$, but also
for the relevant Dirac fermions, which are denoted by $\chi_k(z)$ or $x_k(z)$,
respectively. 
This free fermion description is also convenient in order to determine the 
partition function of the theory and -- by means of the elliptic genus -- to confirm
that it  is a K3 model, see Appendix \ref{sigmamodelspectrum}. In fact, by the results of 
\cite{eoty89}, the usual $\Z_2$-orbifold of every supersymmetric $d=4$-dimensional
torus model has the elliptic genus of K3 and thus
is indeed a K3 model.

%%%%%%%%%%%%%%%%
\subsection{The $\NNN=(4,4)$ supercurrents}
%%%%%%%%%%%%%%%%

The K3 sigma model possesses an ${\mathcal N}=(4,4)$ superconformal symmetry on the world-sheet; 
the relevant
supercharges can be most conveniently 
defined for the underlying supersymmetric $D_4$-torus model.  
With notations as above, we 
define the complex fields
\begin{align}\label{zdef}
\partial Z^{(1)}(z):=&\frac{1}{\sqrt 2} (\partial \phi_1(z)+i\partial \phi_2(z))\ , &
\partial Z^{(1)\,\ast}(z):=&\frac{1}{\sqrt 2} (\partial \phi_1(z)-i\partial \phi_2(z))\ , \nonumber\\
\partial Z^{(2)}(z):=&\frac{1}{\sqrt 2} (\partial \phi_3(z)+i\partial \phi_4(z)) \ ,&
\partial Z^{(2)\,\ast}(z):=&\frac{1}{\sqrt 2} (\partial \phi_3(z)-i\partial \phi_4(z) \ .
% \
% \qu Z^{(1)}(\qu z):=&\frac{1}{\sqrt 2} (\qu\phi_1(\qu z)+i\qu\phi_2(\qu z))\ , &
% \qu Z^{(1)\,\ast}(\qu z):=&\frac{1}{\sqrt 2} (\qu\phi_1(\qu z)-i\qu\phi_2(\qu z))\ , \nonumber\\
%\qu Z^{(2)}(\qu z):=&\frac{1}{\sqrt 2} (\qu\phi_3(\qu z)+i\qu\phi_4(\qu z)) \ ,&
% \qu Z^{(2)\,\ast}(\qu z):=&\frac{1}{\sqrt 2} (\qu\phi_3(\qu z)-i\qu\phi_4(\qu z) \ .\nonumber
\end{align} 
Using the definition of the Dirac fermions $\chi_k(z),\,k=1,\,2$,  see eq.~\eqref{Diracblock1},
the holomorphic $\NNN=4$ supercurrents are then given by
\begin{align}\label{Gpl}
G^{+} (z)& =   \sqrt{2}i \,  (
 \nop{ \chi^*_{1} (z)\partial Z^{(1)}(z)}+
 \nop{ \chi^*_{2}(z)\partial Z^{(2)}(z)}) \ , \vspace*{0.2cm} \\
G^{-} (z)& =   \sqrt{2}i \,  (
 \nop{\chi_{1} (z)\partial  Z^{(1)\,\ast}(z)}+
 \nop{\chi_{2}(z)\partial Z^{(2)\,\ast}(z)})\ ,  \vspace*{0.2cm} \\  
 {G'}^{+} (z)& =   \sqrt{2} \,  (
 -\nop{\chi^*_{1} (z)\partial Z^{(2)\,\ast}(z)}+
 \nop{\chi^*_{2}(z)\partial Z^{(1)\,\ast}(z)}) \ ,\vspace*{0.2cm} \\
{G'}^{-} (z)& =   \sqrt{2} \,  (
 \nop{\chi_{1} (z) \partial Z^{(2)}(z)}-
 \nop{\chi_{2}(z) \partial Z^{(1)}(z)}) \ . \label{Gmin}
\end{align}
Indeed, it is straightforward to check that these fields satisfy the OPEs
\be\label{superpartner}
{1\over\sqrt2} \left( G^+(z)+G^-(z) \right) \psi_k(w)
\sim {j_k(w)\over(z-w)}\ ,\qquad k=1,\ldots,4
\ee
exhibiting $\psi_k(z)$ as superpartner of $j_k(z)$, $k=1,\ldots,4$.
Moreover, \eqref{Gpl} -- \eqref{Gmin} obey the standard OPEs for the $\NNN=4$ supercurrents
\be
\begin{array}{rcl}
G^{\pm}(z)G^{\mp}(w)\sim 
{G'}^{\pm}(z){G'}^{\mp}(w) & \sim & {\displaystyle \frac{4}{(z-w)^3}\pm\frac{4}{(z-w)^2}J^{3,1}(w)} \\[4pt]
& & {\displaystyle +\frac{2}{z-w}(T(w)\pm \partial J^{3,1}(w))}\ ,\\[6pt]
{G'}^+(z)G^+(w) &=& {\displaystyle
\frac{4}{(z-w)^2}J^{+,1}(w)+\frac{2}{(z-w)}\partial J^{+,1}(w)}\ , \\[4pt]
{G'}^-(z)G^-(w)&=& {\displaystyle
\frac{-4}{(z-w)^2}J^{-,1}(w)-\frac{2}{(z-w)}\partial J^{-,1}(w)} \ ,\\[6pt]
G^{\pm}(z)G^{\pm}(w)& \sim & {G'}^{\pm}(z){G'}^{\pm}(w)\sim G^{\pm}(z){G'}^{\mp}(w)\sim 0 \ ,
\end{array}
\ee  where $T$ is the stress-energy tensor and $J^{a,1}$ with $(a=3,+,-)$ are the $\wh{\mathfrak{su}}(2)_1$ currents 
of \eqref{Rsymm} -- \eqref{Rsymm1}. Therefore, the zero modes of these currents generate the ${\rm SU}(2)$ 
R-symmetry group of the $\NNN=4$ algebra.

The free fermion description of our model of Subsection \ref{torferm} allows us
to express the supercurrents \eqref{Gpl} -- \eqref{Gmin} in terms of the Dirac fermions
$\chi_k(z),\, \chi_k^\ast(z),\, k=1,\,2$, and the Majorana fermions $\psi_5(z),\ldots,\, \psi_{12}(z)$ by means
of \eqref{boscurrents}. However, for later use it is more convenient to introduce
Dirac fermions in a completely symmetric way, as opposed to the construction in Subsection~\ref{torferm}. 
Indeed, we extend \eqref{Diracblock1} to  the definitions
\begin{align}\label{Dirac}
\chi_{j}(z)&:=\frac{1}{\sqrt 2} (\psi_{2j-1}(z)+i\psi_{2j}(z))\ ,&  
\chi^*_{j}(z)&:=\frac{1}{\sqrt 2} (\psi_{2j-1}(z)-i\psi_{2j}(z))\ ,\quad
j=1,\,\ldots,6\ ,
\end{align} 
instead of using the fields $x_k$ and $x^*_k$, $k=1,\ldots,4$, of \eqref{Diracblock23}.
Then one checks
\begin{align} 
G^+(z)=\Bigl(\frac{-1-i}{2}\Bigr)\sum_{j=1}^2\Bigl[&\nop{ \chi^*_{j}(z)\chi_{j+2}(z)\chi_{j+4}(z) }+\nop{ \chi^*_{j}(z)\chi^*_{j+2}(z)\chi^*_{j+4}(z) }\Bigr.\nonumber\\
&\Bigl.+i\bigl(\nop{ \chi^*_{j}(z)\chi_{j+2}(z)\chi^*_{j+4}(z) }+\nop{ \chi^*_{j}(z)\chi^*_{j+2}(z)\chi_{j+4}(z) }\bigr)\Bigr]
\label{G+chi}\\ 
G^-(z)=\Bigl(\frac{-1-i}{2}\Bigr)\sum_{j=1}^2\Bigl[&i\bigl(\nop{ \chi_{j}(z)\chi_{j+2}(z)\chi_{j+4}(z) }+\nop{ \chi_{j}(z)\chi^*_{j+2}(z)\chi^*_{j+4}(z) }\bigr)\nonumber\\
&+\nop{ \chi_{j}(z)\chi_{j+2}(z)\chi^*_{j+4}(z) }+\nop{ \chi_{j}(z)\chi^*_{j+2}(z)\chi_{j+4}(z) }\Bigr]\ ,\label{G-chi}
\end{align}
as well as
\begin{align}
{G'}^+(z)=\Bigl(\frac{-1-i}{2}\Bigr)\Bigl[&\nop{ \chi^*_{2}(z)\chi_{3}(z)\chi_{5}(z) }-\nop{ \chi^*_{1}(z)\chi_{4}(z)\chi_{6}(z) }\nonumber\\
&\quad +\nop{ \chi^*_{2}(z)\chi^*_{3}(z)\chi^*_{5}(z) }-\nop{ \chi^*_{1}(z)\chi^*_{4}(z)\chi^*_{6}(z) }\nonumber\\
+i&\bigl(-\nop{ \chi^*_{2}(z)\chi_{3}(z)\chi^*_{5}(z) }+\nop{ \chi^*_{1}(z)\chi_{4}(z)\chi^*_{6}(z) }\nonumber\\
&-\nop{ \chi^*_{2}(z)\chi^*_{3}(z)\chi_{5}(z) }+\nop{ \chi^*_{1}(z)\chi^*_{4}(z)\chi_{6}(z) }\bigr)\Bigr]\ ,\label{Gprime+chi}\\
{G'}^-(z)=\Bigl(\frac{-1-i}{2}\Bigr)\Bigl[&i\bigl(\nop{ \chi_{2}(z)\chi_{3}(z)\chi_{5}(z) }-\nop{ \chi_{1}(z)\chi_{4}(z)\chi_{6}(z) }\nonumber\\
&\quad +\nop{ \chi_{2}(z)\chi^*_{3}(z)\chi^*_{5}(z) }-\nop{ \chi_{1}(z)\chi^*_{4}(z)\chi^*_{6}(z) }\bigr)\nonumber\\
&-\nop{  \chi_{2}(z)\chi_{3}(z)\chi^*_{5}(z) }+\nop{ \chi_{1}(z)\chi_{4}(z)\chi^*_{6}(z) }\nonumber\\
&-\nop{ \chi_{2}(z)\chi^*_{3}(z)\chi_{5}(z) }+\nop{ \chi_{1}(z)\chi^*_{4}(z)\chi_{6}(z) }\Bigr]\ .\label{Gprime-chi}
\end{align}

%%%%%%%%%%%%%
\section{The K3 sigma model as an $\widehat{\mathfrak{su}}(2)^6_1$ RCFT}\label{sec:su2}
%%%%%%%%%%%%%

The K3 model which we described in Section~\ref{sec2} can be 
obtained as a $\Z_2\times\Z_2$-orbifold of the Gepner model $(2)^4$ \cite[Thm.~3.7]{nawe01}.
As such it is a rational CFT. In this section, we give a description of it
as an $\widehat{\mathfrak s \mathfrak u}(2)_1^{6}$ RCFT
which turns out to be useful in order to determine the full symmetry group of this model.

\subsection{Representations of the $\widehat{\mathfrak s \mathfrak u}(2)_{L,1}^6\oplus
\widehat{\mathfrak s \mathfrak u}(2)_{R,1}^6$ current algebra}\label{su21algebra} 

Let us begin by reviewing the representation theory of $\widehat{\mathfrak{su}}(2)_1$.
This algebra possesses only  two irreducible highest weight representations, namely
the vacuum representation $[0]$, whose ground state has conformal weight $0$ and is a singlet under the 
group SU(2) generated by the zero modes of the algebra, and the  representation $[1]$, whose 
ground states have conformal weight ${1\over4}$ and form an SU(2)-doublet. The fusion rules  have 
the structure of a cyclic group of order $2$,
\be 
[0]\times [0]\to [0],\qquad [0]\times [1]\to [1],\qquad [1]\times [1]\to [0]\ .
\ee   
Therefore, the representation content of a model with $\widehat{\mathfrak s\mathfrak u}(2)_{L,1}^6 \oplus \widehat{\mathfrak s\mathfrak u}(2)_{R,1}^6$ (left- and right-moving) affine 
algebra\footnote{See Appendix \ref{appC1} for an overview of a vertex operator construction for 
this affine algebra.} can be encoded in a subgroup 
\be 
\m A\subset \Z_2^6\times \Z_2^6\ ,
\ee whose elements we denote by
\be 
[a_1,\ldots,a_6;b_1,\ldots,b_6]\ ,\qquad a_i,b_i\in \{0,1\}\ .
\ee  
Occasionally, it will be useful to consider $\widehat{\mathfrak s\mathfrak u}(2)_1^6$ as a direct sum of three $\widehat{\mathfrak s\mathfrak o}(4)_1$ algebras.
The algebra $\widehat{\mathfrak s\mathfrak o}(4)_1\cong\widehat{\mathfrak s\mathfrak u}(2)_1^2$ has  four irreducible highest weight representations that we denote by a pair $[ab]$, $a,b\in\{0,1\}$, of 
$\widehat{\mathfrak{su}}(2)_1$ labels.  Equivalently, four Majorana fermions with coupled spin structures yield an
$\widehat{\mathfrak s\mathfrak u}(2)_1^2\cong\widehat{\mathfrak s\mathfrak o}(4)_1$ current algebra
as in \eqref{Rsymm} -- \eqref{Rsymm2}. Thus the 
four representations $[ab]$, $a,b\in\{0,1\}$,
can also be classified by their fermion number $(-1)^{F}$, and their tensor/spinor properties as 
$\wh{\mathfrak{so}}(4)_1$ representations,
\be\label{TSdef}
\begin{array}{c|cc}
& (-1)^F=+1 &\qquad (-1)^F=-1\\
\hline
\text{Tensor (T)}& [00] & [11]\\
\text{Spinor (S)} & [10] & [01]
\end{array}
\ee
where the fermion number of the spinor representations is a matter of convention.
\smallskip

Next we relate this to the free fermion description
of Subsection \ref{torferm}. In contrast to the notations there, we now group  the twelve holomorphic 
(and anti-holomorphic) Majorana fermions  $\psi_k(z)$, $\qu\psi_k(\qu z)$, $k=1,\ldots,\,12$, into Dirac fermions 
according to \eqref{Dirac}. This allows us to construct
an $\widehat{\mathfrak s \mathfrak u}(2)_{L,1}^6$ 
subalgebra of the affine algebra $\widehat{\mathfrak s \mathfrak o}(12)_{ L,1}$ 
generated by these free fermions, where all summands $\widehat{\mathfrak s \mathfrak u}(2)_{ L,1}$
enter completely symmetrically.
The currents generating the first 
$\widehat{\mathfrak s\mathfrak o}(4)_{ L,1}\cong\widehat{\mathfrak s \mathfrak u}(2)_{ L,1}^2$ 
subalgebra in $\widehat{\mathfrak s \mathfrak o}(4)^3_1$ are given in \eqref{Rsymm} -- \eqref{Rsymm2}.
In particular,  the four left-moving Majorana fermions $\psi_1,\ldots,\psi_4$ form a vector representation 
$[11]$ under this first (left) $\widehat{\mathfrak s\mathfrak o}(4)_{ L,1}$, while they are in the singlet 
$[00]$ representation under the second and third  (left) $\widehat{\mathfrak s\mathfrak o}(4)_{ L,1}$ algebras, 
as well as under the right $\widehat{\mathfrak s \mathfrak o}(4)^3_{R,1}$ algebra. 
The four corresponding right-moving Majorana fermions $\overline{\psi_k}(\qu z)$, $k=1,\ldots,4$, 
behave in an analogous manner under  
$\widehat{\mathfrak s \mathfrak o}(4)^3_{R,1}\cong \widehat{\mathfrak s \mathfrak u}(2)_{ R,1}^6$;
symbolically we therefore write
\be \label{su2behaviour}
\psi_1,\,\ldots,\,\psi_4 \to [11\,00\,00;\,00\,00\,00]\ ,\qquad
\overline\psi_1,\,\ldots,\,\overline\psi_4 \to [00\,00\,00;11\,00\,00]\ .
\ee
The remaining fermions are in the singlet representation $[00]$
of the first $\widehat{\mathfrak s\mathfrak o}(4)_{ L,1}
\cong\widehat{\mathfrak s \mathfrak u}(2)_{ L,1}^2$.

We can similarly express the currents $J^{a,k}$, $(a=3,\pm)$, $k=3,\ldots,6$, generating the second and third 
$\widehat{\mathfrak s\mathfrak o}(4)_{L,1}$ subalgebras in 
$\widehat{\mathfrak s \mathfrak o}(4)^3_{L,1}$ analogously to \eqref{Rsymm} -- \eqref{Rsymm2}
in terms of the four Dirac fermions 
$\chi_{3}(z),\,\chi_{4}(z),\,\chi^*_{3}(z),\,\chi^*_{4}(z)$ and $\chi_{5}(z),\,\chi_{6}(z),\,\chi^*_{5}(z),\,\chi^*_{6}(z)$, 
respectively, and likewise for the right-moving currents. Hence, the remaining free fermions transform in the following
representations of  $\widehat{\mathfrak s\mathfrak u}(2)_{L,1}^6\oplus \widehat{\mathfrak s\mathfrak u}(2)_{R,1}^6$,
\be\label{su2behaviour23}
\begin{array}{rlrl} 
\psi_5\ldots,\psi_8&\to [00\,11\,00;\,00\,00\,00]\ , & \qquad
\overline\psi_5\ldots,\overline\psi_8& \to [00\,00\,00;\,00\,11\,00] \ ,\\[0.5em]
\psi_9\ldots,\psi_{12}&\to [00\,00\,11;00\,00\,00]\ ,&
\overline\psi_9\ldots,\overline\psi_{12}&\to [00\,00\,00;\,00\,00\,11]\ .
\end{array}
\ee
The fields and representations discussed so far  belong to the `internal' NS sector
of a free fermion theory which is obtained by fermionisation of a toroidal theory on a $D_6$-torus.
Indeed, in complete analogy to our discussion in Subsections \ref{torferm} and Appendix \ref{sigmamodelspectrum},
the spectrum $\HHH_{D_6{\rm-torus}}$ of the bosonic $D_6$-torus model (see Appendix \ref{app:conv}),
which has charge lattice $\Gamma=\Gamma_{6,6}$ as in \eqref{Ddcharge}, 
is obtained from twelve free left-moving and twelve free right-moving Majorana fermions, all with coupled
spin structures.

%%%%%%%%%%%%%%%%%%%%%%%%%%%%%%
\subsection{Spectrum of the K3 model in terms of representations of 
$\widehat{\mathfrak s \mathfrak u}(2)_1^6\oplus\widehat{\mathfrak s \mathfrak u}(2)_1^6$ }\label{spectrum}
%%%%%%%%%%%%%%%%%%%%%%%%%%%%%%%

In order to obtain the K3 sigma model we now have to perform an orbifold of this
bosonic $D_6$-torus model. In particular, we need to decouple the spin structures of the internal and 
external fermions (in order to describe the supersymmetric $D_4$-torus model), and we have to
perform the usual $\Z_2$-orbifold to obtain from the latter a K3 theory. In terms of the $D_4$-torus model, the
relevant orbifold group is therefore $\Z_2\times\Z_2$  with generators
\be 
g:=(-1)^{\sum\limits_{k=5}^{12}(F_{L,k}+F_{R,k})}  
\qquad \text{and}\qquad  h:=(-1)^{\sum\limits_{k=1}^4(F_{L,k}+F_{R,k})+\sum\limits_{i=9}^{12}(F_{L,k}+F_{R,k})}\ ,
\ee 
where $F_{L,k},\, F_{R,k}$ are the fermion number operators corresponding to 
$\psi_k,\, \qu\psi_k,\, k=1,\,\ldots,\,12$, respectively. Here $g$ is the symmetry  whose orbifold
decouples the spin structures,
while $h$ induces the standard $\Z_2$-orbifold on the ${D_4}$-torus model.
\smallskip

To implement the orbifold procedure, we now introduce the $g$-, $h$- and $gh$-twisted sectors and then 
project onto the invariant states in the untwisted and in the three twisted sectors.
We focus on the NS-NS sector of our model first.
In the $g$-twisted sector, the fields $\psi_k, \overline\psi_k$, $k=5,\ldots, 12$ have integer modes. 
Thus, the $g$-twisted ground states form a representation of the Clifford algebra of the zero modes of these fields, 
i.e.\ they transform in spinor representations of the corresponding $\widehat{\mathfrak s\mathfrak o}(4)_1$ 
algebras. Analogous considerations hold for the $h$- and $gh$-twisted sectors. Therefore, the tensorial 
properties of the various sectors with respect to the left and right-moving $\widehat{\mathfrak s\mathfrak o}(4)_{L,1}^3\oplus \widehat{\mathfrak s\mathfrak o}(4)_{R,1}^3$ algebras are 
\begin{align}\label{tensu}
& \text{untwisted} & [\text{TTT};\text{TTT}] \\
\label{tensg}& g\text{-twisted} & [\text{TSS};\text{TSS}] \\
\label{tensh}& h\text{-twisted} & [\text{STS};\text{STS}] \\
\label{tensgh}& gh\text{-twisted} & [\text{SST};\text{SST}] 
\end{align} 
where $T$ and $S$ denote a tensor or a spinor representation of 
$\widehat{\mathfrak s\mathfrak o}(4)_1$ as in \eqref{TSdef}, respectively. 

Finally, one has to project onto the representations that are invariant under both $g$ and $h$. 
The invariant states have the same fermion numbers with respect to the three sets of fermions, which
allows us to identify the corresponding parity operators with the total fermionic parity $(-1)^{F_L+F_R}$,
\be 
(-1)^{\sum\limits_{k=1}^4 (F_{L,k}+F_{R,k})}
=(-1)^{\sum\limits_{k=5}^8 (F_{L,k}+F_{R,k})}
=(-1)^{\sum\limits_{k=9}^{12} (F_{L,k}+F_{R,k})}
=(-1)^{F_L+ F_R}\ .\label{orbcond}
\ee 
In particular, the space of $(g,h)$-invariant fields contains a \emph{bosonic} subspace 
(i.e.\ with positive total fermion parity)  generated by the 
$\widehat{\mathfrak s \mathfrak u}(2)_{L,1}^6\oplus \widehat{\mathfrak s \mathfrak u}(2)_{R,1}^6$ representations
\begin{align}%\label{NSNSu}
& \text{untw.} &[00\,00\,00;\,00\,00\,00] &&[11\,00\,00;\,11\,00\,00] && [00\,11\,00;\,00\,11\,00] && [00\,00\,11;\,00\,00\,11] &&\nonumber\\
%\label{NSNSg}
& g\text{-tw.} & [00\,10\,10;\,00\,10\,10] && [00\,10\,01;\,00\,10\,01] &&[00\,01\,10;\,00\,01\,10] && [00\,01\,01;\,00\,01\,01]\nonumber\\
%\label{NSNSh}
& h\text{-tw.} & [10\,00\,10;\,10\,00\,10] && [10\,00\,01;10\,00\,01] &&[01\,00\,10;\,01\,00\,10] && [01\,00\,01;\,01\,00\,01]\nonumber\\
%\label{NSNSgh}
& gh\text{-tw.} & [10\,10\,00;\,10\,10\,00] && [10\,01\,00;\,10\,01\,00] &&[01\,10\,00;\,01\,10\,00] && [01\,01\,00;\,01\,01\,00]
\label{NSNS}\end{align}
The representation content of this bosonic subspace corresponds therefore to the 
subgroup $\m A_{\rm bos}$ of $\Z_2^6\times \Z_2^6$,
\be \label{bosgroup}
\m A_{\rm bos}:=\{ [a_1,\ldots,a_6;b_1,\ldots,b_6]\in \Z_2^6\times \Z_2^6 \mid a_i=b_i\ ,\sum_{i=1}^6 a_i\equiv 0\mod 2 \}\ .
\ee
The entire NS-NS spectrum of the orbifold theory is generated by the fusion of these representations 
with one \emph{fermionic} representation (negative total fermion parity),  for example the representation
\be 
[11\,11\,11;\,00\,00\,00]\label{supercurrents}
\ee 
which contains the holomorphic fields of weight ${3\over2}$. The resulting subgroup of $\Z_2^6\times \Z_2^6$ 
describing the entire NS-NS spectrum of the theory is $\m A_{\rm bos}\cup \m A_{\rm ferm}$, where
\be \label{fermgroup}
\m A_{\rm ferm}:=\{ [a_1,\ldots,a_6;b_1,\ldots,b_6]\in \Z_2^6\times \Z_2^6 \mid a_i=b_i+1\ ,\sum_{i=1}^6 a_i\equiv 0\mod 2 \}
\ee 
accounts for the states of negative total fermion parity.
\bigskip

The R-R sector of our model is obtained by 
inverting the tensorial properties of the $\mathfrak{so}(4)_{L,1}^3\oplus \mathfrak{so}(4)_{R,1}^3$
representations with respect to the NS-NS sector, i.e.\ they are given by exchanging 
${\rm S}\leftrightarrow {\rm T}$ in \eqref{tensu} -- \eqref{tensgh}. Since we still need to obey \eqref{orbcond},
the R-R spectrum thus consists of the states
\begin{align}
\label{RRg}& g\text{-twisted} & [10\,00\,00;\,10\,00\,00] && [01\,00\,00;\,01\,00\,00] \\
\label{RRh}& h\text{-twisted} & [00\,10\,00;\,00\,10\,00] && [00\,01\,00;\,00\,01\,00] \\
\label{RRgh}& gh\text{-twisted} & [00\,00\,10;\,00\,00\,10] && [00\,00\,01;\,00\,00\,01] 
\end{align} 
together with all representations that can be obtained by fusion with the NS-NS representations \eqref{NSNS}
and \eqref{supercurrents}.

With this description of the 
entire R-R spectrum of our orbifold, it is then straightforward to calculate the elliptic genus
in terms of $\wh{\mathfrak{su}}(2)^6_1$ characters. This is done in Appendix~\ref{ellipticgenuscalculation},
where we show that the elliptic genus reproduces indeed that of K3.
\medskip

The structure of \eqref{bosgroup} and \eqref{fermgroup} reveals that the spectrum of the orbifold theory is 
invariant under {\em simultaneous} permutations of the six holomorphic and six anti-holomorphic 
$\widehat{\mathfrak s\mathfrak u}(2)_1$ algebras. This is actually  also a symmetry of the OPE, as is 
shown in Appendix~\ref{appC2}.  Hence the group of symmetries of the model is (at least) 
(SU(2)$_L^6\times $ SU(2)$_R^6) : S_6$. Notice that neither the free fermion theory on the bosonic
$D_6$-torus, nor its orbifold by $g$, corresponding to the supersymmetric
sigma model on the $D_4$-torus, 
contain such an $S_6$ symmetry. Therefore, one cannot generate the whole group of 
symmetries of the K3 sigma model just by considering the transformations induced by the 
symmetries of the parent theories.%, which should be contained in a $(SO(4)^3\times SO(4)^3) : S_3$ subgroup ({\bf is this true?}).

Our model contains $64$ holomorphic fields of weight ${3\over2}$, that generate several copies of the 
$\NNN=4$ superconformal algebra and, in particular, the four supercurrents \eqref{Gpl} -- \eqref{Gmin}. 
The corresponding $\widehat{\mathfrak s\mathfrak u}(2)_{L,1}$ algebra 
(whose zero modes generate the SU(2) R-symmetry group) is identified with the first factor of the 
full $\widehat{\mathfrak s\mathfrak u}(2)_{L,1}^6$ affine algebra. Therefore, the 
group of symmetries preserving the four left and four right-moving supercurrents must be a subgroup 
of the stabiliser $({\rm SU}(2)^5\times {\rm SU}(2)^5) : S_5$ of the first 
$\widehat{\mathfrak s\mathfrak u}(2)_{L,1}\oplus\widehat{\mathfrak s\mathfrak u}(2)_{R,1}$ factor.

%%%%%%%%%%%%%%%%
\section{Finite symmetries of the K3 sigma model}\label{sec4}
%%%%%%%%%%%%%%%%

In \cite{Gaberdiel:2011fg}, it was argued that the group $G$ of symmetries of a non-linear sigma model on K3, 
preserving the $\NNN=(4,4)$ superconformal algebra and the four R-R ground states that are charged under the 
R-symmetry, is always a subgroup of the Conway group $Co_0$. Generically, $G$ is a subgroup of 
$\Z_2^{12}:\mathbb{M}_{24}\subset Co_0$, and the only exceptions are given by orbifolds of torus models by cyclic groups 
of order $3$ or $5$ \cite{Gaberdiel:2012um}. 
\smallskip

In this paper we are interested in determining the group $G$ of symmetries of  the specific K3 sigma model 
described so far in three different ways:  as the $\Z_2$-orbifold of the $D_4$-torus model in Subsection \ref{geomD4};
as a free fermion model in Subsection \ref{torferm}; and as an RCFT based on the vertex operator construction of the 
$\widehat{\mathfrak s \mathfrak u}(2)_{L,1}^6\oplus \widehat{\mathfrak s \mathfrak u}(2)_{R,1}^6$ affine algebra
in Section~\ref{sec:su2}. Each of these three descriptions exhibits some of the relevant finite symmetries in a natural
way. Our first aim in this section is to represent all of these symmetries as elements of a subgroup of 
$({\rm SU}(2)_L^6\times {\rm SU}(2)_R^6) : S_6$, see the discussion at the end of the previous section.

We first consider the discrete `geometric' symmetries of the $\Z_2$-orbifold of the $D_4$-torus model 
as a guide, and we use the free fermion description to express these symmetries (and new ones 
discovered in the process) in terms of a subgroup of $({\rm SU}(2)_L^6\times {\rm SU}(2)_R^6) : S_6$. We then  turn to the 
$\widehat{\mathfrak s \mathfrak u}(2)_{L,1}^6\oplus \widehat{\mathfrak s \mathfrak u}(2)_{R,1}^6$ RCFT 
description of the K3 model to express generators of our symmetry group in a form that paves the way 
to the identification of a less obvious $A_5\subset S_6$ symmetry group. This group is 
a factor of the group $G$ of symmetries we are seeking. The full structure of $G$ will then be studied in 
Section~\ref{thegroup}.

\subsection{Symmetries from the $\Z_2$-orbifold  of the $D_4$-torus model}

By construction, the $\Z_2$-orbifold of the bosonic $D_4$-torus model has a geometric interpretation 
on the tetrahedral Kummer surface % \mg{$\wt{\mathbb T_{D_4}/\Z_2}$} 
that is obtained by minimally resolving 
all the singularities of ${\mathbb T_{D_4}/\Z_2}$, where $\mathbb T_{D_4}=\R^4/L_{D_4}$. 
The symmetry group of that Kummer surface was studied in detail in \cite{tawe11};
in particular, the group of holomorphic symplectic automorphisms is the group
${\cal T}_{192}\cong(\Z_2)^4 : A_4$ of order $192$.  The subgroup of type $A_4$ of ${\cal T}_{192}$
is induced by those symmetries of the underlying torus that have a fixed point. The remaining symmetries
in ${\cal T}_{192}$ are generated by including the translational subgroup $(\Z_2)^4$ (half-period shifts) which acts as 
a permutation group
on the twisted ground states and leaves the untwisted sector invariant. 

In this subsection we identify this group of symmetries, as well as some additional non-geometric generators,  
with a subgroup of  $({\rm SU}(2)_L^6\times {\rm SU}(2)_R^6) : S_6$. In fact, we
focus on the action of this group on the holomorphic and antiholomorphic currents generating the 
$\widehat{\mathfrak s \mathfrak u}(2)_{L,1}^6\oplus \widehat{\mathfrak s \mathfrak u}(2)_{R,1}^6$ algebra 
that survives the $\Z_2$-orbifold projection. Actually, we will only identify the group modulo the 
subgroup $\Z_2^6\times\Z_2^6$ of elements of $({\rm SU}(2)_L^6\times {\rm SU}(2)_R^6) : S_6$ that 
act trivially on the currents.

%%%%%%%%%%%%%%%%%%%%%
\subsubsection{Rotations}\label{rotations}
%%%%%%%%%%%%%%%%%%%%%

The group $A_4$ of rotations may be generated by the three following symmetries\footnote{Note that in \cite{tawe11}, a different choice of coordinates was used on the underlying torus. Also note that the minimal number of generators for $A_4$ is two. Indeed, $\gamma_2=\gamma_1^2\gamma_3\gamma_1\gamma_3^2$.}
\be
\gamma_1 = \left( \begin{array}{cccc} 0&1&0&0\\-1&0&0&0\\0&0&0&-1\\0&0&1&0 \end{array} \right),\quad
\gamma_2 = \left( \begin{array}{cccc} 0&0&-1&0\\0&0&0&-1\\1&0&0&0\\0&1&0&0 \end{array} \right),\quad
\gamma_3 = {\textstyle{1\over2}} \left( \begin{array}{cccc} -1&1&1&-1\\-1&-1&1&1\\-1&-1&-1&-1\\1&-1&1&-1 \end{array} \right),
\ee
which act on $\R^4$ and induce symmetries of the $D_4$-torus $\mathbb T_{D_4}$
that descend to the Kummer surface.
Since the $D_4$-torus model  has a non-trivial B-field 
$B$ given in \req{Bfield}, we need to ensure that these
rotations induce symmetries on the conformal field theory. 
This is the case if and only if $\gamma_k^T B \gamma_k=B$
for $k\in\{1,\,2,\,3\}$, and one immediately confirms that this latter condition is indeed obeyed.

We now give a description of the action of $\gamma_1$ and $\gamma_2$ in terms of the 
holomorphic fields in the free fermion 
model; this is sufficient in order to specify the symmetries in the form that is
needed in Section \ref{thegroup}.
With the notations of Section \ref{sec2} and Appendix~\ref{appB}, see
in particular \req{lambdaij}, we observe that the lattice vectors 
$\boldsymbol{l}_{1,2}$, $\boldsymbol{l}_{1,-2}$, $\boldsymbol{l}_{3,4}$, $\boldsymbol{l}_{3,-4}$ form a basis of $\R^4$. The symmetries
$\gamma_1$ and $\gamma_2$ permute the eight lattice vectors $\boldsymbol{l}_{\pm1,\pm2}$, $\boldsymbol{l}_{\pm3,\pm4}$,
inducing the following maps under the identification \req{currentident}
\be
\gamma_1\colon\left\{
\begin{array}{rcl}
\nop{x_1(z)x_2(z)} &\longmapsto& \nop{x_1(z)x_2^\ast(z)}\\
\nop{x_1^\ast(z)x_2^\ast(z)} &\longmapsto& \nop{x_1^\ast(z)x_2(z)}\\
\nop{x_1(z)x_2^\ast(z)} &\longmapsto& \nop{x_1^\ast(z)x_2^\ast(z)}\\
\nop{x_1^\ast(z)x_2(z)} &\longmapsto& \nop{x_1(z)x_2(z)}\\
\nop{x_3(z)x_4(z)} &\longmapsto& \nop{x_3^\ast(z)x_4(z)}\\
\nop{x_3^\ast(z)x_4^\ast(z)} &\longmapsto& \nop{x_3(z)x_4^\ast(z)}\\
\nop{x_3(z)x_4^\ast(z)} &\longmapsto& \nop{x_3(z)x_4(z)}\\
\nop{x_3^\ast(z)x_4(z)} &\longmapsto& \nop{x_3^\ast(z)x_4^\ast(z)}
\end{array}
\right\}\ ,\quad
\gamma_2\colon\left\{
\begin{array}{rcl}
\nop{x_1(z)x_2(z)} &\longmapsto& \nop{x_3(z)x_4(z)}\\
\nop{x_1^\ast(z)x_2^\ast(z)} &\longmapsto& \nop{x_3^\ast(z)x_4^\ast(z)}\\
\nop{x_1(z)x_2^\ast(z)} &\longmapsto& \nop{x_3(z)x_4^\ast(z)}\\
\nop{x_1^\ast(z)x_2(z)} &\longmapsto& \nop{x_3^\ast(z)x_4(z)}\\
\nop{x_3(z)x_4(z)} &\longmapsto&\nop{x_1^\ast(z)x_2^\ast(z)} \\
\nop{x_3^\ast(z)x_4^\ast(z)} &\longmapsto& \nop{x_1(z)x_2(z)}\\
\nop{x_3(z)x_4^\ast(z)} &\longmapsto& \nop{x_1^\ast(z)x_2(z)}\\
\nop{x_3^\ast(z)x_4(z)} &\longmapsto&\nop{x_1(z)x_2^\ast(z)}
\end{array}
\right\}\ .
\ee
These maps are induced by
\be
\begin{array}{rcl}
\gamma_1\colon && x_1\mapsto x_2^\ast\mapsto -x_1^\ast\mapsto -x_2\mapsto x_1,\; 
x_3\mapsto x_4 \mapsto -x_3^\ast\mapsto -x_4^\ast\mapsto x_3 \\
\gamma_2\colon && x_1\mapsto x_3\mapsto x_1^\ast\mapsto x_3^\ast\mapsto x_1,\; 
x_2\mapsto x_4\mapsto x_2^\ast\mapsto x_4^\ast\mapsto x_2\ ,
\end{array}
\ee
from which one obtains the actions on all $(1,0)$-fields in the orbifold.
Equivalently, for $\psi_{{5}},\ldots,\psi_{{12}}$ we have
\be
\begin{array}{rcl}
\gamma_1\colon && \psi_{{5}}\mapsto \psi_{{6}}\mapsto -\psi_{{5}}, \;
\psi_{{7}}\mapsto \psi_{{8}}\mapsto -\psi_{{7}},\;
\psi_{{9}}\leftrightarrow -\psi_{{10}},\; 
\psi_{{11}}\leftrightarrow \psi_{{12}} \\
\gamma_2\colon && \psi_{{5}}\leftrightarrow \psi_{{7}},\; 
\psi_{{6}}\leftrightarrow \psi_{{8}},\;
\psi_{{9}}\mapsto \psi_{{11}}\mapsto -\psi_{{9}},\;
\psi_{{10}}\mapsto \psi_{{12}}\mapsto -\psi_{{10}}\ .
\end{array}
\ee
The action on the superpartners of the four bosonic U(1) currents  $j_k,\; k=1,\ldots,4$, is
\be
\begin{array}{rcl}
\gamma_1\colon &&\psi_1\mapsto -\psi_2\mapsto -\psi_1\ ,%\mapsto \psi_2\mapsto \psi_1 
\qquad \psi_3\mapsto \psi_4\mapsto -\psi_3\ ,\\%\mapsto \psi_4\mapsto \psi_3 
\gamma_2\colon &&\psi_1\mapsto \psi_3\mapsto -\psi_1\ ,%\mapsto \psi_2\mapsto \psi_1 
\qquad \psi_2\mapsto \psi_4\mapsto -\psi_2\ .%\mapsto \psi_4\mapsto \psi_3
\end{array}
\ee
Finally, in terms of the Dirac fermions \eqref{Diracblock1} and \eqref{Dirac}, the $\gamma_1$- and $\gamma_2$-actions read,
\begin{align}
& \chi_{1}\mapsto i\chi_{1}\ , & & \chi_{2}\mapsto -i\chi_{2}\ , & &\chi^*_{1}\mapsto -i\chi^*_{1}\ , 
& & \chi^*_{2}\mapsto i\chi^*_{2}\ ,\\
\gamma_1:\quad& \chi_{3}\mapsto -i\chi_{3}\ , & & \chi_{4}\mapsto -i\chi_{4}\ , & &\chi^*_{3}\mapsto i\chi^*_{3}\  , 
& & \chi^*_{4}\mapsto i\chi^*_{4}\ ,\\
& \chi_{5}\mapsto -i\chi^*_{5}\ , & & \chi_{6}\mapsto i\chi^*_{6}\ , & &\chi^*_{5}\mapsto i\chi_{5}\ , 
& & \chi^*_{6}\mapsto -i \chi_{6}\ .
\end{align} 
\begin{align}
& \chi_{1}\mapsto \chi_{2}\ , & & \chi_{2}\mapsto -\chi_{1}\ , & &\chi^*_{1}\mapsto \chi^*_{2}\ , 
& & \chi^*_{2}\mapsto -\chi^*_{1}\ ,\\
\gamma_2:\qquad& \chi_{3}\mapsto \chi_{4}\ , & & \chi_{4}\mapsto \chi_{3}\ , & &\chi^*_{3}\mapsto \chi^*_{4}\ , 
& & \chi^*_{4}\mapsto \chi^*_{3}\ ,\\
& \chi_{5}\mapsto \chi_{6}\ , & & \chi_{6}\mapsto -\chi_{5}\ , & &\chi^*_{5}\mapsto \chi^*_{6}\ , 
& & \chi^*_{6}\mapsto -\chi^*_{5}\ .
\end{align} 
The transformation induced on the $\widehat{\mathfrak{su}}(2)_1^6$ currents 
by \eqref{Rsymm} -- \eqref{Rsymm2} and the analogous formulas for $k=3,\,\ldots,\,6$ are
\begin{align}\label{g1a}
& J^{3,1}\;,\ J^{\pm,1}\text{ fixed,} & & J^{3,2}\text{ fixed,}\quad J^{\pm,2}\leftrightarrow - J^{\pm,2}\ , \\
\gamma_1:\qquad& J^{3,3}\text{ fixed,}\quad J^{\pm,3}\leftrightarrow - J^{\pm,3}\ , 
& & J^{3,4}\;,\ J^{\pm,4}\text{ fixed,}\label{g1b}\\
& J^{3,5}\leftrightarrow -J^{3,5}\ ,\quad J^{\pm,5}\leftrightarrow  J^{\mp,5}\ , 
& & J^{3,6}\leftrightarrow - J^{3,6}\ ,\quad 
J^{\pm,6}\leftrightarrow - J^{\mp,6}\ .\label{g1c}
\end{align}
\begin{align}\label{g2a}
& J^{3,1}\; ,\ J^{\pm,1}\text{ fixed,} & & J^{3,2}\leftrightarrow -J^{3,2}\ ,\quad J^{\pm,2}\leftrightarrow  J^{\mp,2}\ ,\\
\gamma_2:\qquad& J^{3,3}\text{ fixed,}\quad J^{\pm,3}\leftrightarrow - J^{\pm,3}\ , 
& &  J^{3,4}\leftrightarrow -J^{3,4}\ ,\quad J^{\pm,4}\leftrightarrow  -J^{\mp,4}\ ,\label{g2b}\\
& J^{3,5}\; , \ J^{\pm,5}\text{ fixed,} & & J^{3,6}\leftrightarrow -J^{3,6}\ ,\quad J^{\pm,6}\leftrightarrow  J^{\mp,6}\ .\label{g2c}
\end{align}
The transformations  $\gamma_1,\gamma_2$ form a subgroup $\Z_2^2$ of ${\rm SU}(2)_L^6\times {\rm SU}(2)_R^6$. 
The symmetry $\gamma_3$ has a non-trivial image in $S_6$ 
and will not be needed to generate the group of discrete symmetries we are seeking.

\subsubsection{Half-period shifts and more}\label{halfperiods}

The spectrum of the torus model can be naturally decomposed 
into eigenstates of the zero modes of the 
currents $j_k(z)=i\partial \phi_k(z),\; \qu\jmath_k(\qu z)=i\bar\partial\, \qu \phi_k(\qu z),\, k=1,\ldots,\,4$.  
For the $D_4$-torus model, the possible eigenvalues are given by 
the charge lattice $\Gamma=\Gamma_{4,4}$ of \eqref{Ddcharge}.

In order to construct operators that commute with the orbifold action 
$j_k\mapsto -j_k$ we consider an element 
$({\bf a};\qu {\bf a})\in (\frac{1}{2}\Gamma_{4,4})/\Gamma_{4,4}\cong \Z_2^8$ and 
define its action on the states with charge $({\bf Q};\qu {\bf Q})\in\Gamma_{4,4}$ as 
\be \label{Z28}
e^{2\pi i ({\bf a};\qu {\bf a})\bullet ({\bf Q};\qu {\bf Q})}=(-1)^{2({\bf a};\qu {\bf a})\bullet ({\bf Q};\qu {\bf Q})} \ .
\ee 
In particular, a shift by half a period $\frac{1}{2}{\boldsymbol l}$,  with $\boldsymbol l\in L_{D_4}$, 
corresponds to a symmetry $s_{({\bf a};\qu {\bf a})}$ with 
$({\bf a};\qu {\bf a})=\left[\frac{1}{2}({\bf Q}(0,{\boldsymbol l});\qu {\bf Q}(0,{\boldsymbol l}))\right]
\in(\frac{1}{2}\Gamma_{4,4})/\Gamma_{4,4}$, where we have used \eqref{geometriccharges}. 
The half-period shifts form a  subgroup $\Z_2^4$ of $(\frac{1}{2}\Gamma_{4,4})/\Gamma_{4,4}$ and act by 
multiplication by $(-1)^{{\boldsymbol m}\cdot {\boldsymbol l}}$ on all states of momentum 
${\boldsymbol m}\in L_{D_4}^*$, independently of their winding numbers. The entire group $\Z_2^8$  of symmetries 
\eqref{Z28} is generated by the half-period shifts of the $D_4$-torus model together with the half-period shifts in 
the T-dual  torus model. In the following, we will refer to all these symmetries generically as half-period shifts.
For concreteness we choose a set of generators for
$(\frac{1}{2}\Gamma_{4,4})/\Gamma_{4,4}$ to be of the form
\begin{align*} 
&u:=\frac{1}{2}(\tfrac{1}{2},\tfrac{1}{2},\tfrac{1}{2},\tfrac{1}{2};\;-\tfrac{1}{2},-\tfrac{1}{2},-\tfrac{1}{2},-\tfrac{1}{2})\ ,\\
&v_1:=\frac{1}{2}(1,0,0,0;\;-1,0,0,0)\ ,
&& v_2:=\frac{1}{2}(0,1,0,0;\;0,-1,0,0)\ ,\\ 
&v_3:=\frac{1}{2}(0,0,1,0;\;0,0,-1,0)\ , &&v_4:=\frac{1}{2}(0,0,0,1;\;0,0,0,-1)\ , \\
&w_{12}:=\frac{1}{2}(1,-1,0,0;\;0,0,0,0)\ ,&& w_{23}:=\frac{1}{2}(0,1,-1,0;\;0,0,0,0)\ ,\\ 
&w_{24}:=\frac{1}{2}(0,1,0,-1;\;0,0,0,0)\ , &&w_{14}:=\frac{1}{2}(1,0,0,1;\;0,0,0,0)\ , 
\end{align*}
where we have  the relation $v_1+v_2+v_3+v_4= 0\in({1\over2}\Gamma_{4,4})/\Gamma_{4,4}$.

Let us now describe the action of the elements $s_{({\bf a};\qu {\bf a})}$ on the $\widehat{\mathfrak{su}}(2)_{L,1}^6\oplus \widehat{\mathfrak{su}}(2)_{R,1}^6$ currents, using the free fermion description of the model. In this description, the zero 
mode of  the ${\rm U}(1)$ current $j_k(z) = -i\nop{\psi_{k+4}\psi_{k+8}}(z)$, 
$k=1,\ldots,4$, (see \eqref{boscurrents}) is the generator of rotations in the plane spanned by 
$\psi_{k+4}$ and $\psi_{k+8}$. 
\medskip

$\bullet$ The generators $s_{v_k}$, $k=1,\ldots,4$ act by
\be 
s_{v_k}\colon\qquad 
\psi_{k+4}\mapsto -\psi_{k+4}\ ,\qquad \psi_{k+8}\mapsto -\psi_{k+8}\ ,\qquad 
\overline\psi_{k+4}\mapsto -\overline\psi_{k+4}\ ,\qquad \overline\psi_{k+8}\mapsto -\overline\psi_{k+8} \ ,
\ee 
while all the other fermions $\psi_l,\;\overline\psi_l$ with $l\not\in\{k+4,\, k+8\}$
are fixed by $s_{v_k}$. 
For instance, using  \eqref{Dirac}, one sees that $s_{v_4}$ acts 
on the holomorphic fields by
\be 
s_{v_4}\colon\qquad
\chi_{4}\leftrightarrow \chi^*_{4}\ ,\qquad  \chi_{6}\leftrightarrow\chi^*_{6}\ ,\qquad
\chi_{k}\;,\ \chi_{k}^*\quad \text{ fixed for } k\in\{1,2,3,5\}\ .
\ee 
Thus, the induced action on the $\widehat{\mathfrak s\mathfrak u}(2)_{L,1}^6$ holomorphic currents 
is\footnote{Recall \eqref{Rsymm} - \eqref{Rsymm2} and analogous expressions for $J^{3,k}$ and $J^{\pm, k}$ for $k=3,\ldots,6$.}
\be \label{sv4}
s_{v_4}\colon\qquad
J^{a,3}\leftrightarrow J^{a,4}\ ,\qquad J^{a,5}\leftrightarrow J^{a,6}\ ,\qquad
J^{a,1}\;,\ J^{a,2} \text{ fixed,}\qquad a\in\{3,+,-\}\ .
\ee 
Therefore, $s_{v_4}$ corresponds to a $(34)(56)$ permutation acting simultaneously on the six left and the six right 
${\rm SU}(2)$ factors in  ${\rm SU}(2)_L^6\times {\rm SU}(2)_R^6$.
\medskip

$\bullet$ The symmetry  $s_u$ acts by a simultaneous 90-degree rotation in all planes 
$(i+4,i+8)$, $i=1,\ldots,4$, that is
\be 
s_{u}\colon\qquad
\psi_{i+4}\mapsto \psi_{i+8}\ ,\qquad \psi_{i+8}\mapsto -\psi_{i+4}\ ,\qquad 
\overline\psi_{i+4}\mapsto \overline\psi_{i+8}\ ,\qquad \overline\psi_{i+8}\mapsto -\overline\psi_{i+4}\ ,
\ee 
for all $i=1,\ldots,4$. The induced action on the holomorphic currents is
\be \label{su}
s_{u}\colon\qquad
J^{a,3}\leftrightarrow J^{a,5}\ ,\qquad J^{a,4}\leftrightarrow J^{a,6}\ ,
\qquad J^{a,1}\; ,\ J^{a,2} \text{ fixed,}\qquad a\in\{3,+,-\}\ .
\ee 
Therefore, $s_{u}$ corresponds to a $(35)(46)$ permutation acting simultaneously on the six left and 
the six right ${\rm SU}(2)$ factors in  ${\rm SU}(2)_L^6\times {\rm SU}(2)_R^6$.
\medskip

$\bullet$ The elements $s_{v_i+v_j}$, $1\le i<j\le 4$, correspond to elements in 
${\rm SU}(2)_L^6\times {\rm SU}(2)_R^6$. In particular, $s_{v_1+v_2}$ acts on the holomorphic fields by 
\be
\begin{array}{r}
s_{v_1+v_2}\colon\qquad
\chi_{3}\leftrightarrow -\chi_{3}\ ,\qquad \chi^*_{3}\leftrightarrow -\chi^*_{3}\ ,\qquad
\chi_{5}\leftrightarrow -\chi_{5}\ ,\qquad \chi^*_{5}\leftrightarrow -\chi^*_{5}\ , \\[0.5em]
\chi_{k}\; ,\ \chi_{k}^* \ \text{ fixed for } k\in\{1,2,4,6\}\ .
\end{array}
\ee 
Therefore, the induced transformation on the holomorphic currents is
\be\label{s12} 
s_{v_1+v_2}\colon\qquad
J^{\pm,i}\leftrightarrow -J^{\pm,i}\ ,\quad J^{3,i}\text{ fixed,}\ \ i=3,4,5,6\ ,
\qquad J^{\pm,1}\; ,\ J^{3,1}\; ,\ J^{\pm,2}\; ,\ J^{3,2}  \text{ fixed.}
\ee 
Furthermore, $s_{v_2+v_4}$ acts on the holomorphic fields by 
\be 
s_{v_2+v_4}\colon\qquad
\chi_{3}\leftrightarrow \chi^*_{3}\ ,\qquad \chi_{4}\leftrightarrow \chi^*_{4}\ ,\qquad
\chi_{5}\leftrightarrow \chi^*_{5}\ ,\qquad \chi_{6}\leftrightarrow \chi^*_{6}\ ,
\ee 
so that the induced action on the holomorphic currents is
\be\label{s24} 
s_{v_2+v_4}\colon\qquad
J^{3,i}\leftrightarrow -J^{3,i}\ ,\quad J^{\pm,i}\leftrightarrow J^{\mp,i}\ ,\ \ i=3,4,5,6\ ,
\qquad
J^{\pm,1}\; ,\  J^{3,1}\; ,\  J^{\pm,2}\; ,\  J^{3,2} \text{ fixed.}
\ee 
\medskip

$\bullet$ The symmetries $s_{w_{ij}}$ act on the left-moving currents in the same way as 
$s_{v_i+v_j}$, $1\le i<j\le 4$, while they leave the right-moving currents fixed. In particular, 
on the left-moving currents, $s_{w_{12}}$ 
acts as $s_{v_1+v_2}$ as given in \eqref{s12}, 
while $s_{w_{24}}$  acts as $s_{v_2+v_4}$ according to \eqref{s24}. 
Furthermore, $s_{w_{23}+w_{14}}$ and $s_{w_{12}+w_{24}+w_{14}}$ act trivially on all currents,
so they must correspond to elements in the center of ${\rm SU}(2)^6_L\times {\rm SU}(2)^6_R$; however, 
$w_{23}+w_{14},\, w_{12}+w_{24}+w_{14}\not\equiv0\in({1\over2}\Gamma_{4,4})/\Gamma_{4,4}$.

%%%%%%%%%%%%%%%%%% 
\subsubsection{Quantum symmetry}\label{quantumsymm}
%%%%%%%%%%%%%%%%%%%
Apart from the geometric symmetries, our K3 model has a quantum symmetry $Q$ of order $2$ that acts by 
$-1$ on the twisted sector and fixes the untwisted sector. In the free fermion description, $Q$ acts by $-1$ on the $h$- and $gh$-twisted sectors and trivially on the $g$-twisted and untwisted sectors.
By \eqref{NSNS}, \eqref{supercurrents} and \eqref{RRg} -- \eqref{RRgh} this implies that $Q$ acts by
\be\label{orbireverse}
Q([a_1,\ldots,a_6;b_1,\ldots,b_6])=(-1)^{a_3+a_4+a_5+a_6}[a_1,\ldots,a_6;b_1,\ldots,b_6]\ .
\ee

%%%%%%%%%%%%%%%%%% 
\subsection{Symmetries in the $\widehat{\mathfrak{su}}(2)_{L,1}^6\oplus \widehat{\mathfrak{su}}(2)_{R,1}^6$ RCFT}
\label{symmetries}
%%%%%%%%%%%%%%%%%% 

As is manifest from the  description of the model
as an $\widehat{\mathfrak{su}}(2)_{L,1}^6\oplus \widehat{\mathfrak{su}}(2)_{R,1}^6$ RCFT in Section~\ref{sec:su2}, 
the group of symmetries of our K3 model is $({\rm SU}(2)_L^6\times {\rm SU}(2)_R^6) : S_6$, see 
Appendix~\ref{appC2} for a detailed proof. 
We are ultimately interested in identifying the finite subgroup of $({\rm SU}(2)_L^6\times {\rm SU}(2)_R^6) : S_6$ that 
preserves the $\NNN=(4,4)$ superconformal algebra \eqref{Gpl} -- \eqref{Gmin}, and that fixes the four R-R ground 
states which transform as doublets under the left- and right-moving SU(2) R-symmetries.
In this section, we describe a number of symmetries in terms of the 
$\widehat{\mathfrak{su}}(2)_{L,1}^6\oplus \widehat{\mathfrak{su}}(2)_{R,1}^6$ RCFT.
In Section \ref{thegroup} we then prove that these symmetries generate the symmetry
group $G= \Z_2^8:\mathbb{M}_{20}$.

Some elements in the  group $({\rm SU}(2)_L^6\times {\rm SU}(2)_R^6) : S_6$
of symmetries of the 
$\widehat{\mathfrak s\mathfrak u}(2)_{L,1}^6\oplus \widehat{\mathfrak s\mathfrak u}(2)_{R,1}^6$ 
RCFT obviously leave the four supercharges invariant. In particular, this is true for the 
central elements $t_i, \qu t_i$, $i=1,\ldots,6$, of  ${\rm SU}(2)_L^6\times {\rm SU}(2)_R^6$  
which fix the $\widehat{\mathfrak s\mathfrak u}(2)_{L,1}^6\oplus \widehat{\mathfrak s\mathfrak u}(2)_{R,1}^6$ 
currents and  act on their representations  by
\begin{align}
& t_i ([a_1,\ldots,a_6;b_1,\ldots,b_6])=(-1)^{a_i}[a_1,\ldots,a_6;b_1,\ldots,b_6]\ ,\nonumber\\
& \qu t_i ([a_1,\ldots,a_6;b_1,\ldots,b_6])=(-1)^{b_i}[a_1,\ldots,a_6;b_1,\ldots,b_6]\ .
\label{central}
\end{align}
Since \eqref{G+chi} -- \eqref{Gprime-chi} implies that the holomorphic
supercurrents of the $\NNN=(4,4)$ superconformal algebra transform in the representation
$[11\, 11\,  11;\,  00\,  00\,  00]$, it follows that the
subgroup which fixes the $\NNN=(4,4)$ superconformal algebra is generated by
\be 
t_it_j =\qu t_i\qu t_j\ ,\qquad 1\le i<j\le 6\ ,
\ee
where we recall from \eqref{bosgroup} and \eqref{fermgroup} that the spectrum of the K3 model 
contains only representations with $a_i=b_i$ or with $a_i=b_i+1$, so that $t_it_j$ and $\qu t_i\qu t_j$ 
corresponds to the same symmetry of the K3 model. 
Therefore, we obtain a  group $\Z_2^5$ of symmetries preserving the $\NNN=(4,4)$ algebra. 
\smallskip

In the R-R sector, the elements $t_1t_i$, $1<i\le 6$, act by 
multiplication with $(-1)$ on the four charged R-R ground states in the
representation  $[10\,00\,00; 10\,00\,00]$ in \eqref{RRg}, corresponding to the $\NNN=(4,4)$ supermultiplet
$(\frac{1}{4},\frac{1}{2};\frac{1}{4},\frac{1}{2})$. In order to preserve these states, we should compose 
$t_1t_i$ with the symmetry $(-1)^R$ that acts by multiplication with
$(-1)$ on the R-R sector and trivially on the NS-NS sector. Note, however, that the symmetry
\be 
(-1)^Rt_1t_2t_3t_4t_5t_6
\ee 
acts trivially on all the states in the spectrum. Therefore, the subgroup of $\Z_2^6\times \Z_2^6$ preserving the $\NNN=(4,4)$ superconformal algebra and the `charged' R-R ground states is $\Z_2^4$, and it is generated by
\be t_2t_j\ ,\qquad  j=3,4,5,6\ .
\ee
This group contains the quantum symmetry $Q$ of the $\Z_2$-orbifold of the $D_4$-torus model, 
which by \eqref{orbireverse} is given by
\be \label{Q3456}
Q_{3456}:=Q= t_3t_4t_5t_6\ .
\ee
\smallskip

Let us next consider the symmetries that are induced from the $D_4$-torus model in the affine algebra description.
We focus our attention on those symmetries which in Section \ref{thegroup} are shown
to generate the symmetry group $G= \Z_2^8:\mathbb{M}_{20}$. As we have seen above, the 
half-period shifts $s_{v_4}$ in \eqref{sv4} and $s_u$ in \eqref{su} correspond to the permutations
\be \label{sv4su}
s_{v_4} \wh = (34)(56)\ ,\qquad s_{u}  \wh = (35)(46) 
\ee 
of the currents and of the corresponding representations.

The half-period shifts $s_{v_2+v_4}$ and $s_{v_1+v_2}$ and the rotations $\gamma_1$, $\gamma_2$ act as 
left-right symmetric ${\rm SU}(2)_L^6\times {\rm SU}(2)_R^6$ transformations.  They are determined by their 
action on the currents (see eqs.~\eqref{s12}, \eqref{s24}, \eqref{g1a} -- \eqref{g1c}, \eqref{g2a} -- \eqref{g2c}), 
up to elements in the centre $\Z_2^6\times \Z_2^6$ of ${\rm SU}(2)_L^6\times {\rm SU}(2)_R^6$. It is convenient to 
define the ${\rm SU}(2)$ matrices
\be \label{f4matrices}
0:= \begin{pmatrix} 1& 0\\ 0 & 1\end{pmatrix}\ ,\qquad 1:= \begin{pmatrix} 0& 1\\ -1 & 0\end{pmatrix}\ ,
\qquad 
\omega:= \begin{pmatrix} 0& i\\ i & 0\end{pmatrix}\ ,\qquad \bar\omega := \begin{pmatrix} -i& 0\\ 0 & i\end{pmatrix}\ ,
\ee 
which act by conjugation on $\mathfrak{su}(2)_{1,L}$ and $\mathfrak{su}(2)_{1,R}$.
The unusual notations are motivated by the particular representations
of these matrices which we discover in Section \ref{thegroup}. Here we only note that with
these four matrices,
\be
{\rm SU}(2) = \left\{ a_0\cdot 0 + a_1\cdot 1 + a_2\cdot \omega + a_3\cdot \qu\omega \ \left| \ 
a_0,\,\ldots,\,a_3\in\R,\; a_0^2+\cdots+a_3^2=1 \right. \right\}
\ee
is realised as the group of unit quaternions. We also observe a natural symmetry of
order $3$ on ${\rm SU}(2)$, induced by the cyclic permutation of $(1,\omega,\qu\omega)$.
It is the inner automorphism $\mu(\omega)$ of the quaternion algebra which is given by
conjugation with $\hf(0+1+\omega+\qu\omega)$
\be\label{F4mult}
\begin{array}{rcl}
\fa A\in SU(2)\colon\quad
\mu(\omega)(A)&:=& \Omega^{-1} A \Omega\ ,\quad  \Omega:= \hf(0+1+\omega+\qu\omega)\ ,\\[0.5em]
\Longrightarrow && \mu(\omega)(0,1,\omega,\qu\omega) = (0,\omega,\qu\omega,1)\ .
\end{array}
\ee 
We denote by 
\be \rho_{L,R}: {\rm SU}(2)^6\to {\rm SU}(2)_L^6\times {\rm SU}(2)_R^6
\ee 
the diagonal (i.e.\ left-right symmetric) embedding. Then,
\be\label{sv2v4}
\begin{array}{rlrl} 
s_{v_2+v_4}&=\rho_{L,R}(00\,\omega\omega\,\omega\omega)\ , \qquad
& s_{v_1+v_2}&=\rho_{L,R}(00\,\bar\omega\bar\omega\,\bar\omega\bar\omega)\ ,\\[0.5em]
\gamma_1&=\rho_{L,R}(0\bar\omega\,\bar\omega 0\, \omega 1)\ ,& 
\gamma_2&=\rho_{L,R}(0\omega\,\bar\omega 1 \,0\omega)\ .
\end{array}
\ee
We observe that 
the commutator of any two such elements is in the centre $\Z_2^6\times \Z_2^6$ of 
${\rm SU}(2)_L^6\times {\rm SU}(2)_R^6$. More precisely, it is always a product of an even number of symmetries 
$t_i \qu t_i$, which, by the discussion above, acts trivially on all the states in the theory. Thus,  the 
four symmetries in \eqref{sv2v4} 
effectively generate an abelian subgroup $\Z_2^4$ of the group of symmetries preserving the $\NNN=(4,4)$ superconformal algebra.
\medskip

The geometric symmetries above are left-right symmetric elements of ${\rm SU}(2)_L^6\times {\rm SU}(2)_R^6$. 
However, it is clear that the action of the purely left-moving or purely right-moving ${\rm SU}(2)^6$ will also preserve the 
$\NNN=(4,4)$ superconformal algebra and the four R-R ground states that are charged under the R-symmetry. Two 
examples of such purely left-moving transformations are $s_{w_{12}}$ and $s_{w_{24}}$. Thus, if we define the 
embedding
\be \rho_L:{\rm SU}(2)^6\to {\rm SU}(2)_L^6\times {\rm SU}(2)_R^6
\ee
into the \emph{left} ${\rm SU}(2)_L^6$ factor, we obtain four additional symmetries
\be\label{leftsymm} 
\begin{array}{rlrl}
s_{w_{24}}&=\rho_{L}(00\,\omega\omega\,\omega\omega)\ , \qquad\qquad
& s_{w_{12}}&=\rho_{L}(00\,\bar\omega\bar\omega\,\bar\omega\bar\omega)\ ,\\[0.5em]
\gamma_1^L&=\rho_{L}(0\bar\omega\,\bar\omega 0\, \omega 1)\ ,& 
\gamma_2^L&=\rho_{L}(0\omega\,\bar\omega 1 \,0\omega)\ ,
\end{array}
\ee
that form a subgroup of the left-moving ${\rm SU}(2)_L^6$, and that have no geometric interpretation. 
The commutators of these elements are again in the centre $\Z_2^6$ of the left ${\rm SU}(2)_L^6$, but, in general, 
they act non-trivially on the states of the theory. Therefore, the resulting group is \emph{non-abelian}.

%%%%%%%%%%%%%%%%%%%
\subsection{Further symmetries from other fermionisation choices}\label{s:newtorus}
%%%%%%%%%%%%%%%%%%%%
Apart form the quantum symmetries $t_2t_j$, $j=3,\,4,\,5,\,6$, all symmetries
which we have constructed so far have some geometric origin, for
example by restricting a geometrically induced symmetry to its left-moving part as in 
\eqref{leftsymm}. In this subsection we obtain additional symmetries by making 
different fermionisation choices, thus giving rise to other symmetries. More precisely, we  regroup the summands of our
$\wh{\frak{su}}(2)_{L,1}^6\oplus \wh{\frak{su}}(2)_{R,1}^6$ current algebra
into $\wh{\frak{so}}(4)_{1}$ pieces in a different fashion. To do so, we remark that
taking the orbifold of our K3 model by the order two quantum symmetry 
$Q = Q_{3456}$ of \eqref{Q3456} recovers
the original $D_4$-torus model. On the other hand,  by conjugating $Q$ by elements in 
$({\rm SU}(2)_L^6\times {\rm SU}(2)_R^6):S_6$, we obtain $15$ different symmetries
\be\label{newQ} 
Q_{ijkl}= t_it_jt_kt_l\ ,\qquad 1\le i<j<k<l\le 6\ .
\ee 
The orbifold by any of these symmetries is  a $D_4$-torus model 
equivalent to the one we considered previously.

Furthermore, the five symmetries $Q_{ijkl}$ with $1<i<j<k<l$ preserve the four supercharges as well as the four 
R-R ground states in the $(\frac{1}{4},\frac{1}{2};\frac{1}{4},\frac{1}{2})$ supermultiplet. Thus, we have five 
\emph{different} descriptions of our K3 model (considered as an $\NNN=(4,4)$ superconformal model) as a 
$\Z_2$-orbifold of a $D_4$-torus model. 
Notice that all these torus models are actually equivalent to one another, in the sense that we can identify the fields of 
any two of them so that both the OPEs and the $\NNN=(4,4)$ superconformal algebra are preserved. Nevertheless, these 
descriptions are different in the sense that the fields of the torus orbifold are associated to different fields of the K3 model --- 
in particular, we have a different splitting between the twisted and untwisted sector of the model. As a consequence, the 
symmetries of the K3 model, induced by the geometric symmetries of one underlying $D_4$-torus description, might 
correspond to symmetries mixing twisted and untwisted states in a different  description. In this subsection, we  
show that this is indeed the case.
\smallskip 

Let us consider the orbifold of our K3 model by $Q'=Q_{2345}$ instead of $Q= Q_{3456}$. 
This symmetry commutes with the 
$\widehat{\mathfrak s \mathfrak u}(2)_{L,1}^6\oplus \widehat{\mathfrak s \mathfrak u}(2)_{R,1}^6$ affine algebra, that is 
therefore preserved in the orbifold model. The orbifold model will also contain four left-moving Majorana fermions 
$\widehat\psi_1,\ldots,\widehat\psi_4$ of weight  $({1\over2},0)$ and four right-moving fermions 
$\widehat{\qu\psi}_1,\ldots,\widehat{\qu\psi}_4$ of weight $(0,{1\over2})$, transforming in the representations
\be 
[10\, 00\, 01;\, 00\, 00\, 00]\ ,\qquad [00\, 00\, 00;\, 10\, 00\, 01]\ 
\ee of 
$\widehat{\mathfrak s \mathfrak u}(2)_{L,1}^6\oplus \widehat{\mathfrak s \mathfrak u}(2)_{R,1}^6$, respectively. 
Furthermore, there are $2^4=16$ left-moving and $16$ right-moving currents, transforming in the 
$\widehat{\mathfrak s \mathfrak u}(2)_{L,1}^6\oplus \widehat{\mathfrak s \mathfrak u}(2)_{R,1}^6$ representations
\be [01\, 11\,10;\, 00\, 00\, 00]\ ,\qquad [00\, 00\, 00;\, 01\, 11\, 10]\ ,
\ee respectively. These currents enhance the affine algebra to 
$\widehat{\mathfrak s\mathfrak o}(8)_{L,1}\oplus \widehat{\mathfrak s\mathfrak o}(8)_{R,1}$. 
Via fermionisation, we can describe these currents as bilinears  
in the eight  Majorana fermions $\widehat\psi_5,\ldots,\widehat\psi_{12}$.

Analogously to \eqref{Dirac}, we
arrange  the twelve Majorana fermions into six Dirac fermions $\widehat\chi_{j}$ as
\begin{align}
\widehat\chi_{j}(z)&:=\frac{1}{\sqrt 2} (\widehat\psi_{2j-1}(z)+i\widehat\psi_{2j}(z))\ ,&  
\widehat\chi^*_{j}(z)&:=\frac{1}{\sqrt 2} (\widehat\psi_{2j-1}(z)-i\widehat\psi_{2j}(z))\ ,
\quad j=1,\ldots,6.
\end{align}
Formally, this description is completely analogous to that of the original $D_4$-torus model. 
However, the six Dirac fermions transform in different representations of the 
$\widehat{\mathfrak s \mathfrak u}(2)^6_1$ current algebra. 
Correspondingly, the expression of the  $\widehat{\mathfrak s\mathfrak u}(2)_{L,1}^6$-currents in terms of the 
Dirac fermions is different, namely (for the Cartan torus)
\begin{align}
J^{3,1}(z)  & = \frac{1}{2} (\nop{\widehat\chi^*_{1}(z) \widehat\chi_{1} (z)}+\nop{\widehat\chi^*_{2} (z) \widehat\chi_{2}(z)})\ ,\\
%J^{+,1}(z) &= i:\chi^*_{1}(z) \, \chi^*_{2}(z) : \ , \qquad
%J^{-,1}(z) =  i  :\chi_{1}(z) \, \chi_{2}(z) : \ ,\\
J^{3,6}(z)  & = \frac{1}{2}(\nop{\widehat\chi^*_{1}(z) \widehat\chi_{1} (z)}-\nop{\widehat\chi^*_{2} (z) \widehat\chi_{2}(z)})\ ,\\
J^{3,3}(z)  & = \frac{1}{2} (\nop{\widehat\chi^*_{3}(z) \widehat\chi_{3} (z)}+\nop{\widehat\chi^*_{4} (z) \widehat\chi_{4}(z)})\ ,\\
J^{3,4}(z)  & = \frac{1}{2}(\nop{\widehat \chi^*_{3}(z) \widehat\chi_{3} (z)}-\nop{ \widehat\chi^*_{4} (z) \widehat\chi_{4}(z)})\ ,\\
J^{3,2}(z)  & = \frac{1}{2} (\nop{\widehat\chi^*_{5}(z) \widehat\chi_{5} (z)}+\nop{\widehat\chi^*_{6} (z) \widehat\chi_{6}(z)})\ ,\\
J^{3,5}(z)  & = \frac{1}{2}(\nop{\widehat \chi^*_{5}(z) \widehat\chi_{5} (z)}-\nop{ \widehat\chi^*_{6} (z) \widehat\chi_{6}(z)})\ .
\end{align}

The four supercharges $G^{\pm},{G'}^\pm$ are invariant under $Q'=Q_{2345}$, so they are preserved by the 
orbifold projection and form an $\NNN=(4,4)$ superconformal algebra in the torus model. The explicit expression 
of the supercharge $G^+$ in the free fermion description of the new torus model is found to be (see 
Appendix~\ref{a:newtorus} for the detailed calculation)
\begin{align}
G^+(z)
& =\Bigl(\frac{i-1}{2}\Bigr)\widehat\chi^*_{1}(z)\Bigl[i\widehat\chi^*_{3}\widehat\chi^*_{5}(z)
-
i\widehat\chi_{4}\widehat\chi_{5}(z)-\widehat\chi_{3}\widehat\chi^*_{5}(z)+\widehat\chi^*_{4}\widehat\chi_{5}(z)\Bigr]\nonumber\\[2pt]
& \quad +\Bigl(\frac{i-1}{2}\Bigr)\widehat\chi^*_{2}(z)\Bigl[-i\widehat\chi_{3}\widehat\chi^*_{6}(z)
+i\widehat\chi^*_{4}\widehat\chi_{6}(z)
+\widehat\chi^*_{3}\widehat\chi^*_{6}(z)-\widehat\chi_{4}\widehat\chi_{6}(z)\Bigr]\ .
\end{align}
\smallskip
We want to identify four left-moving currents 
$\wh\jmath_k(z)=i\partial \widehat\phi_k(z)$, $k=1,\ldots,4$, as the superpartners of the Majorana fermions 
$\widehat\psi_k$, $k=1,\ldots, 4$, with respect to the supercharge 
${1\over\sqrt2}(G^++G^-)$. With notations analogous to \eqref{zdef}, we obtain
\begin{align}
\partial \widehat Z_1(z)=&\Bigl(\frac{1+i}{2\sqrt{2}}\Bigr)\Bigl[i\widehat\chi^*_{3}\widehat\chi^*_{5}(z)-
i\widehat\chi_{4}\widehat\chi_{5}(z)-\widehat\chi_{3}\widehat\chi^*_{5}(z)+\widehat\chi^*_{4}\widehat\chi_{5}(z)\Bigr]\ ,\\
\partial \widehat Z_2(z)=&\Bigl(\frac{1+i}{2\sqrt{2}}\Bigr)\Bigl[-i\widehat\chi_{3}\widehat\chi^*_{6}(z)+i\widehat\chi^*_{4}\widehat\chi_{6}(z)
+\widehat\chi^*_{3}\widehat\chi^*_{6}(z)-\widehat\chi_{4}\widehat\chi_{6}(z)\Bigr]\ ,
\end{align}
so that 
\begin{align}
\widehat \jmath_1(z)= i\partial \widehat\phi_1(z)&=\frac{i}{2	}:\bigl(-\widehat\psi_5(z)+\widehat\psi_6(z)+\widehat\psi_7(z)+\widehat\psi_8(z)\bigr)\,\widehat\psi_{9}(z):\ ,\\
\widehat \jmath_2(z)= i\partial \widehat\phi_2(z)&=\frac{i}{2	}:\bigl(\widehat\psi_5(z)-\widehat\psi_6(z)+\widehat\psi_7(z)+\widehat\psi_8(z)\bigr)\,\widehat\psi_{10}(z):\ ,\\
\widehat \jmath_3(z)= i\partial \widehat\phi_3(z)&=\frac{i}{2	}:\bigl(\widehat\psi_5(z)+\widehat\psi_6(z)-\widehat\psi_7(z)+\widehat\psi_8(z)\bigr)\,\widehat\psi_{11}(z):\ ,\\
\widehat \jmath_4(z)= i\partial \widehat\phi_4(z)&=\frac{i}{2	}:\bigl(-\widehat\psi_5(z)-\widehat\psi_6(z)-\widehat\psi_7(z)+\widehat\psi_8(z)\bigr)\,\widehat\psi_{12}(z):\ .
\end{align}
In particular, $ \widehat\jmath_1(z)$ generates the rotations in the plane spanned by $\widehat\psi_{9}$ and 
$\frac{1}{2}(-\widehat\psi_5(z)+\widehat\psi_6(z)+\widehat\psi_7(z)+\widehat\psi_8(z))$. The half-shift symmetry $\alpha$, 
corresponding to a 180-degree rotation in this plane, acts on the Dirac fermions by
\be \widehat\chi_{5}\leftrightarrow -\widehat\chi^*_{5}\ ,\qquad 
\widehat\chi_{1}\; ,\ \widehat\chi^*_{1}\; ,\ \widehat\chi_{2}\; ,\ \widehat\chi^*_{2}\; ,\ 
\widehat\chi_{6}\; ,\ \widehat\chi^*_{6}\text{ fixed,}
\ee 
and
\be\label{equa} \begin{pmatrix}
\widehat\chi_{3}\\ \widehat\chi^*_{3}\\ \widehat\chi_{4} \\ \widehat\chi^*_{4}
\end{pmatrix}\mapsto \frac{1}{2}\begin{pmatrix}
1 & i & -i & 1\\
-i & 1 & 1 & i\\
i & 1 & 1 & -i\\
1 & -i & i & 1
\end{pmatrix}\begin{pmatrix}
\widehat\chi_{3}\\ \widehat\chi^*_{3}\\ \widehat\chi_{4} \\ \widehat\chi^*_{4}
\end{pmatrix}\ .
\ee 
In terms of currents, this corresponds to
\be
\alpha:\left\{ \begin{array}{ll}
J^{3,1}\, ,\; J^{\pm,1}\text{ fixed;} \
J^{3,6}\, ,\; J^{\pm,6}\text{ fixed;} \ 
&J^{3,2}\leftrightarrow -J^{3,5}\; ,\ J^{\pm,2}\leftrightarrow - J^{\mp,5}\ ;\\ 
&\\
J^{+,4}\leftrightarrow - J^{3,3}- \frac{i}{2}(J^{+,3}+J^{-,3})\ ,&\qquad
J^{-,4}\leftrightarrow - J^{3,3}+ \frac{i}{2}(J^{+,3}+J^{-,3})\ ;\\
&\\
J^{3,4}\leftrightarrow -\frac{i}{2}(J^{+,3}-J^{-,3})\ .&
\end{array}\right.
\ee
In terms of $({\rm SU}(2)_L^6\times {\rm SU}(2)_R^6) : S_6$, the half-period shift $\alpha$ that we just found 
corresponds therefore to the permutation 
\be \label{2534}
\alpha^{p,T} = (25)(34)
\ee
of the currents and their representations, followed by a left-right symmetric ${\rm SU}(2)_L^6\times {\rm SU}(2)_R^6$ transformation
\be\label{form} 
\wt\alpha:=
\rho_{L,R}\Biggl(
\begin{pmatrix} 1& 0\\ 0 & 1\end{pmatrix},\,
\begin{pmatrix} 0& 1\\ -1 & 0\end{pmatrix},\,
\begin{pmatrix} \frac{1-i}{2}& \frac{i-1}{2}\\ \frac{1+i}{2}& \frac{1+i}{2}\end{pmatrix},\,
\begin{pmatrix} \frac{-1-i}{2}& \frac{i-1}{2}\\ \frac{1+i}{2}& \frac{i-1}{2}\end{pmatrix},\,
\begin{pmatrix} 0& 1\\ -1 & 0\end{pmatrix},\,
\begin{pmatrix} 1& 0\\ 0 & 1\end{pmatrix}\Biggr)\ .
\ee
Modulo ${\rm SU}(2)_L^6\times {\rm SU}(2)_R^6$ transformations, the symmetry 
$\alpha=\wt\alpha\circ\alpha^{p,T}$ with $\alpha^{p,T}= (25)(34)$
together with $s_{v_4} \wh = (34)(56)$ and $s_u \wh = (35)(46)$ from the half-period shifts 
in the `original' K3 model, generate the subgroup $A_5\subset S_6$. 
This is immediate from the fact that the permutations $(25)(34)$, $(34)(56)$ and $(35)(46)$
are even permutations of the five digits $2,\,3,\,4,\,5,\,6$, where $(34)(56)$ and $(35)(46)$
generate a group of type $\Z_2^2$, while $(34)(56)\circ (25)(34)$ has order $3$ and 
$(35)(46)\circ (25)(34)$ has order $5$. Hence these permutations generate a subgroup of
order at least $4\cdot3\cdot5=|A_5|$ of $A_5$, which must therefore agree with $A_5$.

%%%%%%%%%%%%%%%%%%%%%%%
\section{$\Z_2^8:\mathbb{M}_{20}$ as symmetry group of a K3 model}\label{thegroup}
%%%%%%%%%%%%%%%%%%%%%%%

In \cite{Gaberdiel:2011fg} the ${\cal N}=(4,4)$ preserving symmetries of K3 sigma models 
(fixing the states in the $(\frac{1}{4},\frac{1}{2};\frac{1}{4},\frac{1}{2})$ supermultiplet)  were classified. 
It follows from the proof of the main theorem in \cite{Gaberdiel:2011fg} (see case 1 in Appendix
B.3 of \cite{Gaberdiel:2011fg})
that one of the maximal groups of symmetries is $\Z_2^8:\mathbb{M}_{20}$,
i.e.\ there is no bigger symmetry group that contains $\Z_2^8:\mathbb{M}_{20}$. In the following we  
show that the group of ${\cal N}=(4,4)$ preserving  symmetries of our orbifold model contains
$\Z_2^8:\mathbb{M}_{20}$; together with the above result, this  establishes that the group of 
these symmetries is precisely equal to $\Z_2^8:\mathbb{M}_{20}$.
\bigskip

Recall that the group $\Z_2^{12}:\mathbb{M}_{24}$ is a maximal subgroup of the Conway group $Co_0$, which is the group of 
automorphisms of the Leech lattice, the unique $24$-dimensional even self-dual lattice with no vectors of squared length $2$
\cite[Chapter 10]{Conway}. Furthermore,  $\Z_2^{12}:\mathbb{M}_{24}$ has a standard  $24$-dimensional real representation, 
where $\Z_2^{12}$ acts by certain
changes of signs of the basis vectors $x_1,\ldots,x_{24}$, and $\mathbb{M}_{24}\subset S_{24}$ acts by permutations 
of these vectors. More precisely, 
consider $\Z^{24}\subset\R^{24}$ with $x_1,\ldots,x_{24}\in\Z^{24}$ the standard basis of $\R^{24}$.
Let $\GGG_{24}\subset\F_2^{24}=(\Z/2\Z)^{24}$ denote the extended binary Golay code\footnote{Hereafter, we 
use the less precise term `Golay code' to designate $\GGG_{24}$ as there is no ambiguity.}
$\GGG_{24}$, a $12$-dimensional subspace of $\F_2^{24}$ whose elements have $0$, $8$, $12$, $16$ or $24$
non-zero coordinate entries. Then $g\in\GGG_{24}\cong\Z_2^{12}$ acts by flipping the signs of
those $x_k$ for which $g_k\neq0$. The Mathieu group $\mathbb{M}_{24}$, by definition, is the subgroup of $S_{24}$ 
that preserves $\GGG_{24}\subset\F_2^{24}$.
\smallskip

In this section, we show that the group of symmetries of the K3 model described in Sections \ref{sec2} 
and \ref{sec:su2} is the subgroup of $\Z_2^{12}:\mathbb{M}_{24}$ that fixes four basis vectors (say, $x_1,x_2,x_3,x_4$) 
in the standard representation of that group.
The choice of four arbitrary distinct vectors (a \emph{tetrad}) determines a decomposition of the basis into the 
disjoint union of six tetrads
\be 
\{x_1,\ldots, x_{24}\} =T_1\sqcup T_2\sqcup T_3\sqcup T_4\sqcup T_5\sqcup T_6\ ,
\ee  
where $T_1=\{x_1,x_2,x_3,x_4\}$, such that the union of any two distinct tetrads $T_i\sqcup T_j$, $1\le i< j\le 6$, 
corresponds to an element of length $8$ (\emph{octad}) in the Golay code. The subgroup of $\Z_2^{12}:\mathbb{M}_{24}$ 
that preserves the tetrad $T_1:=\{x_1,x_2,x_3,x_4\}$ pointwise is $\Z_2^8:\mathbb{M}_{20}$. Here $\Z_2^8$ is the subgroup of 
$\Z_2^{12}\cong\GGG_{24}$ whose elements have empty intersection with $T_1$, 
and $\mathbb{M}_{20}\cong \Z_2^4 : A_5$ is the semidirect product of a group $\Z_2^4\subset \mathbb{M}_{24}$  that fixes $T_1$ 
pointwise and all six tetrads setwise, and the group $A_5$ of even permutations of the tetrads $T_2,\ldots, T_6$.

\medskip

Let $G$ be the group of symmetries of our K3 model which is generated by 
(i) the rotations $\gamma_1,\, \gamma_2$ of Subsection \ref{rotations};
(ii) the half-period shifts $s_{v_1+v_2},\, s_{v_2+v_4},\, s_{v_4}$ and $s_u$ of Subsection \ref{halfperiods};
(iii) the central symmetries $t_it_j,\, 1<i<j\leq6$, and the asymmetric symmetries $s_{w_{24}},\, s_{w_{12}},\,
\gamma_1^L$ and $\gamma_2^L$ of Subsection \ref{symmetries};
and finally (iv) the new
symmetry $\alpha$ of Subsection \ref{s:newtorus}. We will
show that the representation of $G$ on the $24$-dimensional space of R-R ground states is 
exactly the standard representation of $\Z_2^8:\mathbb{M}_{20}$. In particular, this  establishes that 
$\Z_2^8:\mathbb{M}_{20}\subseteq G$, and hence by the argument above, that $G$ is actually the full 
${\cal N}=(4,4)$ preserving symmetry group of our orbifold model.

The space of R-R ground states is 
naturally split into six four-dimensional subspaces, where the $i$-th subspace transforms as a 
$({\bf 2},{\bf 2})$-representation under the $i$-th left-right ${\rm SU}(2)_L\times {\rm SU}(2)_R$ factor,
and trivially under the other factors of ${\rm SU}(2)_L^6\times {\rm SU}(2)_R^6$. In each subspace
we choose a basis of states as
\be\label{goodbasis1}
\begin{array}{ll}
|1\rangle := \frac{1}{\sqrt{2}}(|+-\rangle - |-+\rangle)\ , \qquad&
|2\rangle :=\frac{1}{\sqrt{2}}(|++\rangle + |--\rangle)\ ,\\[0.5em] 
|3\rangle := \frac{i}{\sqrt{2}}(|++\rangle - |--\rangle)\ , &
|4\rangle := \frac{i}{\sqrt{2}}( |+-\rangle + |-+\rangle) \ ,
\end{array} 
\ee
where $|\pm\pm\rangle=|\pm\rangle\otimes\qu{|\pm\rangle}$ with $|\pm\rangle,\, \qu{|\pm\rangle}$
the eigenstates of eigenvalues $\pm\frac{1}{2}$ under the $i$-th Cartan generators 
$J^{3,i}_0,\,{\qu J^{3,i}_0}$. Thus the matrices in \eqref{f4matrices} yield the action of
$\rho_L(x),\, \rho_R(x),\, x\in \{0,\, 1,\,\omega,\,\qu\omega\}$ with respect to the bases $(|+\rangle,|-\rangle),\, 
(\qu{|+\rangle},\, \qu{|-\rangle})$.  The six sets of four vectors \eqref{goodbasis1}
will be identified with the six tetrads in the standard representation of $\Z_2^8:\mathbb{M}_{20}$. In particular, 
the tetrad $T_1$ fixed by $\Z_2^8:\mathbb{M}_{20}$ consists of the states in the $(\frac{1}{4},\frac{1}{2};\frac{1}{4},\frac{1}{2})$ 
supermultiplet of the $\NNN=(4,4)$ 
superconformal algebra. 

It is useful to arrange this basis of ground states into an array of six columns and four rows, where each column represents a tetrad and the states in each tetrad are ordered as $|1\rangle$, $|2\rangle$, $|3\rangle$, $|4\rangle$ from top downwards:
\begin{align*}&%\boxed{
\begin{array}{|cccccc|}%{c|c|c|c|c|c}
\hline &&&&&\\[-10pt]
|1\rangle &|1\rangle&|1\rangle&|1\rangle&|1\rangle&|1\rangle\\
|2\rangle &|2\rangle&|2\rangle&|2\rangle&|2\rangle&|2\rangle\\
|3\rangle &|3\rangle&|3\rangle&|3\rangle&|3\rangle&|3\rangle\\
|4\rangle &|4\rangle&|4\rangle&|4\rangle&|4\rangle&|4\rangle\\[3pt]
\hline\end{array} \;\;.\\
&\begin{array}{cccccc}
T_1 & T_2 & T_3 & T_4 & T_5 & T_6\\
\phantom{|4\rangle}& \phantom{|4\rangle}& \phantom{|4\rangle}& \phantom{|4\rangle}& \phantom{|4\rangle}& \phantom{|4\rangle} \end{array}
\end{align*}
This array corresponds to the Miracle Octad Generator (MOG) arrangement of the Golay code 
(see \cite[Chapter 11]{Conway}). 
\smallskip

Our proof that $G\cong \Z_2^8:\mathbb{M}_{20}$ consists of three steps. 
First we  show, using the information gathered in Subsection \ref{symmetries}, that the group $G'$ generated by the 
symmetries $t_it_j$ in the centre of ${\rm SU}(2)_L^6\times {\rm SU}(2)_R^6$, by the half-period shifts 
$s_{v_1+v_2}$, $s_{v_2+v_4}$, and by the rotations $\gamma_1$ and $\gamma_2$, is isomorphic to the 
subgroup $\Z_2^8$ of the Golay code acting by sign changes in the standard representation of $\Z_2^8:\mathbb{M}_{20}$. 
Then we  adjoin to $G'$ the purely left-moving symmetries $s_{w_{12}}$, $s_{w_{24}}$, 
$\gamma_1^L$ and $\gamma_2^L$; we obtain a group $G''$  that is identified with the subgroup 
$\Z_2^8:\Z_2^4$ of $\Z_2^8:\mathbb{M}_{20}$ that fixes each tetrad setwise. Finally, we  show that, by adjoining to 
$G''$ the half-period shifts $s_{v_4}$, $s_u$ and the symmetry $\alpha$, we obtain the group $G\cong \Z_2^8:\mathbb{M}_{20}$.

%%%%%%%%%%%%%%%%%%%%%%%%%%%%%%%%%
\subsection{The subgroup $G'\cong \Z_2^8$ of $\Z_2^8:\mathbb{M}_{20}$}\label{Gprime}
%%%%%%%%%%%%%%%%%%%%%%%%%%%%%%%%%
Let us consider the subgroup $\Z_2^4$ of the centre $\Z_2^6\times \Z_2^6$ of 
${\rm SU}(2)_L^6\times {\rm SU}(2)_R^6$ generated by the ten elements
\be 
t_it_j\ ,\qquad 1<i<j\le 6\ , \ee 
where $t_i$ acts as in \eqref{central}.
These symmetries change the sign of all the states in the tetrads $T_i$ and $T_j$, 
while leaving the other states fixed. We represent the elements of this group pictorially as 

$$
\begin{tikzpicture}
\node (zz2) at (-3,0) {};
\draw  (zz2)+ (0,1.1) node  {$t_2t_5=$};
\draw (zz2) +(1,2.3) rectangle +(5,0);
\draw (zz2)+(1.33,1.9) node {$+$}; 
\draw (zz2)+(1.33,1.4) node {$+$};
\draw (zz2)+(1.33,0.9)  node {$+$};
\draw (zz2)+(1.33,0.4)  node {$+$};
\draw (zz2)+(2,1.9)  node {$-$}; 
\draw (zz2)+(2,1.4)  node {$-$};
\draw (zz2)+(2,0.9)  node {$-$};
\draw (zz2)+(2,0.4)  node {$-$};
\draw (zz2)+(2.66,1.9) node {$+$};
\draw (zz2)+(2.66,1.4) node {$+$}; 
\draw (zz2)+(2.66,0.9) node {$+$};
\draw (zz2)+(2.66,0.4) node {$+$};
\draw (zz2)+(3.33,1.9) node {$+$};
\draw (zz2)+(3.33,1.4) node {$+$}; 
\draw (zz2)+(3.33,0.9) node {$+$};
\draw (zz2)+(3.33,0.4) node {$+$};
\draw (zz2)+(4,1.9) node {$-$};
\draw (zz2)+(4,1.4) node {$-$}; 
\draw (zz2)+(4,0.9) node {$-$};
\draw (zz2)+(4,0.4) node {$-$};
\draw (zz2)+(4.66,1.9) node {$+$};
\draw (zz2)+(4.66,1.4) node {$+$}; 
\draw (zz2)+(4.66,0.9) node {$+$};
\draw (zz2)+(4.66,0.4) node {$+$};
\draw (zz2)+(5.33,1.1) node {,};
\node (zza) at (5,0) {};
\draw  (zza)+ (-0.5,1.1) node  {$Q=t_3t_4t_5t_6=$};
\draw (zza) +(1,2.3) rectangle +(5,0);
\draw (zza)+(1.33,1.9) node {$+$}; 
\draw (zza)+(1.33,1.4) node {$+$};
\draw (zza)+(1.33,0.9)  node {$+$};
\draw (zza)+(1.33,0.4)  node {$+$};
\draw (zza)+(2,1.9)  node {$+$}; 
\draw (zza)+(2,1.4)  node {$+$};
\draw (zza)+(2,0.9)  node {$+$};
\draw (zza)+(2,0.4)  node {$+$};
\draw (zza)+(2.66,1.9) node {$-$};
\draw (zza)+(2.66,1.4) node {$-$}; 
\draw (zza)+(2.66,0.9) node {$-$};
\draw (zza)+(2.66,0.4) node {$-$};
\draw (zza)+(3.33,1.9) node {$-$};
\draw (zza)+(3.33,1.4) node {$-$}; 
\draw (zza)+(3.33,0.9) node {$-$};
\draw (zza)+(3.33,0.4) node {$-$};
\draw (zza)+(4,1.9) node {$-$};
\draw (zza)+(4,1.4) node {$-$}; 
\draw (zza)+(4,0.9) node {$-$};
\draw (zza)+(4,0.4) node {$-$};
\draw (zza)+(4.66,1.9) node {$-$};
\draw (zza)+(4.66,1.4) node {$-$}; 
\draw (zza)+(4.66,0.9) node {$-$};
\draw (zza)+(4.66,0.4) node {$-$};
\draw (zza)+(5.33,1.1) node {.};
\end{tikzpicture}
$$

Next we consider the group $\Z_2^4$ of symmetries that is generated by the geometric symmetries 
$s_{v_2+v_4}$, $s_{v_1+v_2}$, $\gamma_1$ and $\gamma_2$ as given in \eqref{sv2v4}; 
this is a subgroup of left-right symmetric elements of ${\rm SU}(2)_L^6\times {\rm SU}(2)_R^6$.
Note that the matrices $0,1,\omega,\bar\omega$ that were introduced in \eqref{f4matrices} act on 
each tetrad by\footnote{By slight abuse of notation, we use $\rho_{L,R}$ to denote both diagonal embeddings 
${\rm SU}(2)\to {\rm SU}(2)_L\times {\rm SU}(2)_R$ and ${\rm SU}(2)^6\to {\rm SU}(2)_L^6\times {\rm SU}(2)_R^6$, 
and similarly for  $\rho_L$.}
$\rho_{L,R}(0)$, $\rho_{L,R}(1)$, $\rho_{L,R}(\omega)$, $\rho_{L,R}(\bar\omega)\in {\rm SU}(2)_L\times {\rm SU}(2)_R$. 
We write $\F_4:=\{0,1,\omega,\bar\omega\}$, and for $x,\,y\in\F_4$ we define $x+y\in\F_4$ by
\be
\rho_{L,R}(x+y)= \rho_{L,R}(x)\circ\rho_{L,R}(y)\ .
\ee
One checks that the resulting rules of addition 
\be\label{sumF4} 
1+\omega =\omega +1=\bar\omega,\,\, 
1+\bar\omega =\bar\omega+1=\omega,\,\, 
\omega+\bar{\omega}=\bar\omega+\omega=1,\,\, 0+x=x+0=x,\,\, x+x=0
\ee
agree with those of the finite field $\F_4$ with four elements.\footnote{The field $\F_4$ 
can be constructed as the quotient $\F_2[\omega]/(\omega^2+\omega+1)$ of the ring $\F_2[\omega]$ of 
polynomials in the variable $\omega$ with coefficients in $\F_2\cong \Z/2\Z$, modulo the irreducible 
polynomial $\omega^2+\omega+1$. In this description, $\bar\omega$ represents the polynomial $\omega+1$.}
Observe that the inner automorphism $\mu(\omega)$ in
\eqref{F4mult} of the underlying quaternion algebra
corresponds to multiplication by $\omega$ on $\F_4$, and thus it equips $\F_4$ with the multiplication
law of the field with four elements.

Then the elements of the group $\Z_2^4$ that is  generated by $s_{v_2+v_4}$, $s_{v_1+v_2}$, 
$\gamma_1$ and $\gamma_2$ are given in terms of vectors in $\F_4^6$, where 
the group law for the abelian group is given by component-wise sum of the six digits. For example,
\be 
s_{v_2+v_4}\circ\gamma_1=\rho_{L,R}(00\,\omega\omega\,\omega\omega+0\bar\omega\,\bar\omega 0\, \omega 1)=
\rho_{L,R}(0\bar\omega\,1 \omega \,0\bar\omega)\ .
\ee 
Using \eqref{sv2v4} one checks that
the $15$ non-trivial elements of this  group $\Z_2^4$ correspond to all words of the form
\be\label{hexacode} 
00\, xx\, xx\qquad 0x\, 0x\, yz\qquad 0x\, x0\, zy\qquad 
0x\, yz\, 0x\qquad 0x\, zy\, x0\ ,
\ee where $(x,y,z)$ is any cyclic permutation of $(1,\omega,\bar\omega)$,
i.e.\ $(x,y,z)=(x,x\omega,x\qu\omega)$.
Note that all these words have the form
\be\label{hexacoderule}
\begin{array}{rl}
ab\, cd\, ef \quad \text{ where }& a,b,c,d,e,f\in\F_4\,,\ \Phi_{a,b,c}(x):=ax^2+bx+c\ ,\\[0.5em]
&d=\Phi_{a,b,c}(1)\, ,\  e=\Phi_{a,b,c}(\omega)\,,\ f=\Phi_{a,b,c}(\qu\omega)
\end{array}
\ee
with $a=0$.

The basis \eqref{goodbasis1} consists of simultaneous eigenvectors for the (left-right symmetric) matrices 
$\rho_{L,R}(0),\rho_{L,R}(1),\rho_{L,R}(\omega),\rho_{L,R}(\bar\omega)\in {\rm SU}(2)_L\times {\rm SU}(2)_R$, 
with the following eigenvalues
\be\label{hexaGolay} 
\begin{tikzpicture}
\node (0) at (0,0) {};
\draw  (0)+ (0,1.1) node {$\rho_{L,R}(0) =$};
\draw (0) +(1,2.3) rectangle +(1.7,0);
\draw (0)+(1.35,1.9) node {$+$}; 
\draw (0)+(1.35,1.4) node {$+$}; 
\draw (0)+(1.35,0.9) node {$+$}; 
\draw (0)+(1.35,0.4) node {$+$}; 
\draw  (0)+ (2,1) node {,};
\node (1) at (3.5,0) {};
\draw  (1)+ (0,1.1) node  {$\rho_{L,R}(1)=$};
\draw (1) +(1,2.3) rectangle +(1.7,0);
\draw (1)+(1.35,1.9) node {$+$}; 
\draw (1)+(1.35,1.4) node {$+$}; 
\draw (1)+(1.35,0.9) node {$-$}; 
\draw (1)+(1.35,0.4) node {$-$}; 
\draw  (1)+ (2,1) node {,};
\node (om) at (7,0) {};
\draw  (om)+ (0,1.1) node {$\rho_{L,R}(\omega) =$};
\draw (om) +(1,2.3) rectangle +(1.7,0);
\draw (om)+(1.35,1.9) node {$+$}; 
\draw (om)+(1.35,1.4) node {$-$}; 
\draw (om)+(1.35,0.9) node {$+$}; 
\draw (om)+(1.35,0.4) node {$-$}; 
\draw  (om)+ (2,1) node {,};
\node (bo) at (10.5,0) {};
\draw  (bo)+ (0,1.1) node  {$\rho_{L,R}(\bar\omega) =$};
\draw (bo) +(1,2.3) rectangle +(1.7,0);
\draw (bo)+(1.35,1.9) node {$+$}; 
\draw (bo)+(1.35,1.4) node {$-$}; 
\draw (bo)+(1.35,0.9) node {$-$}; 
\draw (bo)+(1.35,0.4) node {$+$}; 
\draw  (bo)+ (2,1) node {.};
\end{tikzpicture}
\ee
Therefore, each element in the  group $\Z_2^4$ acts by sign flips, 
and it is easy to construct the precise eigenvalues using the rules \eqref{hexaGolay}. For example,
by \eqref{sv2v4}
\be
\begin{tikzpicture}
\node (zz2) at (-3,0) {};
\draw  (zz2)+ (0,1.1) node  {$s_{v_2+v_4}=$};
\draw (zz2) +(1,2.3) rectangle +(5,0);
\draw (zz2)+(1.33,1.9) node {$+$}; 
\draw (zz2)+(1.33,1.4) node {$+$};
\draw (zz2)+(1.33,0.9)  node {$+$};
\draw (zz2)+(1.33,0.4)  node {$+$};
\draw (zz2)+(1.33,-0.5) node {$0$};
\draw (zz2)+(2,1.9)  node {$+$}; 
\draw (zz2)+(2,1.4)  node {$+$};
\draw (zz2)+(2,0.9)  node {$+$};
\draw (zz2)+(2,0.4)  node {$+$};
\draw (zz2)+(2,-0.5) node {$0$};
\draw (zz2)+(2.66,1.9) node {$+$};
\draw (zz2)+(2.66,1.4) node {$-$}; 
\draw (zz2)+(2.66,0.9) node {$+$};
\draw (zz2)+(2.66,0.4) node {$-$};
\draw (zz2)+(2.66,-0.5) node {$\omega$};
\draw (zz2)+(3.33,1.9) node {$+$};
\draw (zz2)+(3.33,1.4) node {$-$}; 
\draw (zz2)+(3.33,0.9) node {$+$};
\draw (zz2)+(3.33,0.4) node {$-$};
\draw (zz2)+(3.33,-0.5) node {$\omega$};
\draw (zz2)+(4,1.9) node {$+$};
\draw (zz2)+(4,1.4) node {$-$}; 
\draw (zz2)+(4,0.9) node {$+$};
\draw (zz2)+(4,0.4) node {$-$};
\draw (zz2)+(4,-0.5) node {$\omega$};
\draw (zz2)+(4.66,1.9) node {$+$};
\draw (zz2)+(4.66,1.4) node {$-$}; 
\draw (zz2)+(4.66,0.9) node {$+$};
\draw (zz2)+(4.66,0.4) node {$-$};
\draw (zz2)+(4.66,-0.5) node {$\omega$};
\draw (zz2)+(5.33,1.1) node {,};
\node (zza) at (4.5,0) {};
\draw  (zza)+ (0,1.1) node  {$\gamma_2=$};
\draw (zza) +(1,2.3) rectangle +(5,0);
\draw (zza)+(1.33,1.9) node {$+$}; 
\draw (zza)+(1.33,1.4) node {$+$};
\draw (zza)+(1.33,0.9)  node {$+$};
\draw (zza)+(1.33,0.4)  node {$+$};
\draw (zza)+(1.33,-0.5) node {$0$};
\draw (zza)+(2,1.9)  node {$+$}; 
\draw (zza)+(2,1.4)  node {$-$};
\draw (zza)+(2,0.9)  node {$+$};
\draw (zza)+(2,0.4)  node {$-$};
\draw (zza)+(2,-0.5) node {$\omega$};
\draw (zza)+(2.66,1.9) node {$+$};
\draw (zza)+(2.66,1.4) node {$-$}; 
\draw (zza)+(2.66,0.9) node {$-$};
\draw (zza)+(2.66,0.4) node {$+$};
\draw (zza)+(2.66,-0.5) node {$\bar\omega$};
\draw (zza)+(3.33,1.9) node {$+$};
\draw (zza)+(3.33,1.4) node {$+$}; 
\draw (zza)+(3.33,0.9) node {$-$};
\draw (zza)+(3.33,0.4) node {$-$};
\draw (zza)+(3.33,-0.5) node {$1$};
\draw (zza)+(4,1.9) node {$+$};
\draw (zza)+(4,1.4) node {$+$}; 
\draw (zza)+(4,0.9) node {$+$};
\draw (zza)+(4,0.4) node {$+$};
\draw (zza)+(4,-0.5) node {$0$};
\draw (zza)+(4.66,1.9) node {$+$};
\draw (zza)+(4.66,1.4) node {$-$}; 
\draw (zza)+(4.66,0.9) node {$+$};
\draw (zza)+(4.66,0.4) node {$-$};
\draw (zza)+(4.66,-0.5) node {$\omega$};
\draw (zza)+(5.33,1.1) node {.};
\end{tikzpicture}
\ee

We are now ready to make the connection with the Golay code. A standard construction of the 
Golay code makes use of the hexacode, which is a particular 3-dimensional subspace of $\F_4^6$ 
given by all words $ab\,cd\,ef\in\F_4^6$ which obey \eqref{hexacoderule} \cite[Chapter 11]{Conway}. 
Hence the $15$ elements of $\F_4^6$ of the form \eqref{hexacode}, together with $00\, 00\, 00$, 
are exactly all elements (words) in the hexacode having $0$ as first digit. From each word in the hexacode, 
one can construct various elements of the Golay code, first by using the replacement rules \eqref{hexaGolay},
and then flipping the signs of any even number of columns.\footnote{To be precise,  this way one obtains only half 
of the Golay code, namely those words of `even parity' in MOG terminology, but 
this half contains all the elements with empty intersection with the first tetrad.} We conclude 
that the group  $G'\cong \Z_2^8$ generated by $t_it_j$, $1< i<j\le 6$, together with 
$s_{v_2+v_4}$, $s_{v_1+v_2}$, $\gamma_1$ and $\gamma_2$, is exactly the subgroup of the 
Golay code with empty intersection with the first tetrad. In other words, $G'$ can be identified with the 
normal subgroup $\Z_2^8$ in $\Z_2^8:\mathbb{M}_{20}$,
\be
G'\cong \Z_2^8:=\langle t_it_j, 1 <i <j  \le 6,\, s_{v_2+v_4},\, s_{v_1+v_2},\, 
\gamma_1, \gamma_2\,\rangle \subset \,\Z_2^8:\mathbb{M}_{20}\ .
\ee

\subsection{The group $\Z_2^8:\Z_2^4$ fixing the tetrads setwise}

In this subsection, we enlarge the group  $G'\cong \Z_2^8$ described in the previous subsection by 
adjoining the symmetries $s_{w_{24}}$, $s_{w_{12}}$, $\gamma_1^L$ and $\gamma_2^L$, defined in 
\eqref{leftsymm}. We will show that the resulting group
$G''$ can be identified with the subgroup 
$\Z_2^8:\Z_2^4$ of $\Z_2^8:\mathbb{M}_{20}$ that preserves each tetrad setwise and the first tetrad pointwise.

The action of the \emph{left-moving} matrices $\rho_L(0)$, $\rho_L(1)$, $\rho_L(\omega)$, $\rho_L(\bar\omega)\in 
{\rm SU}(2)\subset {\rm SU}(2)_L\times {\rm SU}(2)_R$ on the states \eqref{goodbasis1} is given by generalised 
permutations. More precisely, each such matrix $\rho_L(M)$ is the composition $\rho_L^s(M)\circ\rho_L^p(M)$ of a 
permutation
\be\label{hexaperm} \begin{tikzpicture} 
\node (0) at (0,0) {};
\draw  (0)+ (0,1.1) node {$\rho_L^p(0) =$};
\draw (0) +(1,2.3) rectangle +(1.7,0);
\draw[fill] (0)+(1.35,1.9) circle [radius=.5pt]; 
\draw[fill] (0)+(1.35,1.4) circle [radius=.5pt];
\draw[fill] (0)+(1.35,0.9) circle [radius=.5pt];
\draw[fill] (0)+(1.35,0.4) circle [radius=.5pt];
\draw  (0)+ (2,1) node {,};
\node (1) at (3.2,0) {};
\draw  (1)+ (0,1.1) node  {$\rho_L^p(1)=$};
\draw (1) +(1,2.3) rectangle +(1.7,0);
\draw[thick,<->] (1)+(1.35,1.9) to [out=300,in=60] +(1.35,1.4);
\draw[thick,<->] (1)+(1.35,0.9) to [out=300,in=60] +(1.35,0.4);
\draw  (1)+ (2,1) node {,};
\node (om) at (6.4,0) {};
\draw  (om)+ (0,1.1) node {$\rho_L^p(\omega) =$};
\draw (om) +(1,2.3) rectangle +(1.7,0);
\draw[thick,<->] (om)+(1.35,1.9) to [out=300,in=60] +(1.35,0.9);
\draw[thick,<->] (om)+(1.35,1.4) to [out=240,in=120] +(1.35,0.4);
\draw  (om)+ (2,1) node {,};
\node (bo) at (9.6,0) {};
\draw  (bo)+ (0,1.1) node  {$\rho_L^p(\bar\omega) =$};
\draw (bo) +(1,2.3) rectangle +(1.7,0);
\draw[thick,<->] (bo)+(1.35,1.9) to [out=300,in=60] +(1.35,0.4);
\draw[thick,<->] (bo)+(1.35,1.4) to [out=240,in=120] +(1.35,0.9);
\draw  (bo)+ (2,1) node {,};
\end{tikzpicture}
\ee
followed by some sign flip
\be 
\begin{tikzpicture}
\node (0) at (0,0) {};
\draw  (0)+ (0,1.1) node {$\rho_L^s(0) =$};
\draw (0) +(1,2.3) rectangle +(1.7,0);
\draw (0)+(1.35,1.9) node {$+$}; 
\draw (0)+(1.35,1.4) node {$+$}; 
\draw (0)+(1.35,0.9) node {$+$}; 
\draw (0)+(1.35,0.4) node {$+$}; 
\draw  (0)+ (2,1) node {,};
\node (1) at (3.2,0) {};
\draw  (1)+ (0,1.1) node  {$\rho_L^s(1)=$};
\draw (1) +(1,2.3) rectangle +(1.7,0);
\draw (1)+(1.35,1.9) node {$+$}; 
\draw (1)+(1.35,1.4) node {$-$}; 
\draw (1)+(1.35,0.9) node {$+$}; 
\draw (1)+(1.35,0.4) node {$-$}; 
\draw  (1)+ (2,1) node {,};
\node (om) at (6.4,0) {};
\draw  (om)+ (0,1.1) node {$\rho_L^s(\omega) =$};
\draw (om) +(1,2.3) rectangle +(1.7,0);
\draw (om)+(1.35,1.9) node {$+$}; 
\draw (om)+(1.35,1.4) node {$-$}; 
\draw (om)+(1.35,0.9) node {$-$}; 
\draw (om)+(1.35,0.4) node {$+$}; 
\draw  (om)+ (2,1) node {,};
\node (bo) at (9.6,0) {};
\draw  (bo)+ (0,1.1) node  {$\rho_L^s(\bar\omega) =$};
\draw (bo) +(1,2.3) rectangle +(1.7,0);
\draw (bo)+(1.35,1.9) node {$+$}; 
\draw (bo)+(1.35,1.4) node {$+$}; 
\draw (bo)+(1.35,0.9) node {$-$}; 
\draw (bo)+(1.35,0.4) node {$-$}; 
\draw  (bo)+ (2,1) node {.};
\end{tikzpicture}
\ee 
Notice that $\rho_L^s$ is different from $\rho_{L,R}$ defined in \eqref{hexaGolay}: 
the two substitutions are related by a cyclic permutation of the symbols $(1,\omega,\bar\omega)$. According to
\eqref{leftsymm} and these rules, the generators $s_{w_{24}}$, $s_{w_{12}}$, $\gamma_1^L$ and $\gamma_2^L$ 
can be represented as
$$
\begin{tikzpicture}
\node (s2) at (3,0) {};
\draw  (s2)+ (-6.5,1.1) node  {$s_{w_{24}}=$};
\draw (s2) +(1,2.3) rectangle +(5,0);
\draw[fill] (s2)+(1.33,1.9) circle [radius=.5pt]; 
\draw[fill] (s2)+(1.33,1.4) circle [radius=.5pt];
\draw[fill] (s2)+(1.33,0.9) circle [radius=.5pt];
\draw[fill] (s2)+(1.33,0.4) circle [radius=.5pt];
\draw (s2)+(1.33,-0.5) node {$0$};
\draw[fill] (s2)+(2,1.9) circle [radius=.5pt]; 
\draw[fill] (s2)+(2,1.4) circle [radius=.5pt];
\draw[fill] (s2)+(2,0.9) circle [radius=.5pt];
\draw[fill] (s2)+(2,0.4) circle [radius=.5pt];
\draw (s2)+(2,-0.5) node {$0$};
\draw[thick,<->] (s2)+(2.66,1.9) to [out=300,in=60] +(2.66,0.9);
\draw[thick,<->] (s2)+(2.66,1.4) to [out=240,in=120] +(2.66,0.4);
\draw (s2)+(2.66,-0.5) node {$\omega$};
\draw[thick,<->] (s2)+(3.33,1.9) to [out=300,in=60] +(3.33,0.9);
\draw[thick,<->] (s2)+(3.33,1.4) to [out=240,in=120] +(3.33,0.4);
\draw (s2)+(3.33,-0.5) node {$\omega$};
\draw[thick,<->] (s2)+(4,1.9) to [out=300,in=60] +(4,0.9);
\draw[thick,<->] (s2)+(4,1.4) to [out=240,in=120] +(4,0.4);
\draw (s2)+(4,-0.5) node {$\omega$};
\draw[thick,<->] (s2)+(4.66,1.9) to [out=300,in=60] +(4.66,0.9);
\draw[thick,<->] (s2)+(4.66,1.4) to [out=240,in=120] +(4.66,0.4);
\draw (s2)+(4.66,-0.5) node {$\omega$};
\node (zz2) at (-3,0) {};
\draw (zz2) +(1,2.3) rectangle +(5,0);
\draw (zz2)+(1.33,1.9) node {$+$}; 
\draw (zz2)+(1.33,1.4) node {$+$};
\draw (zz2)+(1.33,0.9)  node {$+$};
\draw (zz2)+(1.33,0.4)  node {$+$};
\draw (zz2)+(1.33,-0.5) node {$0$};
\draw (zz2)+(2,1.9)  node {$+$}; 
\draw (zz2)+(2,1.4)  node {$+$};
\draw (zz2)+(2,0.9)  node {$+$};
\draw (zz2)+(2,0.4)  node {$+$};
\draw (zz2)+(2,-0.5) node {$0$};
\draw (zz2)+(2.66,1.9) node {$+$};
\draw (zz2)+(2.66,1.4) node {$-$}; 
\draw (zz2)+(2.66,0.9) node {$-$};
\draw (zz2)+(2.66,0.4) node {$+$};
\draw (zz2)+(2.66,-0.5) node {$\omega$};
\draw (zz2)+(3.33,1.9) node {$+$};
\draw (zz2)+(3.33,1.4) node {$-$}; 
\draw (zz2)+(3.33,0.9) node {$-$};
\draw (zz2)+(3.33,0.4) node {$+$};
\draw (zz2)+(3.33,-0.5) node {$\omega$};
\draw (zz2)+(4,1.9) node {$+$};
\draw (zz2)+(4,1.4) node {$-$}; 
\draw (zz2)+(4,0.9) node {$-$};
\draw (zz2)+(4,0.4) node {$+$};
\draw (zz2)+(4,-0.5) node {$\omega$};
\draw (zz2)+(4.66,1.9) node {$+$};
\draw (zz2)+(4.66,1.4) node {$-$}; 
\draw (zz2)+(4.66,0.9) node {$-$};
\draw (zz2)+(4.66,0.4) node {$+$};
\draw (zz2)+(4.66,-0.5) node {$\omega$};
\draw  (s2)+ (0,1.1) node  {$\circ$};
\draw  (s2)+ (5.5,1.1) node  {,};
\end{tikzpicture}
$$
$$
\begin{tikzpicture}
\node (s1) at (3,0) {};
\draw  (s1)+ (-6.5,1.1) node  {$s_{w_{12}}=$};
\draw (s1) +(1,2.3) rectangle +(5,0);
\draw[fill] (s1)+(1.33,1.9) circle [radius=.5pt]; 
\draw[fill] (s1)+(1.33,1.4) circle [radius=.5pt];
\draw[fill] (s1)+(1.33,0.9) circle [radius=.5pt];
\draw[fill] (s1)+(1.33,0.4) circle [radius=.5pt];
\draw[fill] (s1)+(2,1.9) circle [radius=.5pt]; 
\draw[fill] (s1)+(2,1.4) circle [radius=.5pt];
\draw[fill] (s1)+(2,0.9) circle [radius=.5pt];
\draw[fill] (s1)+(2,0.4) circle [radius=.5pt];
\draw[thick,<->] (s1)+(2.66,1.9) to [out=300,in=60] +(2.66,0.4);
\draw[thick,<->] (s1)+(2.66,1.4) to [out=240,in=120] +(2.66,0.9);
\draw[thick,<->] (s1)+(3.33,1.9) to [out=300,in=60] +(3.33,0.4);
\draw[thick,<->] (s1)+(3.33,1.4) to [out=240,in=120] +(3.33,0.9);
\draw[thick,<->] (s1)+(4,1.9) to [out=300,in=60] +(4,0.4);
\draw[thick,<->] (s1)+(4,1.4) to [out=240,in=120] +(4,0.9);
\draw[thick,<->] (s1)+(4.66,1.9) to [out=300,in=60] +(4.66,0.4);
\draw[thick,<->] (s1)+(4.66,1.4) to [out=240,in=120] +(4.66,0.9);
\draw (s1)+(1.33,-0.5) node {$0$};
\draw (s1)+(2,-0.5) node {$0$};
\draw (s1)+(2.66,-0.5) node {$\bar\omega$};
\draw (s1)+(3.33,-0.5) node {$\bar\omega$};
\draw (s1)+(4,-0.5) node {$\bar\omega$};
\draw (s1)+(4.66,-0.5) node {$\bar\omega$};
\node (zz2) at (-3,0) {};
\draw (zz2) +(1,2.3) rectangle +(5,0);
\draw (zz2)+(1.33,1.9) node {$+$}; 
\draw (zz2)+(1.33,1.4) node {$+$};
\draw (zz2)+(1.33,0.9)  node {$+$};
\draw (zz2)+(1.33,0.4)  node {$+$};
\draw (zz2)+(1.33,-0.5) node {$0$};
\draw (zz2)+(2,1.9)  node {$+$}; 
\draw (zz2)+(2,1.4)  node {$+$};
\draw (zz2)+(2,0.9)  node {$+$};
\draw (zz2)+(2,0.4)  node {$+$};
\draw (zz2)+(2,-0.5) node {$0$};
\draw (zz2)+(2.66,1.9) node {$+$};
\draw (zz2)+(2.66,1.4) node {$+$}; 
\draw (zz2)+(2.66,0.9) node {$-$};
\draw (zz2)+(2.66,0.4) node {$-$};
\draw (zz2)+(2.66,-0.5) node {$\bar\omega$};
\draw (zz2)+(3.33,1.9) node {$+$};
\draw (zz2)+(3.33,1.4) node {$+$}; 
\draw (zz2)+(3.33,0.9) node {$-$};
\draw (zz2)+(3.33,0.4) node {$-$};
\draw (zz2)+(3.33,-0.5) node {$\bar\omega$};
\draw (zz2)+(4,1.9) node {$+$};
\draw (zz2)+(4,1.4) node {$+$}; 
\draw (zz2)+(4,0.9) node {$-$};
\draw (zz2)+(4,0.4) node {$-$};
\draw (zz2)+(4,-0.5) node {$\bar\omega$};
\draw (zz2)+(4.66,1.9) node {$+$};
\draw (zz2)+(4.66,1.4) node {$+$}; 
\draw (zz2)+(4.66,0.9) node {$-$};
\draw (zz2)+(4.66,0.4) node {$-$};
\draw (zz2)+(4.66,-0.5) node {$\bar\omega$};
\draw  (s1)+ (0,1.1) node  {$\circ$};\draw  (s2)+ (5.5,1.1) node  {,};
\end{tikzpicture}
$$
$$\begin{tikzpicture}
\node (g1) at (3,0) {};
\draw  (g1)+ (-6.5,1.1) node  {$\gamma_1^L=$};
\draw (g1) +(1,2.3) rectangle +(5,0);
\draw[fill] (g1)+(1.33,1.9) circle [radius=.5pt]; 
\draw[fill] (g1)+(1.33,1.4) circle [radius=.5pt];
\draw[fill] (g1)+(1.33,0.9) circle [radius=.5pt];
\draw[fill] (g1)+(1.33,0.4) circle [radius=.5pt];
\draw[thick,<->] (g1)+(2,1.9) to [out=300,in=60] +(2,0.4);
\draw[thick,<->] (g1)+(2,1.4) to [out=240,in=120] +(2,0.9);
\draw[thick,<->] (g1)+(2.66,1.9) to [out=300,in=60] +(2.66,0.4);
\draw[thick,<->] (g1)+(2.66,1.4) to [out=240,in=120] +(2.66,0.9);
\draw[fill] (g1)+(3.33,1.9) circle [radius=.5pt]; 
\draw[fill] (g1)+(3.33,1.4) circle [radius=.5pt];
\draw[fill] (g1)+(3.33,0.9) circle [radius=.5pt];
\draw[fill] (g1)+(3.33,0.4) circle [radius=.5pt];
\draw[thick,<->] (g1)+(4,1.9) to [out=300,in=60] +(4,0.9);
\draw[thick,<->] (g1)+(4,1.4) to [out=240,in=120] +(4,0.4);
\draw[thick,<->] (g1)+(4.66,1.9) to [out=300,in=60] +(4.66,1.4);
\draw[thick,<->] (g1)+(4.66,0.9) to [out=300,in=60] +(4.66,0.4);
\draw (g1)+(1.33,-0.5) node {$0$};
\draw (g1)+(2,-0.5) node {$\bar\omega$};
\draw (g1)+(2.66,-0.5) node {$\bar\omega$};
\draw (g1)+(3.33,-0.5) node {$0$};
\draw (g1)+(4,-0.5) node {$\omega$};
\draw (g1)+(4.66,-0.5) node {$1$};
\node (zz2) at (-3,0) {};
\draw (zz2) +(1,2.3) rectangle +(5,0);
\draw (zz2)+(1.33,1.9) node {$+$}; 
\draw (zz2)+(1.33,1.4) node {$+$};
\draw (zz2)+(1.33,0.9)  node {$+$};
\draw (zz2)+(1.33,0.4)  node {$+$};
\draw (zz2)+(1.33,-0.5) node {$0$};
\draw (zz2)+(2,1.9)  node {$+$}; 
\draw (zz2)+(2,1.4)  node {$+$};
\draw (zz2)+(2,0.9)  node {$-$};
\draw (zz2)+(2,0.4)  node {$-$};
\draw (zz2)+(2,-0.5) node {$\bar\omega$};
\draw (zz2)+(2.66,1.9) node {$+$};
\draw (zz2)+(2.66,1.4) node {$+$}; 
\draw (zz2)+(2.66,0.9) node {$-$};
\draw (zz2)+(2.66,0.4) node {$-$};
\draw (zz2)+(2.66,-0.5) node {$\bar\omega$};
\draw (zz2)+(3.33,1.9) node {$+$};
\draw (zz2)+(3.33,1.4) node {$+$}; 
\draw (zz2)+(3.33,0.9) node {$+$};
\draw (zz2)+(3.33,0.4) node {$+$};
\draw (zz2)+(3.33,-0.5) node {$0$};
\draw (zz2)+(4,1.9) node {$+$};
\draw (zz2)+(4,1.4) node {$-$}; 
\draw (zz2)+(4,0.9) node {$-$};
\draw (zz2)+(4,0.4) node {$+$};
\draw (zz2)+(4,-0.5) node {$\omega$};
\draw (zz2)+(4.66,1.9) node {$+$};
\draw (zz2)+(4.66,1.4) node {$-$}; 
\draw (zz2)+(4.66,0.9) node {$+$};
\draw (zz2)+(4.66,0.4) node {$-$};
\draw (zz2)+(4.66,-0.5) node {$1$};
\draw  (g1)+ (0,1.1) node  {$\circ$};
\draw  (g1)+ (5.5,1.1) node  {,};
\end{tikzpicture}
$$ 
$$
\begin{tikzpicture}
\node (g2) at (3,0) {};
\draw  (g2)+ (-6.5,1.1) node  {$\gamma_2^L=$};
\draw (g2) +(1,2.3) rectangle +(5,0);
\draw[fill] (g2)+(1.33,1.9) circle [radius=.5pt]; 
\draw[fill] (g2)+(1.33,1.4) circle [radius=.5pt];
\draw[fill] (g2)+(1.33,0.9) circle [radius=.5pt];
\draw[fill] (g2)+(1.33,0.4) circle [radius=.5pt];
\draw[thick,<->] (g2)+(2,1.9) to [out=300,in=60] +(2,0.9);
\draw[thick,<->] (g2)+(2,1.4) to [out=240,in=120] +(2,0.4);
\draw[thick,<->] (g2)+(2.66,1.9) to [out=300,in=60] +(2.66,0.4);
\draw[thick,<->] (g2)+(2.66,1.4) to [out=240,in=120] +(2.66,0.9);
\draw[thick,<->] (g2)+(3.33,1.9) to [out=300,in=60] +(3.33,1.4);
\draw[thick,<->] (g2)+(3.33,0.9) to [out=300,in=60] +(3.33,0.4);
\draw[fill] (g2)+(4,1.9) circle [radius=.5pt]; 
\draw[fill] (g2)+(4,1.4) circle [radius=.5pt];
\draw[fill] (g2)+(4,0.9) circle [radius=.5pt];
\draw[fill] (g2)+(4,0.4) circle [radius=.5pt];
\draw[thick,<->] (g2)+(4.66,1.9) to [out=300,in=60] +(4.66,0.9);
\draw[thick,<->] (g2)+(4.66,1.4) to [out=240,in=120] +(4.66,0.4);
\draw (g2)+(1.33,-0.5) node {$0$};
\draw (g2)+(2,-0.5) node {$\omega$};
\draw (g2)+(2.66,-0.5) node {$\bar\omega$};
\draw (g2)+(3.33,-0.5) node {$1$};
\draw (g2)+(4,-0.5) node {$0$};
\draw (g2)+(4.66,-0.5) node {$\omega$};
\node (zz2) at (-3,0) {};
\draw (zz2) +(1,2.3) rectangle +(5,0);
\draw (zz2)+(1.33,1.9) node {$+$}; 
\draw (zz2)+(1.33,1.4) node {$+$};
\draw (zz2)+(1.33,0.9)  node {$+$};
\draw (zz2)+(1.33,0.4)  node {$+$};
\draw (zz2)+(1.33,-0.5) node {$0$};
\draw (zz2)+(2,1.9)  node {$+$}; 
\draw (zz2)+(2,1.4)  node {$-$};
\draw (zz2)+(2,0.9)  node {$-$};
\draw (zz2)+(2,0.4)  node {$+$};
\draw (zz2)+(2,-0.5) node {$\omega$};
\draw (zz2)+(2.66,1.9) node {$+$};
\draw (zz2)+(2.66,1.4) node {$+$}; 
\draw (zz2)+(2.66,0.9) node {$-$};
\draw (zz2)+(2.66,0.4) node {$-$};
\draw (zz2)+(2.66,-0.5) node {$\bar\omega$};
\draw (zz2)+(3.33,1.9) node {$+$};
\draw (zz2)+(3.33,1.4) node {$-$}; 
\draw (zz2)+(3.33,0.9) node {$+$};
\draw (zz2)+(3.33,0.4) node {$-$};
\draw (zz2)+(3.33,-0.5) node {$1$};
\draw (zz2)+(4,1.9) node {$+$};
\draw (zz2)+(4,1.4) node {$+$}; 
\draw (zz2)+(4,0.9) node {$+$};
\draw (zz2)+(4,0.4) node {$+$};
\draw (zz2)+(4,-0.5) node {$0$};
\draw (zz2)+(4.66,1.9) node {$+$};
\draw (zz2)+(4.66,1.4) node {$-$}; 
\draw (zz2)+(4.66,0.9) node {$-$};
\draw (zz2)+(4.66,0.4) node {$+$};
\draw (zz2)+(4.66,-0.5) node {$\omega$};
\draw  (g2)+ (0,1.1) node  {$\circ$};
\draw  (g2)+ (5.5,1.1) node  {.};
\end{tikzpicture}
$$
Notice that by our analysis of Subsection \ref{Gprime} the sign flips
\be
\begin{array}{lll} 
\rho_{L}^s(00\,\omega\omega\,\omega\omega)=\rho_{L,R}(00\,\bar\omega\bar\omega\,\bar\omega\bar\omega)\ , 
& \qquad&\rho_{L}^s(00\,\bar\omega\bar\omega\,\bar\omega\bar\omega)=\rho_{L,R}(00\, 11\, 11)\ ,\\[0.5em]
\rho_{L}^s(0\bar\omega\,\bar\omega 0\, \omega 1)=\rho_{L,R}(01\,10\,\bar\omega\omega)\ ,& 
&\rho_{L}^s(0\omega\,\bar\omega 1 \,0\omega)=\rho_{L,R}(0\bar\omega\,1\omega\,0\bar\omega)
\end{array} 
\ee
are elements of $G'\cong \Z_2^8$. Therefore, the group generated by $G'$ and by the 
symmetries $s_{w_{24}}$, $s_{w_{12}}$, $\gamma_1^L$ and $\gamma_2^L$ can be equivalently 
obtained by adjoining to $G'$ the pure permutations
$$
\begin{tikzpicture}
\node (ss1) at (0,0) {};
\draw  (ss1)+ (0,1.1) node  {${s_{w_{24}}^p}=$};
\draw (ss1) +(1,2.3) rectangle +(5,0);
\draw[fill] (ss1)+(1.33,1.9) circle [radius=.5pt]; 
\draw[fill] (ss1)+(1.33,1.4) circle [radius=.5pt];
\draw[fill] (ss1)+(1.33,0.9) circle [radius=.5pt];
\draw[fill] (ss1)+(1.33,0.4) circle [radius=.5pt];
\draw[fill] (ss1)+(2,1.9) circle [radius=.5pt]; 
\draw[fill] (ss1)+(2,1.4) circle [radius=.5pt];
\draw[fill] (ss1)+(2,0.9) circle [radius=.5pt];
\draw[fill] (ss1)+(2,0.4) circle [radius=.5pt];
\draw[thick,<->] (ss1)+(2.66,1.9) to [out=300,in=60] +(2.66,0.9);
\draw[thick,<->] (ss1)+(2.66,1.4) to [out=240,in=120] +(2.66,0.4);
\draw[thick,<->] (ss1)+(3.33,1.9) to [out=300,in=60] +(3.33,0.9);
\draw[thick,<->] (ss1)+(3.33,1.4) to [out=240,in=120] +(3.33,0.4);
\draw[thick,<->] (ss1)+(4,1.9) to [out=300,in=60] +(4,0.9);
\draw[thick,<->] (ss1)+(4,1.4) to [out=240,in=120] +(4,0.4);
\draw[thick,<->] (ss1)+(4.66,1.9) to [out=300,in=60] +(4.66,0.9);
\draw[thick,<->] (ss1)+(4.66,1.4) to [out=240,in=120] +(4.66,0.4);
\draw  (ss1)+ (5.5,1.1) node  {$,$};
\draw  (ss1)+(3,-.8) node {$\begin{array}{cccccc}
0 & 0 & \omega & \omega & \omega & \omega\\
\phantom{-} & \phantom{-} & \phantom{-} & \phantom{-} & \phantom{-} & \phantom{-}
\end{array}$};
\node (ss2) at (8,0) {};
\draw  (ss2)+ (0,1.1) node  {${s_{w_{12}}^p}=$};
\draw (ss2) +(1,2.3) rectangle +(5,0);
\draw[fill] (ss2)+(1.33,1.9) circle [radius=.5pt]; 
\draw[fill] (ss2)+(1.33,1.4) circle [radius=.5pt];
\draw[fill] (ss2)+(1.33,0.9) circle [radius=.5pt];
\draw[fill] (ss2)+(1.33,0.4) circle [radius=.5pt];
\draw[fill] (ss2)+(2,1.9) circle [radius=.5pt]; 
\draw[fill] (ss2)+(2,1.4) circle [radius=.5pt];
\draw[fill] (ss2)+(2,0.9) circle [radius=.5pt];
\draw[fill] (ss2)+(2,0.4) circle [radius=.5pt];
\draw[thick,<->] (ss2)+(2.66,1.9) to [out=300,in=60] +(2.66,0.4);
\draw[thick,<->] (ss2)+(2.66,1.4) to [out=240,in=120] +(2.66,0.9);
\draw[thick,<->] (ss2)+(3.33,1.9) to [out=300,in=60] +(3.33,0.4);
\draw[thick,<->] (ss2)+(3.33,1.4) to [out=240,in=120] +(3.33,0.9);
\draw[thick,<->] (ss2)+(4,1.9) to [out=300,in=60] +(4,0.4);
\draw[thick,<->] (ss2)+(4,1.4) to [out=240,in=120] +(4,0.9);
\draw[thick,<->] (ss2)+(4.66,1.9) to [out=300,in=60] +(4.66,0.4);
\draw[thick,<->] (ss2)+(4.66,1.4) to [out=240,in=120] +(4.66,0.9);
\draw  (ss2)+(3,-.8) node {$\begin{array}{cccccc}
0 & 0 & \bar\omega & \bar\omega & \bar\omega & \bar\omega\\
\phantom{-} & \phantom{-} & \phantom{-} & \phantom{-} & \phantom{-} & \phantom{-}
\end{array}$};
\draw  (ss2)+ (5.5,1.1) node  {$,$};
\end{tikzpicture}
$$
$$\begin{tikzpicture}
\node (gg1) at (0,0) {};
\draw  (gg1)+ (0,1.1) node  {${\gamma_1^{L,p}}=$};
\draw (gg1) +(1,2.3) rectangle +(5,0);
\draw[fill] (gg1)+(1.33,1.9) circle [radius=.5pt]; 
\draw[fill] (gg1)+(1.33,1.4) circle [radius=.5pt];
\draw[fill] (gg1)+(1.33,0.9) circle [radius=.5pt];
\draw[fill] (gg1)+(1.33,0.4) circle [radius=.5pt];
\draw[thick,<->] (gg1)+(2,1.9) to [out=300,in=60] +(2,0.4);
\draw[thick,<->] (gg1)+(2,1.4) to [out=240,in=120] +(2,0.9);
\draw[thick,<->] (gg1)+(2.66,1.9) to [out=300,in=60] +(2.66,0.4);
\draw[thick,<->] (gg1)+(2.66,1.4) to [out=240,in=120] +(2.66,0.9);
\draw[fill] (gg1)+(3.33,1.9) circle [radius=.5pt]; 
\draw[fill] (gg1)+(3.33,1.4) circle [radius=.5pt];
\draw[fill] (gg1)+(3.33,0.9) circle [radius=.5pt];
\draw[fill] (gg1)+(3.33,0.4) circle [radius=.5pt];
\draw[thick,<->] (gg1)+(4,1.9) to [out=300,in=60] +(4,0.9);
\draw[thick,<->] (gg1)+(4,1.4) to [out=240,in=120] +(4,0.4);
\draw[thick,<->] (gg1)+(4.66,1.9) to [out=300,in=60] +(4.66,1.4);
\draw[thick,<->] (gg1)+(4.66,0.9) to [out=300,in=60] +(4.66,0.4);
\draw  (gg1)+ (5.5,1.1) node  {$,$};
\draw  (gg1)+(3,-.8) node {$\begin{array}{cccccc}
0 & \bar\omega & \bar\omega & 0 & \omega & 1 \\
\phantom{-} & \phantom{-} & \phantom{-} & \phantom{-} & \phantom{-} & \phantom{-}
\end{array}$};
\node (gg2) at (8,0) {};
\draw  (gg2)+ (0,1.1) node  {${\gamma_2^{L,p}}=$};
\draw (gg2) +(1,2.3) rectangle +(5,0);
\draw[fill] (gg2)+(1.33,1.9) circle [radius=.5pt]; 
\draw[fill] (gg2)+(1.33,1.4) circle [radius=.5pt];
\draw[fill] (gg2)+(1.33,0.9) circle [radius=.5pt];
\draw[fill] (gg2)+(1.33,0.4) circle [radius=.5pt];
\draw[thick,<->] (gg2)+(2,1.9) to [out=300,in=60] +(2,0.9);
\draw[thick,<->] (gg2)+(2,1.4) to [out=240,in=120] +(2,0.4);
\draw[thick,<->] (gg2)+(2.66,1.9) to [out=300,in=60] +(2.66,0.4);
\draw[thick,<->] (gg2)+(2.66,1.4) to [out=240,in=120] +(2.66,0.9);
\draw[thick,<->] (gg2)+(3.33,1.9) to [out=300,in=60] +(3.33,1.4);
\draw[thick,<->] (gg2)+(3.33,0.9) to [out=300,in=60] +(3.33,0.4);
\draw[fill] (gg2)+(4,1.9) circle [radius=.5pt]; 
\draw[fill] (gg2)+(4,1.4) circle [radius=.5pt];
\draw[fill] (gg2)+(4,0.9) circle [radius=.5pt];
\draw[fill] (gg2)+(4,0.4) circle [radius=.5pt];
\draw[thick,<->] (gg2)+(4.66,1.9) to [out=300,in=60] +(4.66,0.9);
\draw[thick,<->] (gg2)+(4.66,1.4) to [out=240,in=120] +(4.66,0.4);
\draw  (gg2)+(3,-.8) node {$\begin{array}{cccccc}%{c|c|c|c|c|c}
 0 & \omega & \bar\omega & 1 & 0 & \omega\\
\phantom{-} & \phantom{-} & \phantom{-} & \phantom{-} & \phantom{-} & \phantom{-}
\end{array}$};
\draw  (gg2)+ (5.5,1.1) node  {$.$};
\end{tikzpicture}
$$
These symmetries have order $2$ and commute with each other, so that they form an abelian  
group $\Z_2^4$ of permutations of the $24$ R-R ground states that preserve each tetrad setwise. Each non-trivial 
element of this group is associated with codewords from the hexacode
of the form \eqref{hexacode} through the rules \eqref{hexaperm}. 
By the results of Subsection~\ref{Gprime}, eq.~\eqref{hexacode} lists all non-zero codewords of the
hexacode whose first entry is zero. Hence by \cite[Ch.~11, Sect.~9]{Conway}, 
the group generated by $s_{w_{24}}^p$, $s_{w_{12}}^p$, $\gamma_1^{L,p}$ and $\gamma_2^{L,p}$
is exactly the subgroup of $\mathbb{M}_{24}$ that preserves the tetrads setwise and fixes the first tetrad pointwise. 
Therefore, the group generated by $G'$ together with $s_{w_{24}}$, $s_{w_{12}}$, $\gamma_1^L$ and 
$\gamma_2^L$ can be identified with the group $\Z_2^8:\Z_2^4$ in $\Z_2^8:\mathbb{M}_{20}$,
\be
G'':= \langle G',s_{w_{24}},\,s_{w_{12}},\, \gamma_1^L,\,\gamma_2^L\,\rangle\cong \Z_2^8:\Z_2^4 \subset \Z_2^8:\mathbb{M}_{20}\ .
\ee
%%%%%%%%%%%%%%%%%%%%%
\subsection{Permutations of the tetrads}
%%%%%%%%%%%%%%%%%%%%%
Finally, in this subsection, we consider the entire group $G$, i.e.\ 
we enlarge $G''$ by the symmetries $s_{v_4}$, $s_u$ and 
$\alpha=\wt\alpha\circ\alpha^{p,T}$ that act as in \eqref{sv4su} and 
\eqref{2534}, \eqref{form}. 
The symmetries $s_{v_4}$ and $s_u$ act by permutations on the $24$ basis vectors \eqref{goodbasis1}
$$
\begin{tikzpicture}
\node (ss3) at (0,0) {};
\draw  (ss3)+ (0,1.1) node  {$ s_{v_4}=$};
\draw (ss3) +(1,2.3) rectangle +(5,0);
\draw[fill] (ss3)+(1.33,1.9) circle [radius=.5pt]; 
\draw[fill] (ss3)+(1.33,1.4) circle [radius=.5pt];
\draw[fill] (ss3)+(1.33,0.9) circle [radius=.5pt];
\draw[fill] (ss3)+(1.33,0.4) circle [radius=.5pt];
\draw[fill] (ss3)+(2,1.9) circle [radius=.5pt]; 
\draw[fill] (ss3)+(2,1.4) circle [radius=.5pt];
\draw[fill] (ss3)+(2,0.9) circle [radius=.5pt];
\draw[fill] (ss3)+(2,0.4) circle [radius=.5pt];
\draw[thick,<->] (ss3)+(3.33,0.9) to  +(2.66,0.9);
\draw[thick,<->] (ss3)+(3.33,0.4) to  +(2.66,0.4);
\draw[thick,<->] (ss3)+(3.33,1.9) to +(2.66,1.9);
\draw[thick,<->] (ss3)+(3.33,1.4) to +(2.66,1.4);
\draw[thick,<->] (ss3)+(4,1.9) to +(4.66,1.9);
\draw[thick,<->] (ss3)+(4.66,0.4) to  +(4,0.4);
\draw[thick,<->] (ss3)+(4,0.9) to  +(4.66,0.9);
\draw[thick,<->] (ss3)+(4.66,1.4) to +(4,1.4);
\draw  (ss3)+ (5.5,1.1) node  {,};
\node (ss4) at (8,0) {};
\draw  (ss4)+ (0,1.1) node  {$ s_{u}=$};
\draw (ss4) +(1,2.3) rectangle +(5,0);
\draw[fill] (ss4)+(1.33,1.9) circle [radius=.5pt]; 
\draw[fill] (ss4)+(1.33,1.4) circle [radius=.5pt];
\draw[fill] (ss4)+(1.33,0.9) circle [radius=.5pt];
\draw[fill] (ss4)+(1.33,0.4) circle [radius=.5pt];
\draw[fill] (ss4)+(2,1.9) circle [radius=.5pt]; 
\draw[fill] (ss4)+(2,1.4) circle [radius=.5pt];
\draw[fill] (ss4)+(2,0.9) circle [radius=.5pt];
\draw[fill] (ss4)+(2,0.4) circle [radius=.5pt];
\draw[thick,<->] (ss4)+(2.66,1.9) to [out=330,in=210] +(4,1.9);
\draw[thick,<->] (ss4)+(2.66,1.4) to [out=330,in=210] +(4,1.4);
\draw[thick,<->] (ss4)+(3.33,1.9) to [out=30,in=150] +(4.66,1.9);
\draw[thick,<->] (ss4)+(3.33,1.4) to [out=30,in=150] +(4.66,1.4);
\draw[thick,<->] (ss4)+(2.66,0.4) to [out=330,in=210] +(4,0.4);
\draw[thick,<->] (ss4)+(2.66,0.9) to [out=330,in=210] +(4,0.9);
\draw[thick,<->] (ss4)+(3.33,0.4) to [out=30,in=150] +(4.66,0.4);
\draw[thick,<->] (ss4)+(3.33,0.9) to [out=30,in=150] +(4.66,0.9);
\draw  (ss4)+ (5.5,1.1) node  {,};
\end{tikzpicture}
$$
while $\alpha$ acts by a permutation followed by a sign flip,
$$
\begin{tikzpicture}
\node (al) at (3,0) {};
\draw  (al)+ (-6.5,1.1) node  {$\alpha=$};
\draw (al) +(1,2.3) rectangle +(5,0);
\draw[fill] (al)+(1.33,1.9) circle [radius=.5pt]; 
\draw[fill] (al)+(1.33,1.4) circle [radius=.5pt];
\draw[fill] (al)+(1.33,0.9) circle [radius=.5pt];
\draw[fill] (al)+(1.33,0.4) circle [radius=.5pt];
\draw[thick,<->] (al)+(2.66,1.9) to +(3.33,1.9);
\draw[thick,<->] (al)+(2.66,1.4) to  +(3.33,0.4);
\draw[thick,<->] (al)+(2.66,0.9) to +(3.33,1.4);
\draw[thick,<->] (al)+(2.66,0.4) to +(3.33,0.9);
\draw[thick,<->] (al)+(2,0.4) to [out=330,in=210] +(4,0.4);
\draw[thick,<->] (al)+(2,0.9) to [out=330,in=210] +(4,0.9);
\draw[thick,<->] (al)+(2,1.4) to [out=30,in=150] +(4,1.4);
\draw[thick,<->] (al)+(2,1.9) to [out=30,in=150] +(4,1.9);
\draw[fill] (al)+(4.66,1.9) circle [radius=.5pt]; 
\draw[fill] (al)+(4.66,1.4) circle [radius=.5pt];
\draw[fill] (al)+(4.66,0.9) circle [radius=.5pt];
\draw[fill] (al)+(4.66,0.4) circle [radius=.5pt];
\draw  (al)+ (0,1.1) node  {$\circ$};
\node (zz2) at (-3,0) {};
\draw (zz2) +(1,2.3) rectangle +(5,0);
\draw (zz2)+(1.33,1.9) node {$+$}; 
\draw (zz2)+(1.33,1.4) node {$+$};
\draw (zz2)+(1.33,0.9)  node {$+$};
\draw (zz2)+(1.33,0.4)  node {$+$};
\draw (zz2)+(2,1.9)  node {$+$}; 
\draw (zz2)+(2,1.4)  node {$+$};
\draw (zz2)+(2,0.9)  node {$-$};
\draw (zz2)+(2,0.4)  node {$-$};
\draw (zz2)+(2.66,1.9) node {$+$};
\draw (zz2)+(2.66,1.4) node {$-$}; 
\draw (zz2)+(2.66,0.9) node {$-$};
\draw (zz2)+(2.66,0.4) node {$+$};
\draw (zz2)+(3.33,1.9) node {$+$};
\draw (zz2)+(3.33,1.4) node {$-$}; 
\draw (zz2)+(3.33,0.9) node {$+$};
\draw (zz2)+(3.33,0.4) node {$-$};
\draw (zz2)+(4,1.9) node {$+$};
\draw (zz2)+(4,1.4) node {$+$}; 
\draw (zz2)+(4,0.9) node {$-$};
\draw (zz2)+(4,0.4) node {$-$};
\draw (zz2)+(4.66,1.9) node {$+$};
\draw (zz2)+(4.66,1.4) node {$+$}; 
\draw (zz2)+(4.66,0.9) node {$+$};
\draw (zz2)+(4.66,0.4) node {$+$};
\draw  (al)+ (5.3,1.1) node  {.};
\end{tikzpicture}
$$
By composing $\alpha$ with the element $\rho_{L,R}(01\,\bar\omega\omega\,10)\in G'\cong \Z_2^8$, 
one obtains a pure permutation
$$
\begin{tikzpicture}
\node (al2) at (5,0) {};
\draw  (al2)+ (0,1.1) node  {${\alpha^p}=$};
\draw (al2) +(1,2.3) rectangle +(5,0);
\draw[fill] (al2)+(1.33,1.9) circle [radius=.5pt]; 
\draw[fill] (al2)+(1.33,1.4) circle [radius=.5pt];
\draw[fill] (al2)+(1.33,0.9) circle [radius=.5pt];
\draw[fill] (al2)+(1.33,0.4) circle [radius=.5pt];
\draw[thick,<->] (al2)+(2.66,1.9) to +(3.33,1.9);
\draw[thick,<->] (al2)+(2.66,1.4) to  +(3.33,0.4);
\draw[thick,<->] (al2)+(2.66,0.9) to +(3.33,1.4);
\draw[thick,<->] (al2)+(2.66,0.4) to +(3.33,0.9);
\draw[thick,<->] (al2)+(2,0.4) to [out=330,in=210] +(4,0.4);
\draw[thick,<->] (al2)+(2,0.9) to [out=330,in=210] +(4,0.9);
\draw[thick,<->] (al2)+(2,1.4) to [out=30,in=150] +(4,1.4);
\draw[thick,<->] (al2)+(2,1.9) to [out=30,in=150] +(4,1.9);
\draw[fill] (al2)+(4.66,1.9) circle [radius=.5pt]; 
\draw[fill] (al2)+(4.66,1.4) circle [radius=.5pt];
\draw[fill] (al2)+(4.66,0.9) circle [radius=.5pt];
\draw[fill] (al2)+(4.66,0.4) circle [radius=.5pt];
\draw  (al2)+ (5.5,1.1) node  {.};
\end{tikzpicture}
$$
Using the explicit description of the involutions of $\mathbb{M}_{24}$ that fix the first tetrad $T_1$ 
(see \cite[Ch.~11, Sect.~9]{Conway}), it is clear that the permutations $s_{u}$, $s_{v_4}$ and ${\alpha^p}$ 
are elements of $\mathbb{M}_{20}\subset \mathbb{M}_{24}$. Hence
\be
G:=\langle G'', s_u, s_{v_4}, \alpha^p\rangle \subseteq \Z_2^8:\mathbb{M}_{20}\ ,
\ee
and every word in the generators $s_u$, $s_{v_4}$ and 
$\alpha^p$ which does not permute the six factors of ${\rm SU}(2)$ is an element of $G'$. 
One also verifies that for every  $g\in G''\cong \Z_2^8:\Z_2^4$, the conjugates 
$s_{v_4}gs_{v_4}^{-1}$, $s_ugs_u^{-1}$ and $\alpha^p g(\alpha^p)^{-1}$  belong to $G''$. 
Therefore, $G''$ is a normal subgroup of $G$ and $G''=G\cap (SU(2)_L^6\times SU(2)_R^6)$.
But since $G''\cong \Z_2^8:\Z_2^4$ and $G/G''\cong A_5\cong \mathbb{M}_{20}/\Z_2^4$, we conclude that  $G=\Z_2^8:\mathbb{M}_{20}$.

%%%%%%%%%%%%%%%%
\section{A special symmetry $g$ of order four in $G=\Z_2^8:\mathbb{M}_{20}$ }\label{sec6}

%%%%%%%%%%%%%%%%

In \cite{Gaberdiel:2012um} it was observed that for those K3 sigma models that
are abelian torus orbifolds, the corresponding quantum symmetry (whose orbifold leads 
back to the torus model) is never an element of $\mathbb{M}_{24}$. This  result was obtained by 
studying the $42$ $Co_0$ conjugacy classes that define  possible symmetries of 
K3 sigma models. Of these $42$ conjugacy classes, $31$ certainly have a trivial 
multiplier as the trace over the $24$-dimensional representation is non-zero; then it 
is consistent to orbifold by the cyclic group that is generated by the relevant symmetry, 
and one can analyse (by calculating the elliptic genus) whether the resulting orbifold is a 
K3 or a toroidal sigma model. It was found that the symmetries which lead to a 
toroidal model do not have a representative in $\mathbb{M}_{24}$, see section~4 of 
\cite{Gaberdiel:2012um}.

For the remaining $11$ $Co_0$ conjugacy classes it was on the other hand not obvious
whether the corresponding symmetry obeys the level-matching condition, i.e.\ whether it
is consistent to orbifold by it. (The level-matching condition is equivalent to the statement
that a certain multiplier phase of the twining genus is trivial.) In all but one case, the elliptic genus 
of the putative orbifold did not make sense (i.e.\ did not agree with either that of K3 or the four-torus),
thus suggesting that the orbifold was in fact inconsistent. However, there was one 
class,  the 4D conjugacy class of $Co_0$, for which the putative orbifold gave rise to the 
elliptic genus of the four-torus, thus indicating that the orbifold may in fact be consistent. 
As we shall see below, this suspicion is indeed correct, as the 4D generator of $Co_0$
can be identified with an order $4$ symmetry of our K3 sigma model.

In addition, this orbifold turns out to induce an
equivalence between  different descriptions of the  $D_4$-torus model underlying
our K3 sigma model, which 
could be relevant in our quest for a field theoretic explanation of Mathieu moonshine.
Indeed, we have already
remarked in Section \ref{s:newtorus} that there are at least fifteen 
different ways in which one may write our model as a $\Z_2$-orbifold of the $D_4$-torus model.
As will be substantiated further down, it turns out that orbifolding by the symmetry of order four that we
identify with the generator of the 4D conjugacy class of $Co_0$  yields a `new' $D^{\rm {new}}_4$-torus model. 
The latter is equivalent to the original $D_4$-torus model as an  $\NNN=(4,4)$ superconformal field theory. 

\subsection{The  $\langle g\rangle$-orbifold of the K3 model}

Let us consider the symmetry  $g$ of our K3 sigma model defined by 
\be \label{gspec}
g:=t_2t_3s_{w_{12}}=\rho_L\Biggl(\begin{pmatrix}
1 & 0\\ 0 & 1
\end{pmatrix}
\begin{pmatrix}
-1 & 0\\ 0 & -1
\end{pmatrix}
\begin{pmatrix}
i & 0\\ 0 & -i
\end{pmatrix}
\begin{pmatrix}
-i & 0\\ 0 & i
\end{pmatrix}
\begin{pmatrix}
-i & 0\\ 0 & i
\end{pmatrix}
\begin{pmatrix}
-i & 0\\ 0 & i
\end{pmatrix}\Biggr)
\ee

$$
\begin{tikzpicture}
\node (s1) at (3,0) {};
\draw  (s1)+ (-5.5,1.1) node  {$=$};
\draw (s1) +(1,2.3) rectangle +(5,0);
\draw[fill] (s1)+(1.33,1.9) circle [radius=.5pt]; 
\draw[fill] (s1)+(1.33,1.4) circle [radius=.5pt];
\draw[fill] (s1)+(1.33,0.9) circle [radius=.5pt];
\draw[fill] (s1)+(1.33,0.4) circle [radius=.5pt];
\draw[fill] (s1)+(2,1.9) circle [radius=.5pt]; 
\draw[fill] (s1)+(2,1.4) circle [radius=.5pt];
\draw[fill] (s1)+(2,0.9) circle [radius=.5pt];
\draw[fill] (s1)+(2,0.4) circle [radius=.5pt];
\draw[thick,<->] (s1)+(2.66,1.9) to [out=300,in=60] +(2.66,0.4);
\draw[thick,<->] (s1)+(2.66,1.4) to [out=240,in=120] +(2.66,0.9);
\draw[thick,<->] (s1)+(3.33,1.9) to [out=300,in=60] +(3.33,0.4);
\draw[thick,<->] (s1)+(3.33,1.4) to [out=240,in=120] +(3.33,0.9);
\draw[thick,<->] (s1)+(4,1.9) to [out=300,in=60] +(4,0.4);
\draw[thick,<->] (s1)+(4,1.4) to [out=240,in=120] +(4,0.9);
\draw[thick,<->] (s1)+(4.66,1.9) to [out=300,in=60] +(4.66,0.4);
\draw[thick,<->] (s1)+(4.66,1.4) to [out=240,in=120] +(4.66,0.9);
\draw  (s1)+ (0,1.1) node  {$\circ$};
\node (zz2) at (-3,0) {};
\draw (zz2) +(1,2.3) rectangle +(5,0);
\draw (zz2)+(1.33,1.9) node {$+$}; 
\draw (zz2)+(1.33,1.4) node {$+$};
\draw (zz2)+(1.33,0.9)  node {$+$};
\draw (zz2)+(1.33,0.4)  node {$+$};
\draw (zz2)+(2,1.9)  node {$-$}; 
\draw (zz2)+(2,1.4)  node {$-$};
\draw (zz2)+(2,0.9)  node {$-$};
\draw (zz2)+(2,0.4)  node {$-$};
\draw (zz2)+(2.66,1.9) node {$-$};
\draw (zz2)+(2.66,1.4) node {$-$}; 
\draw (zz2)+(2.66,0.9) node {$+$};
\draw (zz2)+(2.66,0.4) node {$+$};
\draw (zz2)+(3.33,1.9) node {$+$};
\draw (zz2)+(3.33,1.4) node {$+$}; 
\draw (zz2)+(3.33,0.9) node {$-$};
\draw (zz2)+(3.33,0.4) node {$-$};
\draw (zz2)+(4,1.9) node {$+$};
\draw (zz2)+(4,1.4) node {$+$}; 
\draw (zz2)+(4,0.9) node {$-$};
\draw (zz2)+(4,0.4) node {$-$};
\draw (zz2)+(4.66,1.9) node {$+$};
\draw (zz2)+(4.66,1.4) node {$+$}; 
\draw (zz2)+(4.66,0.9) node {$-$};
\draw (zz2)+(4.66,0.4) node {$-$};
\draw  (s1)+ (5.3,1.1) node  {.};
\end{tikzpicture}
$$
This symmetry  has order $4$, its trace over the $24$-dimensional representation of R-R ground states is $0$,
and its square is the quantum symmetry $Q=t_3t_4t_5t_6$, whose trace over the $24$-dimensional representation 
is $-8$. These properties identify $g$ as an element in the conjugacy class 4D in the Conway group $Co_0$, as discussed 
in Section 4 of \cite{Gaberdiel:2012um}. In the following we want to show that the orbifold by this group 
element is indeed consistent and leads to a toroidal superconformal field theory.
\smallskip

As is explained in Appendix~\ref{ellipticgenuscalculation}, the elliptic genus of our model can be written in terms of 
$\widehat{\mathfrak s \mathfrak u}(2)_1$ characters as in (\ref{ellgenus}). Thus we can calculate the twining genus,
i.e.\ the elliptic genus with the insertion of the group element (\ref{gspec}), by inserting the various operators into
the $\widehat{\mathfrak s \mathfrak u}(2)_1$ traces \eqref{chi10} and \eqref{chi1hf}. With the help of the identities 
\begin{align}\label{insertions}
&\Tr_{[0]}(\left(\begin{smallmatrix}
-1 & 0\\ 0 & -1
\end{smallmatrix}\right)q^{L_0-\frac{1}{24}}y^{J_0})=\frac{\vartheta_3(2\tau,2z)}{\eta(\tau)}\ ,\nonumber\\
&\Tr_{[1]}(\left(\begin{smallmatrix}
-1 & 0\\ 0 & -1
\end{smallmatrix}\right)q^{L_0-\frac{1}{24}}y^{J_0})=-\frac{\vartheta_2(2\tau,2z)}{\eta(\tau)}\ ,\nonumber\\
&\Tr_{[0]}(\left(\begin{smallmatrix}
i & 0\\ 0 & -i
\end{smallmatrix}\right)q^{L_0-\frac{1}{24}}y^{J_0})=\Tr_{[0]}(\left(\begin{smallmatrix}
-i & 0\\ 0 & i
\end{smallmatrix}\right)q^{L_0-\frac{1}{24}}y^{J_0})=\frac{\vartheta_4(2\tau,2z)}{\eta(\tau)}\ ,\\
&\Tr_{[1]}(\left(\begin{smallmatrix}
i & 0\\ 0 & -i
\end{smallmatrix}\right)q^{L_0-\frac{1}{24}}y^{J_0})=-\Tr_{[1]}(\left(\begin{smallmatrix}
-i & 0\\ 0 & i
\end{smallmatrix}\right)q^{L_0-\frac{1}{24}}y^{J_0})=-\frac{\vartheta_1(2\tau,2z)}{\eta(\tau)}\ ,\nonumber
\end{align}
we obtain the twining genera
\be
\begin{array}{rcl}
\phi_{e,g}(\tau,z)&=&\phi_{e,g^3}(\tau,z)\nonumber\\[1em]
&=&\displaystyle\frac{2}{\eta(\tau)^6}\Bigl(\vartheta_2(2\tau,2z)\vartheta_3(2\tau)\vartheta_4(2\tau)^4
-\vartheta_3(2\tau,2z)\vartheta_2(2\tau)\vartheta_4(2\tau)^4\Bigr)\ .
\end{array}
\ee
By the Riemann bilinear identities
\begin{align}
\vartheta_2(2\tau,2z)\, \vartheta_3(2\tau)=&\frac{1}{2}\bigl(\vartheta_2(\tau,z)^2-\vartheta_1(\tau,z)^2\bigr)\label{Riemannone}\\
\vartheta_3(2\tau,2z)\, \vartheta_2(2\tau)=&\frac{1}{2}\bigl(\vartheta_2(\tau,z)^2+\vartheta_1(\tau,z)^2\bigr)
\end{align} 
this can be rewritten as 
\be 
\phi_{e,g}(\tau,z)
=-2\frac{\vartheta_1(\tau,z)^2}{\eta(\tau)^6}\vartheta_4(2\tau)^4
=-2\vartheta_4(2\tau)^4\phi_{-2,1}(\tau,z)\ ,
\ee 
where
\be 
\phi_{-2,1}(\tau,z):=\frac{\vartheta_1(\tau,z)^2}{\eta(\tau)^6}
\ee 
is the standard weak Jacobi form of weight $-2$ and index $1$. Using the modular properties of the theta functions, 
it is easy to see that $\phi_{e,g}$ is a Jacobi form for $\Gamma_0(4)$ with trivial multiplier. Therefore, the 
orbifold of the K3 model by $\langle g \rangle$  is expected to be consistent, since the level matching condition for the twisted sector 
is satisfied.  In fact, $\phi_{e,g}=\phi_{e,g^3}$ equals the $\mathbb{M}_{24}$-twining genus $\phi_{\rm 2B}$ of class 2B, so 
that 
\begin{align}
&\phi_{g,e}=\phi_{g^3,e}=-\phi_{g,g^2}=-\phi_{g^3,g^2}\ ,\\
&\phi_{e,g}=\phi_{e,g^3}=-\phi_{g^2,g}=-\phi_{g^2,g^3}\ ,\\
&\phi_{g,g}=\phi_{g^3,g^3}=-\phi_{g,g^3}=-\phi_{g^3,g}\ .
\end{align} Furthermore, since $g^2$ is the quantum symmetry $Q$ and the orbifold by $\langle Q \rangle$ is a torus model, one has
\be \phi_{e,e}+\phi_{e,g^2}+\phi_{g^2,e}+\phi_{g^2,g^2}=0\ .
\ee It follows that the orbifold of the K3 model by $\langle g \rangle$ has vanishing elliptic genus
\be 
\phi_{\rm orb}(\tau,z)=\frac{1}{4}\sum_{i,j=0}^3 \phi_{g^i,g^j}(\tau,z)=0\ ,
\ee
so that it defines a torus model, as predicted in \cite{Gaberdiel:2012um}.

%%%%%%%%%%%%%%%%%%% 
\subsection{The  $D^{\rm new}_4$-torus model and the interpretation of the orbifold action}
%%%%%%%%%%%%%%%%%%%%

The preceding arguments show that the $\langle g \rangle$-orbifold of our K3 model, denoted ${\rm K3}/\langle g \rangle$,
is a torus model. Since we have the full K3 model 
under control, we can  work out not just the elliptic genus of the orbifold by $\langle g \rangle$, but the full partition function.
This then allows us to determine the actual torus model. As we shall see it is again 
a $D_4$-torus model, which we call the $D^{\rm {new}}_4$-torus model, although it is equivalent to the original one as an 
$\NNN=(4,4)$ superconformal field theory.

The calculation of the partition function is somewhat technical, and we only sketch some of the relevant steps in 
Appendix~\ref{orbpart}. The final answer, eq.~(\ref{partfinal}), is however rather simple, and it agrees precisely
with the partition function of the ${D_4}$-torus model. Since this is the only torus model with this
partition function, 
as one confirms by observing that the underlying bosonic torus model is the only torus model
at central charge 
$(c,\qu c)=(4,4)$ which possesses a current algebra of dimension $28$, 
it follows that the model agrees with the $D_4$-torus model.

We can therefore write schematically
\be
{\rm K3}/\langle g \rangle = D^{{\rm new}}_4\ ,\qquad {\rm K3}/\langle Q \rangle = D_4\ , \qquad
D^{\rm {new}}_4\, \sim \, D_4 \ , 
\ee
and note that, since $g^2=Q$, it is possible to construct the $\langle g \rangle$-orbifold of the K3 model in two steps. 
The first one yields
\be
{\rm K3}/\langle g^2 \rangle = D_4 \ , 
\ee
while the second step involves taking the orbifold of the $D_4$-torus model by the order two symmetry $\bar{g}$  
induced by $g$ on that model; thus we have 
\be
D_4/\langle \bar{g} \rangle  = D_4^{{\rm new}}\ . 
\ee
These steps can be clearly identified at the level of the partition function, as shown in Appendix~\ref{orbpart}. 
\smallskip

Since the group $\langle g \rangle$ is abelian, we can reverse the orbifold ${\rm K3}/\langle g \rangle = D^{{\rm new}}_4$, 
and hence write our K3 sigma model as a $\mathbb{Z}_4$-orbifold of the  $D^{{\rm new}}_4$-torus model. 
Denoting by $\tilde{g}$ the generator of that orbifold, we thus have 
\be
D^{{\rm new}}_4/\langle \tilde{g} \rangle = {\rm K3}\ .
\ee
It is  natural to seek an interpretation of $\tilde{g}$, i.e.\ of the generator of the `quantum symmetry' associated 
to $g$ as a symmetry of the $D_4^{\rm new}$-torus model.

As before, we may perform also the $\langle \tilde{g} \rangle$-orbifold in two steps. The $\langle \tilde{g}^2 \rangle $-orbifold 
of the  $D^{{\rm new}}_4$-torus model yields our 
original $D_4$-torus model, i.e.
\be
D^{{\rm new}}_4/\langle \tilde{g}^2 \rangle = D_4 \ ,
\ee
and $\tilde{g}$ induces on it the usual $\mathbb{Z}_2$-orbifold action, 
i.e.\ the one  described in Section~\ref{sec2}. In fact, $\tilde{g}^2$ turns out to agree
with the ${\rm T}$-duality generator of the $D^{{\rm new}}_4$-torus model, i.e.\
with the operator that inverts the signs of all left-moving oscillators, with a trivial action on the right-movers. 
This is known to be a symmetry of a $D_4$-torus model with extended $\mathfrak s\mathfrak o(8)_1$-symmetry, 
as follows from the analysis of 
\cite{Lauer:1990tm}.

Thus we have shown that our K3 sigma model can also be written as an asymmetric $\mathbb{Z}_4$ torus orbifold,
and that the corresponding quantum symmetry is the one associated to the 4D
conjugacy class of $Co_0$.

%%%%%%%%%%%%%%%%
\section{Conclusions}\label{sec:concl}
%%%%%%%%%%%%%%%%

In this paper we have considered a superconformal field theory that describes a K3 sigma
model with one of the  largest maximal symmetry groups, namely $\mathbb{Z}_2^8 : \mathbb{M}_{20}$. In particular,
we have found different descriptions for this model: as a $\mathbb{Z}_2$-orbifold of the $D_4$-{torus model}, as a theory of 
$12$ left- and right-moving Majorana fermions, and as a rational conformal field theory based on the chiral algebra
$\widehat{\mathfrak{su}}(2)_1^{\oplus 6}$.  By combining these different viewpoints various properties
of this model have become manifest. This may prove  useful in order to understand the origin of the 
$\mathbb{M}_{24}$ symmetry in the elliptic genus of K3. 

A result of our work is a very explicit description of all symmetries of the sigma model 
on the tetrahedral Kummer surface. In \cite{tawe12b} two of us have highlighted a 
$45$-dimensional vector space of states $V_{45}^{\rm CFT}$ that are generic to all 
standard $\Z_2$-orbifold CFTs on K3 and which govern the massive leading  order of the elliptic
genus of K3.
On $V_{45}^{\rm CFT}$, the combined geometric symmetries of all such theories generate an action
of the maximal subgroup \mbox{$\Z_2^4 : A_8$} of $\mathbb{M}_{24}$, as is shown in \cite{tawe12b}. 
Using the description of symmetries for the special model studied in the present paper, 
it will be possible to investigate
the action on $V_{45}^{\rm CFT}$
for symmetries that are `non-geometric' from the viewpoint of the tetrahedral
Kummer surface.

Recall from \cite{nawe01} that our K3 sigma model also possesses a Gepner-type description
as $\Z_2\times\Z_2$-orbifold of the well-known model $(2)^4$, and that this implies invariance
under Greene-Plesser mirror symmetry of our model \cite{Greene:1990ud}. This symmetry is not contained in the group 
$\Z_2^8 : \mathbb{M}_{20}$ investigated in the present paper, as mirror symmetry is an automorphism
of the $\NNN=(4,4)$ superconformal algebra, but it does not preserve it pointwise. It might
be interesting to determine the action of mirror symmetry in this model.

There are also other special points in the moduli space of 
K3 sigma models {that would be interesting to construct}. For example, there should be
a K3 sigma model with $\mathbb{M}_{21}$ symmetry group, and it would  be very interesting
to find an explicit description for it. However, it is clear that it cannot be a standard torus orbifold.

The $\mathbb{Z}_2^8 : \mathbb{M}_{20}$ theory we have {considered} should possess an interesting
exactly marginal deformation that breaks the symmetry group to $\mathbb{M}_{20}$.  One should
expect that, at least generically, this deformation will break the large chiral symmetry of our K3 sigma model 
to  the ${\cal N}=(4,4)$ superconformal algebra. Thus the resulting deformed models should possess
an $\mathbb{M}_{20}$ symmetry while at the same time exhibiting only the `minimal' number of BPS 
states.\footnote{Note that the appearance of fermionic BPS states
--- that contribute with the `wrong' sign to the elliptic genus --- is always associated to an extension of 
the chiral algebra, because
under spectral flow any `fermionic' BPS state transforming in the $(h=\frac{1}{4},l=\tfrac{1}{2})$ 
representation
of the right-movers is mapped to the right-moving NS vacuum, see \cite{Ooguri:1989,Wendland:2000} for details.}
These deformed models may therefore play an important role in understanding the algebraic 
reasons underlying Mathieu Moonshine.

%%%%%%%%%%%
\newpage
\appendix

\section{Conventions and notations for torus models}\label{app:conv}

In this appendix we fix our conventions and notations for supersymmetric torus models.
A real $d$-dimensional torus may be described as $\mathbb{T}=\R^d/ L$, where $L \subset\R^d$
is a lattice of maximal rank. We denote by $L^\ast \subset \R^d$ its dual lattice, using the standard
Euclidean metric to define inner products and to identify $\R^d$ with $(\R^d)^\ast$. We shall usually use 
standard Cartesian coordinates, with ${\bf e}_1,\ldots, {\bf e}_d\in\R^d$
the standard basis of $\R^d$. In order to describe a torus theory one must also fix a Kalb-Ramond B-field in terms
of a skew-symmetric $d\times d$-matrix $B$ with real entries. The field content of the
\emph{bosonic} torus model is then generated by
\begin{itemize}
\item 
$d$ real left-moving U(1)-currents $j_k(z)=i\partial \phi_k(z),\,k=1,\ldots,d$, which obey the OPEs
\begin{equation}\label{u1ope}
j_k(z)j_l(w)\sim {\delta_{kl}\over (z-w)^2} \ .
\end{equation}
The notation for the right-moving currents is analogous, with\footnote{The choice of relative sign for the 
bosons $\phi_k,\, \qu\phi_k$ is of course a matter of convention.}
$\qu\jmath_k(\qu z)={i\qu\partial\,\qu\phi_k(\qu z)}$.
The mode expansions of the left-moving currents are 
\begin{equation}
j_k(z)=\sum_{n \in \mathbb{Z}}a_n^{(k)}z^{-n-1}\ , \qquad \hbox{with} \qquad
[a_m^{(k)},a_n^{(l)}]=m\delta^{kl}\delta_{m,-n} \ .
\end{equation}
\item
Winding-momentum fields  associated with vectors  
$(\boldsymbol{m},\boldsymbol{l})\in L^\ast\oplus L$.
In order to define them, we set
\begin{equation}\label{geometriccharges}
{{\bf Q}}(\boldsymbol{m},\boldsymbol{l}) := {1\over\sqrt2}(\boldsymbol{m}-B\boldsymbol{l}+\boldsymbol{l})\ , \;
{{\bf \qu Q}}(\boldsymbol{m},\boldsymbol{l}) := {1\over\sqrt2}(\boldsymbol{m}-B\boldsymbol{l}-\boldsymbol{l})
\quad\text{ for } (\boldsymbol{m},\boldsymbol{l})\in L^\ast\oplus L.
\end{equation}
The charges $({{\bf Q}}(\boldsymbol{m},\boldsymbol{l}); {{\bf \qu Q}}(\boldsymbol{m},\boldsymbol{l}))$
form an even, selfdual lattice 
\be\label{chargelattice}
\Gamma:= \left\{ ({{\bf Q}}(\boldsymbol{m},\boldsymbol{l}); {{\bf \qu Q}}(\boldsymbol{m},\boldsymbol{l}))
\mid (\boldsymbol{m},\boldsymbol{l})\in L^\ast\oplus L \right\} \subset\R^{d,d}
\ee
of signature $(d,d)$ 
with quadratic form given by
\be\label{quadraticform}
({\bf Q};\qu{\bf Q})\bullet ({\bf Q}^\prime;\qu{\bf Q}^\prime)
:= {\bf Q}\cdot{\bf Q}^\prime- \qu{\bf Q}\cdot\qu{\bf Q}^\prime
\qquad \fa ({\bf Q};\qu{\bf Q}),\, ({\bf Q}^\prime;\qu{\bf Q}^\prime)\in\Gamma.
\ee
We also  introduce operators $c_{\lambda}$ for each $\lambda\in\Gamma$ which obey
\be 
c_\lambda c_\mu = \epsilon(\lambda,\mu)c_{\lambda+\mu}\quad \text{ for all } \lambda,\,\mu\in\Gamma
\ee 
with a suitable $2$-cocycle $\epsilon(\lambda,\mu)\in\{\pm 1\}$. In other words,\footnote{If the charge
lattice $\Gamma$ is even, then the factor $(-1)^{\lambda^2\mu^2}$ in \eqref{local} is trivial. 
However, in Appendix \ref{appC1} we need more general charge lattices $\Gamma$ that are
only required to be integral and selfdual.}
\be\label{local}
\fa  \lambda,\,\mu,\,\nu\in\Gamma\colon\qquad\left\{
\begin{array}{rcl}
\epsilon(\lambda,\mu)&=&(-1)^{\lambda^2\mu^2} (-1)^{\lambda\cdot\mu} \, \epsilon(\mu,\lambda)\ ,\\[0.5em]
\epsilon(\lambda,\mu)\epsilon(\lambda+\mu,\nu) &=& \epsilon(\lambda, \mu+\nu)\epsilon(\mu,\nu)\ .
\end{array}\right.
\ee
Then for $({\bf Q};\qu{\bf Q})\in\Gamma$,  the fields
\begin{equation}\label{vdef}
V_{({\bf Q};\qu{\bf Q})}(z,{\qu z})
:=\nop{ \exp\left[ i\sum_{k=1}^d{Q}_k\phi_k(z) 
+ i\sum_{k=1}^d{ \qu Q}_k \qu\phi_k(\qu z)\right] }\; c_{({\bf Q};\qu{\bf Q})}
\end{equation}
obey the OPEs 
\be
\begin{array}{rl}
j_k(z) V_{({\bf Q};\qu{\bf Q})}(w, {\qu w})\  \sim \ 
& \displaystyle{Q_k\over z-w} V_{({\bf Q};\qu{\bf Q})}(w,{\qu w})\ , \qquad k=1,\ldots, d, \quad
\lambda:=({\bf Q};\qu{\bf Q})\in\Gamma,
\\[0.75em]
\qu\jmath_k(\qu z) V_{({\bf Q};\qu{\bf Q})}(w,{\qu w}) \ \sim \ 
&\displaystyle{ {\qu Q_k}\over \qu z-\qu w} V_{({\bf Q};\qu{\bf Q})}(w,{\qu w})\ ,  \qquad k=1,\ldots, d,\quad
\lambda^\prime:=({\bf Q}^\prime;\qu{\bf Q}^\prime)\in\Gamma,
\\[0.75em]
V_{\lambda}(z, {\qu z} ) V_{\lambda^\prime}(w, {\qu w}) \ \sim \ 
&\displaystyle
\eps(\lambda,\lambda^\prime)
(z-w)^{{\bf {Q}}\cdot{\bf {Q}}^\prime}(\qu z-\qu w)^{{\bf {\qu Q}}\cdot{\bf {\qu Q}}^\prime} 
V_{\lambda+\lambda^\prime}(w, {\qu w})\ ,
\end{array}
\ee
where $\sim$ only indicates the most singular terms,
and $V_{({\bf Q};\qu{\bf Q})}(z, {\qu z})$ has conformal dimension $(h;\qu h)=({{\bf Q}^2\over2};{{\bf\qu Q}^2\over2})$.
Here $V_{(0;0)}(z, {\qu z})$ is the vacuum field. 
\end{itemize}
The toroidal model is uniquely determined by its charge lattice $\Gamma\subset\R^{d,d}$ by means of 
\eqref{u1ope}, \eqref{chargelattice} -- \eqref{vdef}, independently of its geometric interpretation on the torus
${\mathbb T}=\R^d/L$ with B--field $B$. 
Different choices for the cocycle $\epsilon$ satisfying \eqref{local} are related by a redefinition of the fields 
$V_\lambda$, see, for example, \cite{gool84} for details. \\

In the corresponding \emph{supersymmetric} torus model, in addition to the fields listed
above, we adjoin
\begin{itemize}
\item
$d$ `external' free Majorana fermions $\psi_k(z), \,k=1,\ldots,d$, 
which are related to the U(1)-currents by world-sheet supersymmetry and which obey the OPEs
\begin{equation}\label{Majope}
\psi_k(z)\psi_l(w)\sim {\delta_{kl}\over (z-w)}\ .
\end{equation}
The right-moving Majorana fermions are denoted by $\qu{\psi}_k(\qu z)$. The mode expansions of the 
left-moving fermions are 
\begin{equation}
\psi_k(z)=\sum_{n \in \mathbb{Z}+\nu}\psi_n^{(k)}z^{-n-\hf} \ , \qquad \hbox{where} \qquad 
\{\psi_m^{(k)},\psi_n^{(l)}\}=\delta^{kl}\delta_{m,-n} \ , 
\end{equation}
with  $\nu=0$ and $\nu=\frac{1}{2}$ in the Ramond and Neveu-Schwarz sector, respectively. 
\end{itemize}
If $d$ is even, then the external fermions comprise the U(1)-currents $J(z),\,\qu J(\qu z)$ of a 
left- and a right-moving $\NNN=2$ superconformal algebra at central charge $c={3d\over2}=\qu c$,
where we choose
\be\label{U1SCA}
J(z):= i \sum_{k=1}^d \nop{\psi_{2k-1}(z)\psi_{2k}(z)}\ ,\qquad
\qu J(\qu z):= i \sum_{k=1}^d \nop{\qu \psi_{2k-1}(\qu z)\qu \psi_{2k}(\qu z)}\ .
\ee

Then the partition function of the supersymmetric torus model is
\begin{eqnarray}
Z(\tau, z) &=& 
\Tr
\left({\textstyle{1\over2}}\left(1+(-1)^{F_L+F_R}\right)
y^{J_0} \qu y^{\qu J_0}\, q^{L_0-\frac{c}{24}}\bar q^{{\qu L_0}-\frac{{\qu c}}{24}}  \right)\nonumber\\
&=& {1\over2}\sum_{k=1}^4 \left| {\vartheta_k(\tau,z)\over\eta(\tau)}\right|^d \cdot
\sum_{({\bf Q};\qu{\bf Q})\in\Gamma} {q^{{\bf Q}^2/2}\;\qu q^{\qu{\bf Q}^2/2}\over\left|\eta(\tau)\right|^{2d}},
\quad y=e^{2\pi iz},\; q=e^{2\pi i\tau}.\label{toruspartitionfunction}
\end{eqnarray}
Here, the trace is taken over the full Hilbert space of the theory,
$F_L+F_R$ is the total fermion number operator accounting for the external fermions,
$J_0,\; \qu J_0$ denote the zero modes of the U(1)-currents of the left- and the right-moving 
$\NNN=2$ superconformal algebras as in \eqref{U1SCA}, and the central charges are 
$c=\qu c = {3d\over2}$. Moreover,
$\eta(\tau)$ is the Dedekind eta function and $\vartheta_k(\tau,z),\, k=1,\ldots, 4$, are
the Jacobi theta functions described in Appendix \ref{Jacobigym}.
The first factor in the partition function \eqref{toruspartitionfunction} accounts for the external
fermions, while the second factor accounts for the contributions from the bosonic torus model.
If $\Gamma=\Gamma_{d,d}$ as in \eqref{Ddcharge}, its partition function can be written
in terms of theta functions as 
\be\label{so(d)pf}
\sum_{({\bf Q};\qu{\bf Q})\in\Gamma_{d,d}} {q^{{\bf Q}^2/2}\;\qu q^{\qu{\bf Q}^2/2}\over\left|\eta(\tau)\right|^{2d}}
=
\hf \sum_{k=2}^4 \left| {\vartheta_k(\tau)\over\eta(\tau)}\right|^{2d}\ ,
\ee
thus suggesting a free fermionic description of this bosonic torus model.
%%%%%%%%

%%%%%%%%%%%%%%%%%
\section{Fermionisation of the bosonic $D_4$-torus model}\label{appB}
%%%%%%%%%%%
In this appendix we provide some details on the fermionisation procedure needed in
Subsection \ref{torferm}. Consider eight left-moving Majorana fermions $\psi_k(z)$ 
with right-moving partners $\qu\psi_k(\qu z)$,  where $k=5,\ldots,12$. For each of them, we have two 
twist-fields $\eta_k^\pm(z, \qu z)$ with conformal dimension $({1\over16};{1\over16})$
\be \label{rcircl}
\begin{array}{rclrcl}
\psi_k(z)\, \eta_k^\pm(w, \qu w) &\sim&\displaystyle {\zeta^{\pm1}\over\sqrt2} {\eta_k^\mp(w, \qu w)\over(z-w)^{1/2}}\ ,
& \quad\quad\qu\psi_k(\qu z)\, \eta_k^\pm(w, \qu w) &\sim&  \displaystyle{\zeta^{\mp1}\over\sqrt2} 
{\eta_k^\mp(w, \qu w)\over(\qu z-\qu w)^{1/2}}\ ,
\\[10pt]
\eta_k^+(z, \qu z)\, \eta_k^-(w, \qu w) &\sim& \multicolumn{4}{l}{\displaystyle
{1\over\sqrt2|z-w|^{1/4}} \left( {\zeta}(z-w)^{1/2} \psi_k(w) 
+ {\zeta^{-1}} (\qu z-\qu w)^{1/2} \qu\psi_k(\qu w) \right)}\ ,
\end{array}
\ee
where $\zeta\in\C$ is a primitive eighth root of unity with $\zeta^2=i$, (see e.g.\ \cite[(12.67), (12.68)]{dms96}).
In addition,
\be
\eta_k^\pm(z,\qu z)\, \eta_k^\pm(w, \qu w) 
\sim |z-w|^{-1/4} \pm {i\over2} \nop{\psi_j(w)\qu\psi_j(\qu w)}  |z-w|^{3/4}.
\ee
In  \eqref{Diracblock23}, we introduce the four Dirac fermions
\be
\begin{array}{rclrcll}
x_k &:=& {1\over\sqrt2} ( \psi_{k+4} + i \psi_{k+8} )\ , \quad 
&x_k^\ast &:=& {1\over\sqrt2} ( \psi_{k+4} - i \psi_{k+8} )\ , \\[0.5em]
\qu x_k &:=& {1\over\sqrt2} ( \qu\psi_{k+4} + i \qu\psi_{k+8} )\ , \quad 
&\qu x_k^\ast &:=& {1\over\sqrt2} ( \qu\psi_{k+4} - i \qu\psi_{k+8} )\ ,
& k\in\{1,\ldots,4\},
\end{array}
\ee
where according to \eqref{Diracfermionope},
\be
x_k(z) x_k^\ast(w)\sim {1\over z-w}\sim x_k^\ast(z) x_k(w)\ .
\ee
Bosonisation amounts to the observation that the fields
\be
j_k(z) := \nop{x_k(z) x_k^\ast(z)}= - i\nop{\psi_{k+4}(z)\psi_{k+8}(z)}
\ee
obey \req{u1ope}, such that $j_k(z)=i\partial\phi_k(z)$ (and analogously on the right hand side)
allows us to identify, up to appropriate cocycle factors,
\begin{eqnarray}\label{fermionisationrules}
i \nop{x_k(z)\qu x_k^\ast(\qu z)} &=& \nop{ \exp\left( i\phi_k(z)-i\qu\phi_k(\qu z)\right)}\ , \nonumber\\
i \nop{x_k^\ast(z)\qu x_k(\qu z)} &=& \nop{ \exp\left( -i\phi_k(z)+i\qu\phi_k(\qu z)\right)}\ , \nonumber\\
\nop{\eta_k^+\eta_{k+4}^+}(z,\qu z) &=& {\textstyle{1\over\sqrt2 }} \left[
\nop{ \exp\left( {i\over2}\phi_k(z){-}{i\over2}\qu\phi_k(\qu z)\right)}
+ \nop{ \exp\left( -{i\over2}\phi_k(z){+}{i\over2}\qu\phi_k(\qu z)\right)}\right]\nonumber\\
&=& {\textstyle{1\over\sqrt2 }} \left[ \nop{\xi_k^+\qu\xi_k^+}(z,\qu z) +\nop{\xi_k^-\qu\xi_k^-}(z,\qu z) \right]
\ ,\\
\nop{\eta_k^-\eta_{k+4}^-}(z,\qu z) &=& {\textstyle{1\over\sqrt2i}} \left[
\nop{ \exp\left( {i\over2}\phi_k(z){-}{i\over2}\qu\phi_k(\qu z)\right)}
- \nop{ \exp\left( -{i\over2}\phi_k(z){+}{i\over2}\qu\phi_k(\qu z)\right)}\right]\nonumber\\
&=& {\textstyle{1\over\sqrt2i }} \left[ \nop{\xi_k^+\qu\xi_k^+}(z,\qu z) -\nop{\xi_k^-\qu\xi_k^-}(z,\qu z) \right]\ .\nonumber
\end{eqnarray}
Here, we have formally introduced the `meromorphic factors' $\xi^\pm_1(z),\ldots, \xi^\pm_4(z)$
with $\xi_k^\pm(z) := \nop{\exp\left( \pm{i\over2}\phi_k(z)\right)}$. By democratically
distributing phases between the holomorphic and antiholomorphic part,
\begin{eqnarray}\label{isingope}
\xi_k^\pm(z)\xi_k^\mp(w)&\sim& (z-w)^{-1/4} \left(1\pm{{1\over2}} (z-w) \nop{x_k(w)x_k^\ast(w)} \right) \ ,\nonumber\\
\xi_k^+(z)\xi_k^+(w)&\sim& x_k(w)(z-w)^{1/4}\zeta\ ,\nonumber\\
\xi_k^-(z)\xi_k^-(w)&\sim& x_k^\ast(w)(z-w)^{1/4}\zeta\ ,\\
\nop{x_k(z)x_k^\ast(z)}\, \xi_k^\pm(w)&\sim& \pm {{1\over2}\xi_k^\pm(w)\over z-w}\ ,\nonumber
\end{eqnarray}
such that for the $c=\qu c=1$ theory of the free Dirac fermion $x_k(z)$, the
two R-R-ground states are created by $\xi_k^\pm(z)\qu \xi_k^\pm(\qu z)$.

In Subsection~\ref{geomD4} we describe the bosonic $d=4$-dimensional
$D_4$-torus model with charge lattice \eqref{Ddcharge}.
Its left-moving $\wh{\mathfrak{so}}(8)_1$ current algebra is generated by the U(1)-currents
$j_1(z),\,\ldots,\,j_4(z)$ together with the twenty-four $(1,0)$-fields $V_{({\bf Q}_{\pm j,\pm k};0)}(z)$ specified
by \eqref{windingcurrents}. All winding-momentum fields can be generated from the $({1\over2},{1\over2})$-fields
$V_{({\bf Q};{\bf\qu Q})}(z,\qu z)$ listed in \eqref{bosgen} by taking OPEs with holomorphic currents. 
Using \eqref{fermionisationrules} we can thus give a complete list of generating fields in terms
of the free fermion data.
One checks that indeed the following identifications are compatible 
with the respective OPEs:
\begin{itemize}
\item 
four $(1,0)$-fields generating the Cartan subalgebra of $\widehat{\mathfrak s\mathfrak o}(8)_1$:
\begin{equation}\label{bosonization}
j_k(z) =- i\nop{\psi_{{k+4}}(z)\psi_{{k+8}}(z)} = \nop{x_k(z) x_k^\ast(z)} \ , \quad k\in\{1,\ldots,4\}\ ;
\end{equation}
\item twenty-four $(1,0)$-fields corresponding to the  roots of $D_4$
for $1\le j< k\le 4$,
\begin{equation}\label{currentident}
\begin{array}{rclrcr}
V_{\boldsymbol{m}_{j,k},\boldsymbol{l}_{j,k}}(z) &=&  \nop{x_j(z) x_k(z)}\ ,&
V_{\boldsymbol{m}_{-j,-k},\boldsymbol{l}_{-j,-k}}(z) &=&  \nop{x_j^\ast(z) x_k^\ast(z)}\ ,\\[5pt]
V_{\boldsymbol{m}_{j,-k},\boldsymbol{l}_{j,-k}}(z)&=&  \nop{x_j(z) x_k^\ast(z)}\ ,&
V_{\boldsymbol{m}_{-j,k},\boldsymbol{l}_{-j,k}}(z)&=&  \nop{x_j^\ast(z)x_k(z) }\ ;
\end{array}
\end{equation}
\item
for the $({1\over2},{1\over2})$-fields $V_{({\bf Q};{\bf\qu Q})}(z, {\qu z})$,
\be\label{halfhalffermion}
\begin{array}{rclrclrcl}
\mbox{if } {\bf Q} &=& {\bf e}_j, & {\bf\qu Q} &=& {\bf e}_k\colon
& V_{({\bf Q};{\bf\qu Q})}(z, {\qu z}) &=& i\nop{ x_j(z)\qu x_k^\ast(\qu z)}\,\,(16\,{\rm fields})\\[5pt]
&&& {\bf\qu Q} &=& -{\bf e}_k\colon
& V_{({\bf Q};{\bf\qu Q})} (z,{\qu z})&=& i\nop{ x_j(z)\qu x_k(\qu z)}\,\,(16\,{\rm fields})\\[5pt]
\mbox{if } {\bf Q} &=& -{\bf e}_j, & {\bf\qu Q} &=& {\bf e}_k\colon
& V_{({\bf Q};{\bf\qu Q})}(z, {\qu z}) &=& i\nop{ x_j^\ast(z)\qu x_k^\ast(\qu z)}\,\,(16\,{\rm fields})\\[5pt]
&&& {\bf\qu Q} &=& -{\bf e}_k\colon
& V_{({\bf Q};{\bf\qu Q})}(z,{\qu z}) &=& i\nop{ x_j^\ast(z)\qu x_k(\qu z)}\,\,(16\,{\rm fields})\\[5pt]
\multicolumn{3}{l}{\mbox{if } \eps_j,\,\delta_k\in\{\pm\},}\\
{\bf Q} &=& {1\over2}\sum\limits_{j=1}^4\eps_j{\bf e}_j, & 
{\bf\qu Q} &=& {1\over2}\sum\limits_{k=1}^4\delta_k{\bf e}_k\colon
\qquad
& V_{({\bf Q};{\bf\qu Q})}(z,{\qu z}) &=& \nop{ \prod\limits_{j=1}^4\xi_j^{\eps_j}(z)\prod\limits_{k=1}^4\qu\xi_k^{\delta_k}(\qu z)}\\[5pt]
&&&&&&&& (128\,{\rm fields}).
\end{array}
\ee
\end{itemize}
%

%%%%%%%%%%%%%%%%%%%%%%%%%%%%%%%%%%%%%%
\section{Partition function from the free fermion description}\label{sigmamodelspectrum}
%%%%%%%%%%%%%%%%%%%%%%%%%%%%%%%%%%%%%%

The free fermion description of  Subsection \ref{torferm} allows us to construct the content of the 
bosonic torus model in terms of eight free left-moving and eight free right-moving Majorana fermions,
all with coupled spin structures. From that point of view, the first two terms in (\ref{D4torus}) (the vacuum 
and vector representations of $\wh{\mathfrak{s0}}(8)_{1}$) come from 
the free fermion states with Neveu-Schwarz boundary conditions, while the last two terms (the spinor representations of $\wh{\mathfrak{s0}}(8)_{1}$) are 
accounted for in terms with Ramond boundary conditions. In both cases
only those states are included where the total (left- and right-moving) fermion number is even. In terms of
partition functions, this then just amounts to the statement that 
\begin{eqnarray}
\Tr_{({\cal H}_{L,0} \otimes {\cal H}_{R,0}) \oplus ({\cal H}_{L,v} \otimes {\cal H}_{R,v})}
\left(q^{L_0-\frac{d}{24}}\bar q^{\qu L_0-\frac{d}{24}}  \right)
&=& \hf \left( \left| {\vartheta_3(\tau)\over\eta(\tau)}\right|^{2d} + \left| {\vartheta_4(\tau)\over\eta(\tau)}\right|^{2d}\right),\\
\Tr_{({\cal H}_{L,s} \otimes {\cal H}_{R,s}) \oplus ({\cal H}_{L,c} \otimes {\cal H}_{R,c})}
\left(q^{L_0-\frac{d}{24}}\bar q^{\qu L_0-\frac{d}{24}}  \right)
&=& \hf\cdot  \left| {\vartheta_2(\tau)\over\eta(\tau)}\right|^{2d} ,
\end{eqnarray}
with $d=4$,
as one confirms by means of the product formulas for the Jacobi theta functions given in
\eqref{anh_th_deffkt}. %

The superpartners $\psi_k(z),\, k=1,\ldots,4$, of the four left-moving bosonic currents 
$j_k(z)$ (together with their analogs in the right-moving sector) 
are uncorrelated with ${\cal H}_{D_4 {\rm -torus}}$, i.e.\ they contribute a tensor factor
$\HHH_{\rm ferm}$ to the space of states $\HHH_{\rm ferm}\otimes{\cal H}_{D_4 {\rm -torus}}$ of our 
supersymmetric torus theory. 
As is explained in Appendix \ref{app:conv}, the U(1)-currents \eqref{U1SCA} of the left- and the right-moving 
$\NNN=2$ superconformal algebras in this model are obtained from $\HHH_{\rm ferm}$, such that
the decomposition of $\HHH_{\rm ferm}\otimes{\cal H}_{D_4 {\rm -torus}}$ into NS-NS and R-R sectors
in the usual sense is governed by $\HHH_{\rm ferm}$. So, for example, the R-R ground states of the 
full $D_4$-torus model come from the sector where the
$\psi_k,\, k=1,\ldots,4$, (and their right-moving counterparts) have Ramond boundary conditions, 
while $\psi_k,\, k=5,\ldots,12$, (and the corresponding right-movers)  have Neveu-Schwarz boundary conditions. 
Since there are four fermionic zero modes $\psi_{k,0},\, k=1,\ldots,4$, 
there are four left- and four right-moving ground states in this sector, which account for the sixteen R-R
ground states of the supersymmetric torus theory. These states have conformal weight $h=\qu h=\frac{1}{4}$.
\smallskip

The $\Z_2$-orbifolding described in Subsection \ref{geomD4}
acts as $\psi_k \mapsto -\psi_k$ for $k=1,\ldots,4$, while its action on the fermions
$\psi_k,\, k=5,\ldots, 12$, as established in Subsection \ref{torferm}, leaves 
$\psi_k,\, k=5,\ldots,8$, invariant and flips the sign of $\psi_k, k=9,\ldots,12$. 
The action on the right-moving fermions is identical.
Thus in the orbifold model, 
the twelve free (left-moving)\footnote{The right-moving fermions are always in the 
same sectors as the left-moving  ones.} fermions organise themselves naturally into three groups of four 
fermions with coupled boundary conditions
each, $(\,[\psi_1,\ldots,\psi_4], [\psi_5,\ldots,\psi_8], [\psi_9,\ldots,\psi_{12}]\,)$, where 
in the untwisted sector of the orbifold theory, also the 
boundary conditions  of the last two groups are coupled. 
The boundary conditions in this sector are therefore
\be
({\rm NS},{\rm NS},{\rm NS})\quad({\rm R}, {\rm NS}, {\rm NS})\quad({\rm NS}, {\rm R}, {\rm R})\quad ({\rm R}, {\rm R}, {\rm R} )\  .
\ee
We note that only eight of the sixteen R-R ground states of the $D_4$-torus model survive orbifolding, 
and as explained above they appear in the sector
$({\rm R}, {\rm NS}, {\rm NS})$.
\smallskip

In the twisted sector the roles of NS and R are reversed for the eight fermions that are affected by the 
orbifold, so that the boundary conditions  in the twisted sector are
\be
({\rm R}, {\rm NS}, {\rm R})\quad({\rm NS}, {\rm NS}, {\rm R})\quad
({\rm R}, {\rm R}, {\rm NS})\quad ({\rm NS}, {\rm R}, {\rm NS}) \ .
\ee
The sixteen twisted R-R ground states appear in the sectors 
$({\rm NS}, {\rm NS}, {\rm R})$ and 
$({\rm NS}, {\rm R}, {\rm NS})$.  The full partition function therefore equals
\begin{eqnarray}\label{orbipf}
Z_{\mathbb T_{D_4}/\Z_2}(\tau, z) 
&= &
{1\over2}\left(
\hf \sum_{k=2}^4 \left| {\vartheta_k(\tau)\over\eta(\tau)}\right|^{8}
+ \left| {\vartheta_3(\tau)\vartheta_4(\tau)\over\eta(\tau)^2}\right|^4\right.\\
&&\left.\qquad+ \left| {\vartheta_2(\tau)\vartheta_3(\tau)\over\eta(\tau)^2}\right|^4
+ \left| {\vartheta_2(\tau)\vartheta_4(\tau)\over\eta(\tau)^2}\right|^4
\right)\cdot {1\over2}\sum_{k=1}^4 \left| {\vartheta_k(\tau,z)\over\eta(\tau)}\right|^4\ .\nonumber
\end{eqnarray}
The R-R-sector with the inclusion of the total fermion number operator that only acts on the
fermions from $\HHH_{\rm ferm}$ contributes to this by
\begin{eqnarray}
Z_{\mathbb T_{D_4}/\Z_2}^{\wt{\rm R}}(\tau, z) 
&=& {1\over2}\left(
 \hf \sum_{k=2}^4 \left| {\vartheta_k(\tau)\over\eta(\tau)}\right|^{8}
\cdot \left| {\vartheta_1(\tau,z)\over\eta(\tau)}\right|^4
+ \left| {\vartheta_3(\tau)\vartheta_4(\tau)\over\eta^2(\tau)}\right|^4
\cdot \left| {\vartheta_2(\tau,z)\over\eta(\tau)}\right|^4 \right.\nonumber\\
&&
\left.\;\; +   
\left| {\vartheta_2(\tau)\vartheta_3(\tau)\over\eta^2(\tau)}\right|^4
\cdot \left| {\vartheta_4(\tau,z)\over\eta(\tau)}\right|^4 
+ 
\left| {\vartheta_2(\tau)\vartheta_4(\tau)\over\eta^2(\tau)}\right|^4
\cdot \left| {\vartheta_3(\tau,z)\over\eta(\tau)}\right|^4 
\right)\ .
\end{eqnarray}
By definition, the elliptic genus of any $\NNN=(2,2)$ SCFT  %that is invariant  under spectral flow 
is
\be\label{K3elldef}
\phi(\tau,z)
=\Tr_{\rm RR}
\left((-1)^{F_L+F_R}y^{J_0}\, q^{L_0-\frac{c}{24}}\bar q^{\qu L_0-\frac{\qu c}{24}}  \right)\ .
\ee
It can thus be obtained from $Z^{\wt{\rm R}}(\tau, z)$ by inserting $\qu y=1$ and
leaving $y$ untouched. Hence for our $\Z_2$-orbifold model,
\be\label{K3ellformula}
\phi(\tau,z)
= {2\vartheta_2(\tau,z)^2\vartheta_3(\tau)^2\vartheta_4(\tau)^2\over\eta(\tau)^6} 
+ {2\vartheta_4(\tau,z)^2\vartheta_2(\tau)^2\vartheta_3(\tau)^2\over\eta(\tau)^6}
+{2\vartheta_3(\tau,z)^2\vartheta_2(\tau)^2\vartheta_4(\tau)^2\over\eta(\tau)^6}\ ,
\ee
which agrees with the elliptic genus of K3, 
thus confirming that our orbifold model is indeed a K3 theory.

%%%%%%%%%%%
\section{ Properties of $\widehat{\mathfrak{su}}(2)_{L,1}^n\oplus \widehat{\mathfrak{su}}(2)_{R,1}^n$ RCFTs}
\label{su2fermionisation}

In this appendix, we collect some properties of the special toroidal models at
central charge $c=\qu c=n$ with $n\in\mathbb N$ which enjoy
an extended $\widehat{\mathfrak{su}}(2)_{L,1}^n\oplus \widehat{\mathfrak{su}}(2)_{R,1}^n$ symmetry.
%%%%%%%%%%%
%%%%%%%%%%%%%%%%
\subsection{Vertex operator construction for 
$\widehat{\mathfrak{su}}(2)_{L,1}^n\oplus \widehat{\mathfrak{su}}(2)_{R,1}^n$}
\label{appC1}
%%%%%%%%%%%%%%%%

For $n\in{\mathbb N}$,
the $\widehat{\mathfrak{su}}(2)_{L,1}^n\oplus \widehat{\mathfrak{su}}(2)_{R,1}^n$ affine algebra can be 
realised as a model of $n$ free bosons 
$Y^k(z,\bar z):=Y^k(z)+\qu Y^k(\qu z)$, $k=1,\ldots,n$, compactified on  
an $n$-dimensional real torus. The Cartan generators are chosen as
\be 
{J^{3,k}}(z) := \frac{i}{\sqrt{2}}\partial Y^k(z)\ ,\qquad  
{\overline  J^{3,k}}(\qu z) := \frac{i}{\sqrt{2}}\bar \partial Y^k(\qu z)\ ,\qquad k=1,\ldots, n\ ,
\ee 
in accord with \eqref{su2generators}.
Now consider any CFT with central charges $c=\qu c=n$ that possesses an
$\widehat{\mathfrak{su}}(2)_{L,1}^n\oplus \widehat{\mathfrak{su}}(2)_{R,1}^n$ current algebra.
As explained in Appendix~\ref{app:conv}, the remaining field content of the model is generated
by winding-momentum fields $V_{({\bf Q};{\bf\qu Q})}(z,\qu z)$ as in \eqref{vdef} with charge vectors
$({\bf Q};{\bf\qu Q})$ in the charge lattice $\Gamma\subset\R^{n,n}$ with quadratic form 
\eqref{quadraticform}, where $n=d$. The simple structure of the representations of $\wh{\mathfrak{su}}(2)_1$
as discussed in Subsection \ref{su21algebra} allows us to elucidate further the form of the
charge lattice $\Gamma$ for the  models with 
$\widehat{\mathfrak{su}}(2)_{L,1}^n\oplus \widehat{\mathfrak{su}}(2)_{R,1}^n$ current algebra.

For the vacuum representation $[0\cdots 0;\,0\cdots 0]$ of 
$\widehat{\mathfrak{su}}(2)_{L,1}^{n}\oplus \widehat{\mathfrak{su}}(2)_{R,1}^{n}$, 
the charge vectors take values in the lattice 
\be 
\Gamma_0 := \sqrt{2} (\Z^{n}\oplus\Z^{n})\subset \R^{n,n}\ .
\ee
More generally, the primary field for a representation 
$[a_1\cdots a_{n};b_1\cdots b_{n}]$ of 
$\widehat{\mathfrak{su}}(2)_{L,1}^{n}\oplus \widehat{\mathfrak{su}}(2)_{R,1}^{n}$ 
corresponds to $V_{\lambda}(z,\bar z)$ with
\be 
\lambda= \frac{1}{\sqrt{2}} (a_1,\ldots,a_{n}; b_1,\ldots,b_{n}) 
\in \Gamma_0^*{=} \frac{1}{\sqrt{2}}(\Z^{n}\oplus \Z^{n})\ ,
\ee 
and all the states in this representation have momenta in the translated lattice $\lambda +\Gamma_0$.

Consistency of our model requires that the charge lattice $\Gamma$ is an even integral selfdual lattice  with
\be\label{Gammas} 
\Gamma_0 \subset \Gamma \subset \Gamma_0^*\ .
\ee 
In order to include fermionic states in this description, one  drops the condition that the lattice 
$\Gamma$ is even and includes vectors $\lambda$ with odd $\lambda^2$. 
In this case, the $\Z_2$-graded locality condition for the vertex operators is satisfied by requiring \eqref{local},
where the factor $(-1)^{\lambda^2\mu^2}$ is trivial if $\Gamma$ is even, i.e.\ if the model  is purely bosonic. 
\medskip

Summarising, the spectrum of the theory is completely determined by specifying the 
charge lattice $\Gamma$ or, equivalently, the abelian group
\be 
{\mathcal A:=} \Gamma/\Gamma_0 \subset \Gamma_0^*/\Gamma_0\cong \Z_2^{n}\times \Z_2^{n}\ ,
\ee 
which, in the notation of  Subsection~\ref{su21algebra}, 
simply yields the subgroup ${\mathcal A} \subset \Z_2^{n}\times \Z_2^{n}$ describing 
the representation content of the model with respect to the  
$\widehat{\mathfrak{su}}(2)_{L,1}^{n}\oplus \widehat{\mathfrak{su}}(2)_{R,1}^{n}$ algebra. 

To understand the representation content of the theory more explicitly, 
recall that the  $\widehat{\mathfrak s\mathfrak u}(2)_k$-character
\be 
\ch_{k,\frac{a}{2}}(\tau,z):=\Tr_{[a]_k}(q^{L_0-\frac{c}{24}}\, y^{2J_0})\ ,\qquad a\in \{0,\ldots,k\}\ , 
\ee
of the highest weight representation $[a]$, $a=0,\ldots,k$, is given by
\be \label{su2levelk}
\ch_{k,s}(\tau,z)=\frac{\Theta_{2(k+2),2s+1}(\tau,z)-\Theta_{2(k+2),-2s-1}(\tau,z)}{\Theta_{4,1}(\tau,z)-\Theta_{4,-1}(\tau,z)}\ ,
\ee
where
\be
\Theta_{n,l}(\tau,z)=\sum_{m\in \Z} q^{\frac{n}{2}(m+\frac{l}{n})^2} y^{nm+l}\ .
\ee
In particular, in terms of the Jacobi theta functions discussed in Appendix \ref{Jacobigym},
\begin{align}
\ch_{1,0}(\tau,z)&=\Tr_{[0]}(q^{L_0-\frac{1}{24}}y^{J_0})=\frac{\sum_{n\in\Z} q^{n^2}y^{2n}}{\eta(\tau)}=\frac{\vartheta_3(2\tau,2z)}{\eta(\tau)}\ ,\label{chi10}\\
\ch_{1,\frac{1}{2}}(\tau,z)&=\Tr_{[1]}(q^{L_0-\frac{1}{24}}y^{J_0})=\frac{\sum_{n\in\Z} q^{(n+\frac{1}{2})^2}y^{2n+1}}{\eta(\tau)}=\frac{\vartheta_2(2\tau,2z)}{\eta(\tau)}\ \label{chi1hf}
\end{align}
are the building blocks of the characters that occur in an 
$\widehat{\mathfrak{su}}(2)_{L,1}^{n}\oplus \widehat{\mathfrak{su}}(2)_{R,1}^{n}$ theory. 
For example, if $n=1$, then 
\be
\Gamma_{\wh{\mathfrak{su}}(2)_1^2}
:=\left\{ \left. {1\over\sqrt2}(a;b) \right| a,\, b\in\Z,\quad a+b\equiv 0\mod 2\right\}
\ee
is the charge lattice of the only consistent 
$\widehat{\mathfrak{su}}(2)_{L,1}^1\oplus \widehat{\mathfrak{su}}(2)_{R,1}^1$ model,
and its partition function is
\be\label{su2parti}
\Tr  \left( y^{J_0} \qu y^{\qu J_0}\, q^{L_0-\frac{1}{24}}\bar q^{{\qu L_0}-\frac{1}{24}}  \right)
=
\left| {\vartheta_3(2\tau,2z)\over\eta(\tau)} \right|^2 + \left| {\vartheta_2(2\tau,2z)\over\eta(\tau)} \right|^2,
\ee
where $J_0,\;\qu J_0$ denote the zero modes of the $U(1)$-currents $2J^{3,1}(z),\; 2\qu J^{3,1}(\qu z)$,
respectively.

%%%%%%%%%%%%%%%%
\subsection{$S_6$ as a symmetry group of the
$\widehat{\mathfrak s\mathfrak u}(2)^6_{L,1}\oplus \widehat{\mathfrak s\mathfrak u}(2)^6_{R,1}$ RCFT}\label{appC2}
%%%%%%%%%%%%%%%%
In Subsection \ref{spectrum}, we  consider the spectrum of our K3 model in terms of representations of the 
$\widehat{\mathfrak s\mathfrak u}(2)^6_{L,1}\oplus \widehat{\mathfrak s\mathfrak u}(2)^6_{R,1}$  
affine algebra. According to \eqref{NSNS} and \eqref{supercurrents}, 
the spectrum is symmetric under a group $S_6$ that permutes simultaneously 
the various $\widehat{\mathfrak s\mathfrak u}(2)_1$ factors in the left and right sectors, 
and  this symmetry preserves the fusion rules. In this appendix, 
we prove that these transformations are compatible with the OPEs of the primary fields of the current 
algebra, and therefore define genuine symmetries of the CFT. 

In order to do so, we make use of the vertex operator construction of the 
$\widehat{\mathfrak s\mathfrak u}(2)^6_{L,1}\oplus \widehat{\mathfrak s\mathfrak u}(2)^6_{R,1}$ 
affine algebra given in Appendix \ref{appC1}. In this description, 
\eqref{bosgroup} and \eqref{fermgroup} imply that the charge lattice of both the 
bosonic and the fermionic NS-NS sector is
\be \label{su212lattice}
\Gamma_{\wh{\mathfrak{su}}(2)_1^{6} \oplus \wh{\mathfrak{su}}(2)_1^{6}}
=\Bigl( \bigcup_{v \in \m A_{\rm bos}} \frac{v}{\sqrt{2}}+\sqrt{2}(\Z^6\oplus\Z^6)\Bigr)\cup \Bigl(\bigcup_{v \in \m A_{\rm ferm}} \frac{v}{\sqrt{2}}+\sqrt{2}(\Z^6\oplus\Z^6)\Bigr)\ .
\ee 
A  basis for this integral, selfdual lattice is given by
\begin{align}
\alpha_1&=\frac{1}{\sqrt{2}}(1\,0\,0\,0\,0\,1;1\,0\,0\,0\,0\,1)\ , &
\beta_1&=\frac{1}{\sqrt{2}}(2\,0\,0\,0\,0\,0;0\,0\,0\,0\,0\,0)\ , \\
\alpha_2&=\frac{1}{\sqrt{2}}(0\,1\,0\,0\,0\,1;0\,1\,0\,0\,0\,1)\ ,&
\beta_2&=\frac{1}{\sqrt{2}}(0\,2\,0\,0\,0\,0;0\,0\,0\,0\,0\,0)\ ,\\
\alpha_3&=\frac{1}{\sqrt{2}}(0\,0\,1\,0\,0\,1;0\,0\,1\,0\,0\,1)\ ,&
\beta_3&=\frac{1}{\sqrt{2}}(0\,0\,2\,0\,0\,0;0\,0\,0\,0\,0\,0)\ ,\\
\alpha_4&=\frac{1}{\sqrt{2}}(0\,0\,0\,1\,0\,1;0\,0\,0\,1\,0\,1)\ ,&
\beta_4&=\frac{1}{\sqrt{2}}(0\,0\,0\,2\,0\,0;0\,0\,0\,0\,0\,0)\ ,  \\
\alpha_5&=\frac{1}{\sqrt{2}}(0\,0\,0\,0\,1\,1;0\,0\,0\,0\,1\,1)\ ,&
\beta_5&=\frac{1}{\sqrt{2}}(0\,0\,0\,0\,2\,0;0\,0\,0\,0\,0\,0)\ ,\\
\alpha_6&=\frac{1}{\sqrt{2}}(0\,0\,0\,0\,0\,2;0\,0\,0\,0\,0\,2)\ ,&
\xi&=\frac{1}{\sqrt{2}}(1\,1\,1\,1\,1\,1;0\,0\,0\,0\,0\,0)\ .
\end{align}

To construct the winding-momentum fields $V_\lambda(z,\qu z)$ with
$\lambda\in\Gamma_{\wh{\mathfrak{su}}(2)_1^{6}\oplus \wh{\mathfrak{su}}(2)_1^{6}}$, we need to specify a
cocycle $\epsilon$ satisfying \eqref{local}. One checks that
such a cocycle can be defined by
\begin{align}\label{epsilon}
\epsilon(\alpha_i,\alpha_j)&=+1, & 
\epsilon(\beta_k,\beta_l)&=+1, &
\epsilon(\alpha_i,\beta_k)&=(-1)^{\delta_{ik}}, \\
\epsilon(\beta_k,\alpha_i)&=+1, &
\epsilon(\alpha_i,\xi)&=+1, &
\epsilon(\beta_k,\xi)&=+1, \\
\epsilon(\xi,\alpha_i)&=-1,& 
\epsilon(\xi,\beta_k)&=-1 ,&
\epsilon(\xi,\xi)&=+1 ,
\end{align} 
together with the `linearity conditions'
\be\label{linear} 
\fa \lambda,\lambda',\mu,\mu'\in\Gamma_{\wh{\mathfrak{su}}(2)_1^{12}}\colon
\qquad\left\{
\begin{array}{rcl}
\epsilon(\lambda,\mu)&=&(-1)^{\lambda^2\mu^2} (-1)^{\lambda\cdot\mu} \, \epsilon(\mu,\lambda),\\[0.5em]
\epsilon(\lambda+\lambda',\mu)&=&\epsilon(\lambda,\mu)
\epsilon(\lambda',\mu)\ ,\\[0.5em] 
\epsilon(\lambda,\mu+\mu')&=&\epsilon(\lambda,\mu)
\epsilon(\lambda,\mu')\ .
\end{array}\right.
\ee 
Indeed, \eqref{epsilon} -- \eqref{linear} imply
\be
\epsilon(0,\mu) = 1 = \epsilon(\lambda,0),\qquad
\epsilon(\lambda,\mu) = \epsilon(-\lambda,\mu) = \epsilon(\lambda,-\mu)
\quad\text{ for all }\lambda,\,\mu\in\Gamma_{\wh{\mathfrak{su}}(2)_1^{6} \oplus \wh{\mathfrak{su}}(2)_1^{6}}\ ,
\ee
and thereby \eqref{local}.

The group of automorphisms of the lattice 
$\Gamma_{\wh{\mathfrak{su}}(2)_1^{6} \oplus \wh{\mathfrak{su}}(2)_1^{6}}$
contains the permutation group $S_6$, with $\pi\in S_6$ acting by
\be 
\pi({Q}_1,\ldots,{ Q}_6; {\qu Q}_1,\ldots,{\qu Q}_6)=
({ Q}_{\pi(1)},\ldots,{ Q}_{\pi(6)}; {\qu Q}_{\pi(1)},\ldots,{\qu Q}_{\pi(6)})
\quad \fa ({\bf Q};{\bf\qu Q})\in\Gamma_{\wh{\mathfrak{su}}(2)_1^{6} \oplus \wh{\mathfrak{su}}(2)_1^{6}} 
\ .
\ee  
Since this permutation leaves the cocycle invariant,
\be 
\fa\lambda,\mu\in\Gamma_{\wh{\mathfrak{su}}(2)_1^{6} \oplus \wh{\mathfrak{su}}(2)_1^{6}} \colon\qquad
\epsilon(\pi(\lambda),\pi(\mu))=\epsilon(\lambda,\mu)\ ,
\ee 
the obvious action on the vertex operators
\be 
\pi(J^{{a,k}}(z)):=J^{{a,\pi(k)}}(z)\ ,\qquad 
\pi({\qu J^{{a,k}}}(\bar z)):={\qu{J}^{{a,\pi(k)}}}(\bar z)\ ,
\qquad \pi(V_\lambda(z,\bar{z})):=V_{\pi(\lambda)}(z,\bar{z})\ 
\ee 
for all $a\in\{3,\,\pm\},\, k=1,\,\ldots,\,6,\, \lambda\in\Gamma_{\wh{\mathfrak{su}}(2)_1^{6} \oplus \wh{\mathfrak{su}}(2)_1^{6}}$,
defines also a symmetry of the OPE. The model furthermore has an 
${\rm SU}(2)_L^6\times {\rm SU}(2)_R^6$  symmetry, generated by the zero modes of the currents, 
which includes all the other symmetries induced by automorphisms of the lattice. Therefore, the OPE is 
preserved by an $({\rm SU}(2)_L^6\times {\rm SU}(2)_R^6) : S_6$ group of transformations.

%%%%%%%%%%%%%%%%
\subsection{Elliptic genus for the 
$\widehat{\mathfrak s\mathfrak u}(2)^6_{L,1}\oplus \widehat{\mathfrak s\mathfrak u}(2)^6_{R,1}$ RCFT}
\label{ellipticgenuscalculation}
%%%%%%%%%%%%%%%%

In this appendix,  we derive the elliptic genus of our model,
which we have already determined in \eqref{K3ellformula}, but this time using the 
$\widehat{\mathfrak s \mathfrak u}(2)_1$ description.

To do so, we need to determine those
R-R states that are BPS with respect to the right $\NNN=4$ superconformal algebra.
These states are all contained in the representations \eqref{RRg} -- \eqref{RRgh} together 
with the representations
\begin{align}
\label{RRgmin}& g\text{-twisted} & [01\,11\,11;\,10\,00\,00] && [10\,11\,11;\,01\,00\,00] \\
\label{RRhmin}& h\text{-twisted} & [11\,01\,11;\,00\,10\,00] && [11\,10\,11;\,00\,01\,00] \\
\label{RRghmin}& gh\text{-twisted} & [11\,11\,01;\,00\,00\,10] && [11\,11\,10;\,00\,00\,01] 
\end{align} 
obtained by fusion of \eqref{RRg} -- \eqref{RRgh} with the representation \eqref{supercurrents}.

With this preparation it is now straightforward to 
calculate the elliptic genus  by exploiting the ${\rm SU}(2)_L^6\times {\rm SU}(2)_R^6$ 
symmetry of the model.
We use the definition \eqref{K3elldef} of the elliptic genus $\phi(\tau,z)$, where
$J_0=2J_0^{3,1}$ with $J_0^{3,1}$
the zero mode of the current \eqref{Rsymm}  in the first factor of  the left affine algebra 
$\widehat{\mathfrak s\mathfrak u}(2)_{L,1}^6$. 
The elliptic genus $\phi(\tau,z)$
of our model is therefore the sum of the characters of  
$\widehat{\mathfrak s\mathfrak u}(2)_{L,1}^6$ for the representations \eqref{RRg} -- \eqref{RRgh} 
and \eqref{RRgmin} -- \eqref{RRghmin}, namely, using 
\eqref{chi10} and \eqref{chi1hf},
\begin{align}\label{ellgenus}
\phi(\tau,z)= &2\bigl(\ch_{1,\frac{1}{2}}(\tau,z)\ch_{1,0}(\tau,0)^5+5\bigr.
\ch_{1,0}(\tau,z)\ch_{1,\frac{1}{2}}(\tau,0)\ch_{1,0}(\tau,0)^4\nonumber\\
&\bigl. -\ch_{1,0}(\tau,z)\ch_{1,\frac{1}{2}}(\tau,0)^5-5\ch_{1,\frac{1}{2}}(\tau,z)\ch_{1,0}(\tau,0)
\ch_{1,\frac{1}{2}}(\tau,0)^4\bigr)\ .
\end{align} This formula exactly reproduces the elliptic genus of K3, as expected. Here, the sign 
$(-1)^{F_L+F_R}$ is positive for the representations \eqref{RRg} -- \eqref{RRgh},  and negative for 
\eqref{RRgmin} -- \eqref{RRghmin}.  The overall factor $2$ takes into account the fact that in the spectrum of the model, 
each of these representations of $\widehat{\mathfrak s\mathfrak u}(2)_{L,1}^6$ is tensored with two distinct right-moving 
states of conformal weight ${1\over4}$ that form a doublet under the diagonal  right-moving ${\rm SU}(2)$-symmetry.

\bigskip

Notice that only the twisted sectors of our $\Z_2\times\Z_2$-orbifold
contribute to the elliptic genus, since the untwisted R-R sector contains no BPS states 
(the right conformal weight of these states is at least ${3\over4}$). 
This is not in contradiction with the description of the theory in terms of a non-linear sigma model on 
the resolution of $\mathbb T_{D_4}/\Z_2$, because the untwisted sector in the $\Z_2$-orbifold 
of the $D_4$-torus model corresponds to the untwisted sector together with the $g$-twisted sector of the 
$\Z_2\times \Z_2$-orbifold of the free fermion theory describing the bosonic $D_6$-torus model.

%%%%%%%%%%%
\section{Fermionisation  of the supercharges}\label{appD}
%%%%%%%%%%%

%%%%%%%%%%%
\subsection{The supercharges in terms of the bosonic $D_6$-torus model}\label{appD1}
%%%%%%%%%%%
In Section~\ref{sec:su2} we construct our K3 model as a $\Z_2\times\Z_2$-orbifold of a 
free fermion model with $\widehat{\mathfrak{so}}(12)_{L,1}\oplus\widehat{\mathfrak{so}}(12)_{R,1}$ symmetry, i.e.\ 
as a $\Z_2\times\Z_2$-orbifold of a bosonic
$D_6$-torus model. This model
can be described in terms of the vertex operator construction for
the $\widehat{\mathfrak{su}}(2)_{L,1}^6\oplus\widehat{\mathfrak{su}}(2)_{R,1}^6$ RCFTs in 
Appendix~\ref{appC1} by means of a charge lattice
$\widetilde{\Gamma}\subset \Gamma_0^\ast$, see \eqref{Gammas}. 
A basis for $\wt\Gamma$ is given by
\begin{align}\label{gammahat1}
\gamma_1&=\frac{1}{\sqrt{2}}(1\,1\,0\,0\,0\,0;0\,0\,0\,0\,0\,0)\ , &
\gamma_2&=\frac{1}{\sqrt{2}}(1\, {-1}\,0\,0\,0\,0;0\,0\,0\,0\,0\,0)\ , \\
\gamma_3&=\frac{1}{\sqrt{2}}(0\,0\,1\,1\,0\,0;0\,0\,0\,0\,0\,0)\ ,&
\gamma_4&=\frac{1}{\sqrt{2}}(0\,0\,1\,{-1}\,0\,0;0\,0\,0\,0\,0\,0)\ ,\label{gammahat2}\\
\gamma_5&=\frac{1}{\sqrt{2}}(0\,0\,0\,0\,1\,1;0\,0\,0\,0\,0\,0)\ ,&
\gamma_6&=\frac{1}{\sqrt{2}}(0\,0\,0\,0\,1\, {-1};0\,0\,0\,0\,0\,0)\ ,\label{gammahat3}
\end{align} 
and similarly for the right-moving $\qu{\gamma}_1,\ldots,\qu{\gamma}_6$. 
The associated vertex operators correspond to the holomorphic free fermions
\begin{align}\label{chiV}
& \chi^*_{i}(z) = \widehat V_{\gamma_i}(z)\ ,  &
& \chi_{i}(z) = \widehat V_{-\gamma_i}(z)\ ,  
%& \chi^*_{2}(z) = \widehat V_{\gamma_2}(z)\ ,  &
%& \chi_{2}(z) = \widehat V_{-\gamma_2}(z)\ ,  \\
%& \bar\psi^{(1)}(z) = \widehat V_{\gamma_3}(z)\ ,  &
%& \psi^{(1)}(z) = \widehat V_{-\gamma_3}(z)\ ,  &
%& \bar\psi^{(2)}(z) = \widehat V_{\gamma_4}(z)\ , &
%& \psi^{(2)}(z) = \widehat V_{-\gamma_4}(z)\ ,  \\
%& \bar\eta^{(1)}(z) = \widehat V_{\gamma_5}(z)\ ,  &
%& \eta^{(1)}(z) = \widehat V_{-\gamma_5}(z)\ ,  &
%& \bar\eta^{(2)}(z) = \widehat V_{\gamma_6}(z)\ , &
%& \eta^{(2)}(z) = \widehat V_{-\gamma_6}(z)\ , 
\end{align} 
for $i=1,\ldots,6$, and analogously the right-moving fermions correspond to 
$\widehat V_{\pm{\qu\gamma_1}},\ \ldots,\ \widehat V_{\pm{\qu\gamma_6}}$. 
The lattice $\widetilde{\Gamma}$ is the orthogonal sum 
$\widetilde{\Gamma}_L\oplus \widetilde{\Gamma}_R$ of a purely `left-moving' (that is, 
${\bf\qu Q}=0$)  and a purely `right-moving' (${\bf Q}=0$) lattice. From now on, we will focus on the 
left-moving component $\widetilde{\Gamma}_L$.

\noindent
For a generic vector $\sum\limits_{i=1}^6 n_i \gamma_i\in \widetilde{\Gamma}_L$, we define the 
holomorphic vertex operator
\be 
\widehat V_{ n_1 \gamma_1+\cdots+n_6 \gamma_6}(z) :=
\nop{\widehat V_{n_1\gamma_1}(z)\,\widehat V_{n_2\gamma_2}(z)\,\cdots \,\widehat V_{n_6\gamma_6}(z)}\ ,
\ee 
where we set
\be \label{Vngamma}
\widehat V_{n_i\gamma_i}(z):=
\begin{cases}:\partial^{n_i-1}\chi^*_{i}(z)\,\cdots \partial\chi^*_{i}(z) \,\chi^*_{i}(z): & n_i>0\ ,\\
1 & n_i=0\ ,\\
:\partial^{-n_i-1}\chi_{i}(z)\,\cdots \partial\chi_{i}(z) \,\chi_{i}(z): & n_i<0\ .
\end{cases} 
\ee 
This definition amounts to a choice of phases whose compatibility with our previous
choices is ensured below in \eqref{fieldident} -- \eqref{phasedef} by implementing appropriate
phase factors $c(\lambda)$. The cocycle $\widetilde{\epsilon}$ determining the OPE of these fields is 
defined by
\be 
\widetilde{\epsilon}(\gamma_i, \gamma_j)
:=
\begin{cases}+1 & \text{for } i\le j\ ,\\
-1 & \text{for } i> j\ ,\end{cases}
\ee 
together with  linearity conditions analogous to \eqref{linear}. 
\smallskip

Let us make the connection between the fields of the bosonic $D_6$-torus model
and the fields of its $\Z_2\times \Z_2$-orbifold, i.e.\ the K3 model.
The holomorphic fields surviving the orbifold projection are the six currents 
$J^{3,k}(z)$, $k=1,\,\ldots,6$, and the winding-momentum fields $\widehat V_\lambda(z)$ for 
$\lambda\in \Gamma_{\wh{\mathfrak{su}}(2)_1^{6}\oplus\wh{\mathfrak{su}}(2)_1^{6}}\cap \widetilde\Gamma_L$
with $\Gamma_{\wh{\mathfrak{su}}(2)_1^{6}\oplus\wh{\mathfrak{su}}(2)_1^{6}}$ as in \eqref{su212lattice}. 
The latter fields are related to the fields $V_\lambda(z)$ in the K3 model by some field redefinition
\be\label{fieldident}
V_\lambda(z)=c(\lambda)\widehat V_\lambda(z)\ ,\qquad \lambda\in \Gamma_{\wh{\mathfrak{su}}(2)_1^{6}\oplus\wh{\mathfrak{su}}(2)_1^{6}}
\cap {\widetilde\Gamma_L}\ ,
\ee 
where $c(\lambda)\in \C^*$ satisfies
\be\label{soddi} 
\epsilon(\lambda,\mu)=\frac{c(\lambda)c(\mu)}{c(\lambda+\mu)}\, {\widetilde\epsilon}(\lambda,\mu)\ ,\qquad  
\lambda,\mu\in {\Gamma_{\wh{\mathfrak{su}}(2)_1^{6}\oplus\wh{\mathfrak{su}}(2)_1^{6}}} \cap {\widetilde\Gamma_L}\ ,
\ee  
with the cocycle $\epsilon$ given in \eqref{epsilon} -- \eqref{linear}.
A basis for $\Gamma_{\wh{\mathfrak{su}}(2)_1^{6}\oplus\wh{\mathfrak{su}}(2)_1^{6}} \cap {\widetilde\Gamma_L}$ is
\begin{align}
&\beta_1=\gamma_1+\gamma_2\ ,&
&\beta_2=\gamma_1-\gamma_2\ ,&
&\beta_3=\gamma_3+\gamma_4\ ,\label{labase1}\\
&\beta_4=\gamma_3-\gamma_4\ ,&
&\beta_5=\gamma_5+\gamma_6\ ,&
&\xi=\gamma_1+\gamma_3+\gamma_5\ ,\label{labase2}
\end{align}
and a choice for $c(\lambda)$ satisfying \eqref{soddi} is given by
\begin{align}\label{cchoice}
c(0)=1\ ,\qquad
&c(\lambda+2\beta_i)=c(\lambda),\quad i=1,\ldots,5\ ,&
&c(\lambda+2\xi)=c(\lambda)\ ,\\
&c(\lambda+\beta_1+\beta_2)=c(\lambda)\ ,&
&c(\lambda+\beta_3+\beta_4)=c(\lambda)\ ,&
\end{align}
for all $\lambda \in \Gamma_{\wh{\mathfrak{su}}(2)_1^{6}\oplus\wh{\mathfrak{su}}(2)_1^{6}}\cap {\widetilde\Gamma}$, as well as
\begin{align}
&c(\beta_1)=c(\beta_3)=c(\beta_5)=c(\xi)=i\ ,\\ 
&c(\beta_1+\beta_3)=c(\beta_3+\beta_5)=c(\beta_1+\beta_5)=-1\ ,\\
&c(\beta_1+\beta_3+\beta_5)=-i\ ,\\
&c(\xi+\beta_1)=c(\xi+\beta_3)=c(\xi+\beta_5)=1\ ,\\ 
&c(\xi+\beta_1+\beta_3)=c(\xi+\beta_3+\beta_5)=c(\xi+\beta_1+\beta_5)=-i\ ,\\
&c(\xi+\beta_1+\beta_3+\beta_5)=-1\ .\label{phasedef}
\end{align}
One checks that \eqref{cchoice} -- \eqref{phasedef} determine $c(\lambda)$ fully
for all $\lambda\in \Gamma_{\wh{\mathfrak{su}}(2)_1^{6}\oplus\wh{\mathfrak{su}}(2)_1^{6}}\cap {\widetilde\Gamma_L}$. 
Moreover, by means of the linearity conditions \eqref{linear} which hold for both $\epsilon$ and 
${\widetilde\epsilon}$, checking \eqref{soddi} amounts to proving
\be
c(\lambda+\mu)c(\lambda+\nu)c(\mu+\nu) = c(\lambda)c(\mu)c(\nu)c(\lambda+\mu+\nu)
\quad\text{ for all } \lambda,\,\mu,\,\nu\in\Gamma_{\wh{\mathfrak{su}}(2)_1^{6}\oplus\wh{\mathfrak{su}}(2)_1^{6}}\cap{\widetilde\Gamma_L}\ ,
\ee
an identity which is confirmed by direct calculation using \eqref{cchoice} -- \eqref{phasedef}.
Finally, introducing the shorthand notation 
\be 
V_{\pm\pm\pm\pm\pm\pm}(z):=V_{\frac{1}{\sqrt{2}}(\pm 1,\pm 1,\pm 1,\pm 1,\pm 1,\pm 1;0,0,0,0,0,0)}(z)
\ee
for the fields of weight $({3\over2},0)$,
we can express
the four holomorphic ${\cal N}=4$ supercurrents in terms of the lattice
$\widetilde\Gamma$. Using \eqref{G+chi} -- \eqref{Gprime-chi} along with 
\eqref{chiV} -- \eqref{Vngamma} and \eqref{fieldident}, we obtain
\begin{align}\label{superc1}
G^+(z)=\Bigl(\frac{{i-1}}{2}\Bigr)\Bigl[&V_{++++++}(z)+V_{++----}(z)-iV_{+-+-+-}(z)-
iV_{+--+-+}(z)\\
&+iV_{++++--}(z)+iV_{++--++}(z)+V_{+-+--+}(z)+V_{+--++-}(z)\Bigr]\notag ,\\
%&V_{-,-,-,-,-,-}-i V_{-,-,-,-,+,+}-i
%   V_{-,-,+,+,-,-}+V_{-,-,+,+,+,+}\\
%   &-i
%   V_{-,+,-,+,-,+}-V_{-,+,-,+,+,-}-V_{-,+,+,-,-,+}-i
%   V_{-,+,+,-,+,-}\\
\label{superc2}
G^-(z)=\Bigl(\frac{{-i-1}}{2}\Bigr)\Bigl[&V_{------}(z)+V_{--++++}(z)-iV_{-+-+-+}(z)-
iV_{-++-+-}(z)\\
&-iV_{----++}(z)-iV_{--++--}(z)-V_{-+-++-}(z)-V_{-++--+}(z)\Bigr]\notag \ ,
\end{align}
as well as 
\begin{align}
%&   -i V_{+,-,-,-,-,-}-V_{+,-,-,-,+,+}-V_{+,-,+,+,-,-}-i
%   V_{+,-,+,+,+,+}\\
%   &-V_{+,+,-,+,-,+}+i V_{+,+,-,+,+,-}+i
%   V_{+,+,+,-,-,+}-V_{+,+,+,-,+,-}\\
\label{superc3}
{G'}^+(z)=\Bigl(\frac{{-i-1}}{2}\Bigr)\Bigl[&V_{+-----}(z)+V_{+-++++}(z)-iV_{++-+-+}(z)-i
V_{+++-+-}(z)\\
&-iV_{+---++}(z)-iV_{+-++--}(z)-V_{++-++-}(z)-V_{+++--+}(z)\Bigr] \notag ,\\
%&   i V_{-,-,-,+,-,+}-V_{-,-,-,+,+,-}-V_{-,-,+,-,-,+}+i
%   V_{-,-,+,-,+,-}\\
%   &-V_{-,+,-,-,-,-}-i V_{-,+,-,-,+,+}-i
%   V_{-,+,+,+,-,-}-V_{-,+,+,+,+,+}\\
\label{superc4}
{G'}^-(z)=\Bigl(\frac{{-i-1}}{2}\Bigr)\Bigl[&iV_{-+++++}(z)+iV_{-+----}(z)+
V_{--+-+-}(z)+V_{---+-+}(z)\\
&-V_{-+++--}(z)-V_{-+--++}(z)+iV_{--+--+}(z)+iV_{---++-}(z)\Bigr]\notag .
\end{align}
%%%%%%%%%%%%%%%%%%%%%%%%%
\subsection{Free fermion model for the $D'_4$-torus model}\label{a:newtorus}
%%%%%%%%%%%%%%%%%%%%%%%%%

In the $D'_4$-torus model of Subsection \ref{s:newtorus}, 
obtained from orbifolding our K3 model by the group $\langle Q'\rangle$, with $Q'=Q_{2345}$, the fermions 
$\widehat \chi_1,\ldots,\widehat \chi_6$ are identified with different fields in the K3 model.
In terms of the lattice description of the
$\widehat{\mathfrak s \mathfrak u}(2)^6_{L,1}\oplus\widehat{\mathfrak s \mathfrak u}(2)^6_{R,1}$ RCFTs, 
we can identify
\begin{align}
&\widehat{ \chi}^*_{i}(z) = \widehat V_{{\wh{\gamma_i}}}(z)\ ,  &
& \widehat{\chi}_{i}(z) = \widehat V_{-{\wh{\gamma_i}}}(z)\ ,  
\end{align}
where
\begin{align}
{\wh{\gamma}_1}&=\frac{1}{\sqrt{2}}(1\,0\,0\,0\,0\,1;0\,0\,0\,0\,0\,0)\ , &
{\wh{\gamma}_2}&=\frac{1}{\sqrt{2}}(1\, 0\,0\,0\,0\, {-1};0\,0\,0\,0\,0\,0)\ , \\
{\wh{\gamma}_3}&=\frac{1}{\sqrt{2}}(0\,0\,1\,1\,0\,0;0\,0\,0\,0\,0\,0)\ ,&
{\wh{\gamma}_4}&=\frac{1}{\sqrt{2}}(0\,0\,1\,{-1}\,0\,0;0\,0\,0\,0\,0\,0)\ ,\\
{\wh{\gamma}_5}&=\frac{1}{\sqrt{2}}(0\,1\,0\,0\,1\,0;0\,0\,0\,0\,0\,0)\ ,&
{\wh{\gamma}_6}&=\frac{1}{\sqrt{2}}(0\,1\,0\,0\,{-1}\, 0;0\,0\,0\,0\,0\,0)\ ,
\end{align}
which is clearly different from \eqref{gammahat1} -- \eqref{gammahat3}. However, we can 
still proceed analogously to Appendix~\ref{appD1}.

The holomorphic fields $V_\lambda(z)$ of the K3 model are preserved by the orbifold projection, and 
are identified with the fields $\widehat V_\lambda(z)$ in the new torus model by
\be
V_\lambda(z)={\wh{c}}(\lambda)\, \widehat V_\lambda(z)\ ,\qquad \lambda\in 
\Gamma_{\wh{\mathfrak{su}}(2)_1^{6}\oplus\wh{\mathfrak{su}}(2)_1^{6}} \cap {\widetilde\Gamma_L}\ ,
\ee 
for a suitable phase ${\wh{c}}(\lambda)$. 
The basis \eqref{labase1} -- \eqref{labase2} for the lattice 
$\Gamma_{\wh{\mathfrak{su}}(2)_1^{6}\oplus\wh{\mathfrak{su}}(2)_1^{6}} \cap \widetilde\Gamma_L$ is given by
\begin{align}
&\beta_1=\wh{\gamma}_1+{\wh{\gamma}_2}\ ,&
&\beta_2={\wh{\gamma}_5}+{\wh{\gamma}_6}\ ,&
&\beta_3={\wh{\gamma}_3}+{\wh{\gamma}_4}\ ,\\
&\beta_4={\wh{\gamma}_3}-{\wh{\gamma}_4}\ ,&
&\beta_5={\wh{\gamma}_5}-{\wh{\gamma}_6}\ ,&
&\xi={\wh{\gamma}_1}+{\wh{\gamma}_3}+{\wh{\gamma}_5}\ ,
\end{align}
in terms of the vectors $\wh{\gamma}_i$, and ${\wh{c}}(\lambda)$ can be chosen such that
\begin{align}
{\wh c(0)=1},\quad
&{\wh{c}}(\lambda+2\beta_i)={\wh{c}}(\lambda),\quad i=1,\ldots,5\ ,&
&{\wh{c}}(\lambda+2{\xi})={\wh{c}}(\lambda)\ ,\\
&{\wh{c}}(\lambda+{\beta_2}+{\beta_5})={\wh{c}}(\lambda)\ ,&
&{\wh{c}}(\lambda+{\beta_3}+{\beta_4})={\wh{c}}(\lambda)\ ,&
\end{align}
for all $\lambda \in \Gamma_{\wh{\mathfrak{su}}(2)_1^{6}\oplus\wh{\mathfrak{su}}(2)_1^{6}}\cap {\widetilde\Gamma_L}$, as well as
\begin{align}
&{\wh{c}}({\beta_1})={\wh{c}}({\beta_2})={\wh{c}}({\beta_3})={\wh{c}}({\xi})=i\ ,\\ 
&{\wh{c}}({\beta_1}+{\beta_2})={\wh{c}}({\beta_1}+{\beta_3})={\wh{c}}({\beta_2}+{\beta_3})=-1\ ,\\
&{\wh{c}}({\beta_1}+{\beta_2}+{\beta_3})=-i\ ,\\
&{\wh{c}}({\xi}+{\beta_1})={\wh{c}}({\xi}+{\beta_2})={\wh{c}}({\xi}+{\beta_3})=1\ ,\\ 
&{\wh{c}}({\xi}+{\beta_1}+{\beta_2})={\wh{c}}({\xi}+{\beta_1}+{\beta_3})=
{\wh{c}}({\xi}+{\beta_2}+{\beta_3})=-i\ ,\\
&{\wh{c}}({\xi}+{\beta_1}+{\beta_2}+{\beta_3})=-1\ .
\end{align}
Consistency follows by  arguments  analogous to those already given in Appendix \ref{appD1}.
The supercharges are then 
\begin{align*}
G^+(z)=\Bigl(\frac{{i-1}}{2}\Bigr)& \Bigl[V_{++++++}(z)+V_{++----}(z)-iV_{+-+-+-}(z)-
iV_{+--+-+}(z)\\
&+iV_{++++--}(z)+iV_{++--++}(z)+V_{+-+--+}(z)+V_{+--++-}(z)\Bigr]\notag\\
=\Bigl(\frac{{i-1}}{2}\Bigr)& \Bigl[V_{{\xi}}(z)+V_{-{\xi}+{\beta_1}+{\beta_2}}(z)-iV_{-{\xi}+{\beta_1}+{\beta_3}+{\beta_5}}(z)-
iV_{{\xi}-{\beta_2}-{\beta_3}-{\beta_5}}(z)\\
& \hspace*{-0.3cm}+iV_{-{\xi}+{\beta_1}+{\beta_2}+{\beta_3}+{\beta_4}}(z)+iV_{{\xi}-{\beta_3}-{\beta_4}}(z)+V_{{\xi}-{\beta_2}-{\beta_4}-{\beta_5}}(z)+V_{-{\xi}+{\beta_1}+{\beta_4}+{\beta_5}}(z)\Bigr]\notag\\
=\Bigl(\frac{{i-1}}{2}\Bigr)& \Bigl[i\widehat V_{{\xi}}(z)-i\widehat V_{-{\xi}+{\beta_1}+{\beta_2}}(z)+i\widehat V_{-{\xi}+{\beta_1}+{\beta_3}+{\beta_5}}(z)-
i\widehat V_{{\xi}-{\beta_2}-{\beta_3}-{\beta_5}}(z)\\
&+\widehat V_{-{\xi}+{\beta_1}+{\beta_2}+{\beta_3}+{\beta_4}}(z)-\widehat V_{{\xi}-{\beta_3}-{\beta_4}}(z)+\widehat V_{{\xi}-{\beta_2}-{\beta_4}-{\beta_5}}(z)-\widehat V_{-{\xi}+{\beta_1}+{\beta_4}+{\beta_5}}(z)\Bigr]\notag\\
=\Bigl(\frac{{i-1}}{2}\Bigr)&\Bigl[i\widehat V_{{\wh{\gamma}_1}+{\wh{\gamma}_3}+{\wh{\gamma}_5}}(z)-i\widehat V_{{\wh{\gamma}_2}-{\wh{\gamma}_3}+{\wh{\gamma}_6}}(z)+i\widehat V_{{\wh{\gamma}_2}+{\wh{\gamma}_4}-{\wh{\gamma}_6}}(z)-
i\widehat V_{{\wh{\gamma}_1}-{\wh{\gamma}_4}-{\wh{\gamma}_5}}(z)\\
&+\widehat V_{{\wh{\gamma}_2}+{\wh{\gamma}_3}+{\wh{\gamma}_6}}(z)-\widehat V_{{\wh{\gamma}_1}-{\wh{\gamma}_3}+{\wh{\gamma}_5}}(z)+\widehat V_{{\wh{\gamma}_1}+{\wh{\gamma}_4}-{\wh{\gamma}_5}}(z)-\widehat V_{{\wh{\gamma}_2}-{\wh{\gamma}_4}-{\wh{\gamma}_6}}(z)\Bigr]\notag\\
=\Bigl(\frac{{i-1}}{2}\Bigr)&\Bigl[i\widehat{\chi}^*_{1}\widehat{\chi}^*_{3}\widehat{\chi}^*_{5}(z)-i\widehat{\chi}^*_{2}\widehat{\chi}_{3}\widehat{\chi}^*_{6}(z)+i\widehat{\chi}^*_{2}\widehat{\chi}^*_{4}\widehat{\chi}_{6}(z)-
i\widehat{\chi}^*_{1}\widehat{\chi}_{4}\widehat{\chi}_{5}(z)\\
&+\widehat{\chi}^*_{2}\widehat{\chi}^*_{3}\widehat{\chi}^*_{6}(z)-\widehat{\chi}^*_{1}\widehat{\chi}_{3}\widehat{\chi}^*_{5}(z)+\widehat{\chi}^*_{1}\widehat{\chi}^*_{4}\widehat{\chi}_{5}(z)-\widehat{\chi}^*_{2}\widehat{\chi}_{4}\widehat{\chi}_{6}(z)\Bigr]\notag\\
=\Bigl(\frac{{i-1}}{2}\Bigr)&\widehat{\chi}^*_{1}(z)\Bigl[i\widehat{\chi}^*_{3}\widehat{\chi}^*_{5}(z)-
i\widehat{\chi}_{4}\widehat{\chi}_{5}(z)-\widehat{\chi}_{3}\widehat{\chi}^*_{5}(z)+\widehat{\chi}^*_{4}\widehat{\chi}_{5}(z)\Bigr]\\
+\Bigl(\frac{{i-1}}{2}\Bigr)&\widehat{\chi}^*_{2}(z)\Bigl[-i\widehat{\chi}_{3}\widehat{\chi}^*_{6}(z)+i\widehat{\chi}^*_{4}\widehat{\chi}_{6}(z)
+\widehat{\chi}^*_{3}\widehat{\chi}^*_{6}(z)-\widehat{\chi}_{4}\widehat{\chi}_{6}(z)\Bigr] \ , \notag
\end{align*}
with similar expressions for the other supercharges.

\section{Theta function identities}\label{Jacobigym}

\subsection{Definitions and useful identities}
In this appendix we fix our conventions for the various modular functions that we shall use. 
We shall always use the parametrisation
$q:=e^{2\pi i\tau}$ and $y:=e^{2\pi iz}$. The Dedekind eta function is defined as 
\be
\eta(\tau):=q^{1/24}\prod_{n=1}^\infty\left(1-q^n\right) \ ,
\ee
while the Jacobi theta functions have product formula presentations of the form
\begin{align}\label{anh_th_deffkt}
\vartheta_1(\tau,z) & := 
i\sum_{n=-\infty}^\infty (-1)^n q^{{1\over2}(n-{1\over2})^2} y^{n-{1\over2}}
&=& iq^{{1\over8}} y^{-{1\over2}} \prod_{n=1}^\infty (1-q^n)(1-q^{n-1}y)(1-q^{n}y^{-1}),\nonumber\\
\vartheta_2(\tau,z) & := 
\sum_{n=-\infty}^\infty q^{{1\over2}(n-{1\over2})^2} y^{n-{1\over2}}
&=&q^{{1\over8}} y^{-{1\over2}} \prod_{n=1}^\infty (1-q^n)(1+q^{n-1}y)(1+q^{n}y^{-1}),\nonumber\\
\vartheta_3(\tau,z) & := 
\sum_{n=-\infty}^\infty q^{{n^2\over2}} y^{n}
&=&\prod_{n=1}^\infty (1-q^n)(1+q^{n-{1\over2}}y)(1+q^{n-{1\over2}}y^{-1}),
\\
\vartheta_4(\tau,z) & := 
\sum_{n=-\infty}^\infty (-1)^n q^{{n^2\over2}} y^{n}
&=&\prod_{n=1}^\infty (1-q^n)(1-q^{n-{1\over2}}y)(1-q^{n-{1\over2}}y^{-1}).    \nonumber
\end{align}
We always use the shorthand $\vartheta_k(\tau):=\vartheta_k(\tau,0)$, $k=1,\ldots,4$.
\subsection{Partition function of the $\langle g \rangle$-orbifold of our K3 model}\label{orbpart}
Using the $\wh{\mathfrak{su}}(2)_1$ characters \eqref{chi10}, \eqref{chi1hf}, 
one confirms that our K3-model has partition function 
\begin{eqnarray}
Z_{e,e}(\tau,z) &=& \left( {|\vartheta_3(2\tau,2z)|^2+|\vartheta_2(2\tau,2z)|^2\over |\eta(\tau)|^2} \right)
\cdot
\left( {|\vartheta_3(2\tau)|^2+|\vartheta_2(2\tau)|^2\over |\eta(\tau)|^2} \right)^5\nonumber\\
&=&
{1\over2} \left(\sum_{k=2}^4 \left| {\vartheta_k(\tau)\over\eta(\tau)}\right|^4\right)^2 \cdot
{1\over2} \sum_{k=1}^4 \left| {\vartheta_k(\tau,z)\over\eta(\tau)}\right|^4,
\end{eqnarray}
in agreement with \eqref{orbipf}. 

\renewcommand{\baselinestretch}{1.2}
Let $g$ denote the special symmetry \eqref{gspec} of our K3 model.
Since $Q=g^2$ is the quantum symmetry which reverses the usual $\Z_2$-orbifold of
the $D_4$-torus model, orbifolding our K3 model by $g^2$ yields
the partition function of the $D_4$-torus model obtained from \eqref{toruspartitionfunction} and 
\eqref{so(d)pf}:
{\footnotesize
\be\label{d4torusmodelpf}
{1\over2}\left( Z_{e,e}(\tau,z) + Z_{e,g^2}(\tau,z) + Z_{g^2,e}(\tau,z) + Z_{g^2,g^2}(\tau,z) \right)
=
{1\over2} \sum_{k=2}^4 \left| {\vartheta_k(\tau)\over\eta(\tau)}\right|^8 \cdot
{1\over2} \sum_{k=1}^4 \left| {\vartheta_k(\tau,z)\over\eta(\tau)}\right|^4.
\ee}
Using the action of $g$ on the $\wh{\mathfrak{su}}(2)_1$ characters
as in \eqref{insertions}, we have 
worked out the remaining twisted twining characters of the K3 sigma model. Explicitly
we find
{\footnotesize
\begin{eqnarray}
Z_{e,g}(\tau,z) &=& Z_{e,g^3}(\tau,z)\nonumber\\
&=&
\left( {|\vartheta_3(2\tau,2z)|^2+|\vartheta_2(2\tau,2z)|^2\over |\eta(\tau)|^2} \right)
\cdot
\left( {|\vartheta_3(2\tau)|^2-|\vartheta_2(2\tau)|^2\over |\eta(\tau)|^2} \right)
\cdot
\left( {\vartheta_4(2\tau)^2\qu{\vartheta_3(2\tau)}\over |\eta(\tau)|^2} \right)^4
\nonumber\\
&=& 
{1\over8}  
{\vartheta_3(\tau)^2\;\vartheta_4(\tau)^2\over |\eta(\tau)|^4}
\left(\qu{\vartheta_3(\tau)}^4+\qu{\vartheta_4(\tau)}^4 + 2\qu{\vartheta_3(\tau)}^2\;\qu{\vartheta_4(\tau)}^2\right) \\
&&\cdot 
{\left(-\vartheta_1(\tau,z)^2\;\qu{\vartheta_2(\tau,z)}^2
-\vartheta_2(\tau,z)^2\;\qu{\vartheta_1(\tau,z)}^2
+\vartheta_3(\tau,z)^2\;\qu{\vartheta_4(\tau,z)}^2
+\vartheta_4(\tau,z)^2\;\qu{\vartheta_3(\tau,z)}^2\right)\over |\eta(\tau)|^4},
\nonumber\\
Z_{g,e}(\tau,z) &=& Z_{g^3,e}(\tau,z)\nonumber\\
&=& 
{1\over8}  {\vartheta_3(\tau)^2\;\vartheta_2(\tau)^2\over |\eta(\tau)|^4}
\left(\qu{\vartheta_3(\tau)}^4+\qu{\vartheta_2(\tau)}^4 + 2\qu{\vartheta_3(\tau)}^2\;\qu{\vartheta_2(\tau)}^2\right)
\cdot\\
&&\cdot
{\left(\vartheta_1(\tau,z)^2\;\qu{\vartheta_4(\tau,z)}^2
+\vartheta_4(\tau,z)^2\;\qu{\vartheta_1(\tau,z)}^2
+\vartheta_3(\tau,z)^2\;\qu{\vartheta_2(\tau,z)}^2
+\vartheta_2(\tau,z)^2\;\qu{\vartheta_3(\tau,z)}^2\right)\over |\eta(\tau)|^4},
\nonumber\\
Z_{g,g}(\tau,z) &=& Z_{g^3,g^3}(\tau,z)\nonumber\\
&=& 
{1\over8}  
{\vartheta_2(\tau)^2\;\vartheta_4(\tau)^2\over |\eta(\tau)|^4}
\left(\qu{\vartheta_4(\tau)}^4-\qu{\vartheta_2(\tau)}^4 - 2i\qu{\vartheta_2(\tau)}^2\;\qu{\vartheta_4(\tau)}^2\right)
\cdot \\
&&\cdot
{\left(-\vartheta_1(\tau,z)^2\;\qu{\vartheta_3(\tau,z)}^2
+\vartheta_3(\tau,z)^2\;\qu{\vartheta_1(\tau,z)}^2
+\vartheta_4(\tau,z)^2\;\qu{\vartheta_2(\tau,z)}^2
-\vartheta_2(\tau,z)^2\;\qu{\vartheta_4(\tau,z)}^2\right)\over |\eta(\tau)|^4},
\nonumber\\
Z_{g,g^2}(\tau,z) &=& Z_{g^3,g^2}(\tau,z)\nonumber\\
&=& 
{1\over8}  
{\vartheta_2(\tau)^2\;\vartheta_3(\tau)^2\over |\eta(\tau)|^4}
\left(\qu{\vartheta_3(\tau)}^4+\qu{\vartheta_2(\tau)}^4 - 2\qu{\vartheta_3(\tau)}^2\;\qu{\vartheta_2(\tau)}^2\right)
\cdot\\
&&\cdot {\left(\vartheta_1(\tau,z)^2\;\qu{\vartheta_4(\tau,z)}^2
+\vartheta_4(\tau,z)^2\;\qu{\vartheta_1(\tau,z)}^2
+\vartheta_3(\tau,z)^2\;\qu{\vartheta_2(\tau,z)}^2
+\vartheta_2(\tau,z)^2\;\qu{\vartheta_3(\tau,z)}^2\right)\over |\eta(\tau)|^4},
\nonumber\\
Z_{g,g^3}(\tau,z) &=& Z_{g^3,g}(\tau,z)\nonumber\\
&=& 
{1\over8}  
{\vartheta_2(\tau)^2\;\vartheta_4(\tau)^2\over |\eta(\tau)|^4}
\left(\qu{\vartheta_4(\tau)}^4-\qu{\vartheta_2(\tau)}^4 + 2i\qu{\vartheta_2(\tau)}^2\;\qu{\vartheta_4(\tau)}^2\right)
\cdot\\
&&\cdot {\left(-\vartheta_1(\tau,z)^2\;\qu{\vartheta_3(\tau,z)}^2
+\vartheta_3(\tau,z)^2\;\qu{\vartheta_1(\tau,z)}^2
+\vartheta_4(\tau,z)^2\;\qu{\vartheta_2(\tau,z)}^2
-\vartheta_2(\tau,z)^2\;\qu{\vartheta_4(\tau,z)}^2\right)\over |\eta(\tau)|^4},
\nonumber\\
Z_{g^2,g}(\tau,z) &=& Z_{g^2,g^3}(\tau,z)\nonumber\\
&=& 
{1\over8}  
 {\vartheta_3(\tau)^2\;\vartheta_4(\tau)^2\over |\eta(\tau)|^4}
\left(\qu{\vartheta_3(\tau)}^4+\qu{\vartheta_4(\tau)}^4 - 2\qu{\vartheta_3(\tau)}^2\;\qu{\vartheta_4(\tau)}^2\right)
\cdot\\
&&\cdot{\left(-\vartheta_1(\tau,z)^2\;\qu{\vartheta_2(\tau,z)}^2
-\vartheta_2(\tau,z)^2\;\qu{\vartheta_1(\tau,z)}^2
+\vartheta_3(\tau,z)^2\;\qu{\vartheta_4(\tau,z)}^2
+\vartheta_4(\tau,z)^2\;\qu{\vartheta_3(\tau,z)}^2\right)\over |\eta(\tau)|^4}.
\nonumber
\end{eqnarray}}

\renewcommand{\baselinestretch}{1.2}
The $g$-orbifold of our K3-model can be obtained in two steps.
First, one performs the $\Z_2$-orbifold by $Q=g^2$ to recover the original $D_4$-torus
model with partition function \eqref{d4torusmodelpf}. Then, one performs another 
$\Z_2$-orbifold of the $D_4$-torus model, where the $\Z_2$-action is given by the symmetry $\qu g$
induced by $g$. The three non-trivial sectors of this $\Z_2$-orbifold  thus are

{\footnotesize
\begin{eqnarray}
\orbox{\qu g}{1}
&=&
{1\over2}\left( Z_{e,g}(\tau,z) + Z_{e,g^3}(\tau,z) + Z_{g^2,g}(\tau,z) + Z_{g^2,g^3}(\tau,z) \right)\nonumber\\
&=& 
{1\over4}  
{\vartheta_3(\tau)^2\;\vartheta_4(\tau)^2\over |\eta(\tau)|^4}
\left(\qu{\vartheta_3(\tau)}^4+\qu{\vartheta_4(\tau)}^4 \right)\cdot
\\
&&\cdot {\left(-\vartheta_1(\tau,z)^2\;\qu{\vartheta_2(\tau,z)}^2
-\vartheta_2(\tau,z)^2\;\qu{\vartheta_1(\tau,z)}^2
+\vartheta_3(\tau,z)^2\;\qu{\vartheta_4(\tau,z)}^2
+\vartheta_4(\tau,z)^2\;\qu{\vartheta_3(\tau,z)}^2\right)\over |\eta(\tau)|^4},
\nonumber\\
\orbox{1}{\qu g}
&=&
{1\over2}\left( Z_{g,e}(\tau,z) + Z_{g^3,e}(\tau,z) + Z_{g,g^2}(\tau,z) + Z_{g^3,g^2}(\tau,z) \right)\nonumber\\
&=& 
{1\over4}  
{\vartheta_2(\tau)^2\;\vartheta_3(\tau)^2\over |\eta(\tau)|^4}
\left(\qu{\vartheta_2(\tau)}^4+\qu{\vartheta_3(\tau)}^4 \right)\cdot
\\
&&\cdot {\left(\vartheta_1(\tau,z)^2\;\qu{\vartheta_4(\tau,z)}^2
+\vartheta_4(\tau,z)^2\;\qu{\vartheta_1(\tau,z)}^2
+\vartheta_3(\tau,z)^2\;\qu{\vartheta_2(\tau,z)}^2
+\vartheta_2(\tau,z)^2\;\qu{\vartheta_3(\tau,z)}^2\right)\over |\eta(\tau)|^4},
\nonumber\\
\orbox{\qu g}{\qu g}
&=&
{1\over2}\left( Z_{g,g}(\tau,z) + Z_{g^3,g^3}(\tau,z) + Z_{g,g^3}(\tau,z) + Z_{g^3,g}(\tau,z) \right)\nonumber\\
&=& 
{1\over4} 
{\vartheta_2(\tau)^2\;\vartheta_4(\tau)^2\over |\eta(\tau)|^4}
\left(\qu{\vartheta_4(\tau)}^4-\qu{\vartheta_2(\tau)}^4 \right)\cdot\\
&&\cdot 
{\left(-\vartheta_1(\tau,z)^2\;\qu{\vartheta_3(\tau,z)}^2
+\vartheta_3(\tau,z)^2\;\qu{\vartheta_1(\tau,z)}^2
+\vartheta_4(\tau,z)^2\;\qu{\vartheta_2(\tau,z)}^2
-\vartheta_2(\tau,z)^2\;\qu{\vartheta_4(\tau,z)}^2\right)\over |\eta(\tau)|^4} \ .
\nonumber
\end{eqnarray}}

\noindent We find that the full partition function  then is
{\footnotesize
\begin{eqnarray}
&&\hspace*{-2em}
{1\over 4}\sum_{a,b\in\{e,g,g^2,g^3\}} Z_{a,b}(\tau,z)\nonumber\\
&=&
\qu{\vartheta_1(\tau,z)}^2
\left[
{|\vartheta_2(\tau)|^2  \left( \vartheta_4(\tau,z)^2\;  \vartheta_3(\tau)^2
- \vartheta_3(\tau,z)^2\;  \vartheta_4(\tau)^2\right) \over4|\eta(\tau)|^6}\right.\nonumber\\
&&\left.
+ {|\vartheta_3(\tau)|^2\left( 
\vartheta_4(\tau,z)^2\;  \vartheta_2(\tau)^2
-\vartheta_2(\tau,z)^2\;  \vartheta_4(\tau)^2\right)\over4|\eta(\tau)|^6}
+ {|\vartheta_4(\tau)|^2\left( 
\vartheta_3(\tau,z)^2\;  \vartheta_2(\tau)^2
-\vartheta_2(\tau,z)^2\;  \vartheta_3(\tau)^2\right)\over4|\eta(\tau)|^6}  \right]\nonumber\\
&&+ \qu{\vartheta_2(\tau,z)}^2
\left[
{|\vartheta_2(\tau)|^2  \left( \vartheta_3(\tau,z)^2\;  \vartheta_3(\tau)^2
- \vartheta_4(\tau,z)^2\;  \vartheta_4(\tau)^2\right) \over4|\eta(\tau)|^6}\right.\nonumber\\
&&\left.
+ {|\vartheta_3(\tau)|^2\left( 
\vartheta_3(\tau,z)^2\;  \vartheta_2(\tau)^2
-\vartheta_1(\tau,z)^2\;  \vartheta_4(\tau)^2\right)\over4|\eta(\tau)|^6}
+ {|\vartheta_4(\tau)|^2\left( 
\vartheta_4(\tau,z)^2\;  \vartheta_2(\tau)^2
-\vartheta_1(\tau,z)^2\;  \vartheta_3(\tau)^2\right)\over4|\eta(\tau)|^6}  \right]\nonumber\\
&&+ \qu{\vartheta_3(\tau,z)}^2
\left[
{|\vartheta_2(\tau)|^2  \left( \vartheta_2(\tau,z)^2\;  \vartheta_3(\tau)^2
+ \vartheta_1(\tau,z)^2\;  \vartheta_4(\tau)^2\right) \over4|\eta(\tau)|^6}\right.\nonumber\\
&&\left.
+ {|\vartheta_3(\tau)|^2\left( 
\vartheta_4(\tau,z)^2\;  \vartheta_4(\tau)^2
+\vartheta_2(\tau,z)^2\;  \vartheta_2(\tau)^2\right)\over4|\eta(\tau)|^6}
+ {|\vartheta_4(\tau)|^2\left( 
\vartheta_4(\tau,z)^2\;  \vartheta_3(\tau)^2
-\vartheta_1(\tau,z)^2\;  \vartheta_2(\tau)^2\right)\over4|\eta(\tau)|^6}  \right]\nonumber\\
&&+ \qu{\vartheta_4(\tau,z)}^2
\left[
{|\vartheta_2(\tau)|^2  \left( \vartheta_1(\tau,z)^2\;  \vartheta_3(\tau)^2
+ \vartheta_2(\tau,z)^2\;  \vartheta_4(\tau)^2\right) \over4|\eta(\tau)|^6}\right.\nonumber\\
&&\left.
+ {|\vartheta_3(\tau)|^2\left( 
\vartheta_3(\tau,z)^2\;  \vartheta_4(\tau)^2
+\vartheta_1(\tau,z)^2\;  \vartheta_2(\tau)^2\right)\over4|\eta(\tau)|^6}
+ {|\vartheta_4(\tau)|^2\left( 
\vartheta_3(\tau,z)^2\;  \vartheta_3(\tau)^2
-\vartheta_2(\tau,z)^2\;  \vartheta_2(\tau)^2\right)\over4|\eta(\tau)|^6}  \right]\nonumber\\
%&\stackrel{\eqref{anh_th_vierpotlad}}{=}&
& = & 
{1\over2} \sum_{k=2}^4 \left| {\vartheta_k(\tau)\over\eta(\tau)}\right|^8 \cdot
{1\over2} \sum_{k=1}^4 \left| {\vartheta_k(\tau,z)\over\eta(\tau)}\right|^4 \ , \label{partfinal}
\end{eqnarray}
}

\noindent which again agrees with the partition function \eqref{d4torusmodelpf} of the $D_4$-torus model.
%%%%%%%%%%%
%%%%%%%%

\acknowledgments
We thank the Simons Center for Geometry and Physics at Stony Brook University and the organisers of the 
2013 programme on `Mock modular forms, moonshine, and string theory' for providing  an inspiring environment 
where this work was finalised. 
A.T., R.V.\ and K.W.\ also thank the Heilbronn Institute and the International Centre for Mathematical Sciences in Edinburgh 
as well as the organisers of the 2012 Heilbronn Day and Workshop on 
`Algebraic geometry, modular forms and applications to physics', where part of this work was done. 
The research of M.R.G.\ is partially supported by a grant from the Swiss National Science Foundation. 
A.T.\ acknowledges a Leverhulme Research Fellowship RF/2012-335, and
K.W.\ acknowledges an ERC Starting Independent Researcher Grant StG No. 204757-TQFT.

\end{document}